\documentclass[aps, prl, reprint, floatfix, superscriptaddress, twocolumn, showpacs]{revtex4-1}
\usepackage{color,graphicx,calc,url}
\usepackage{dcolumn}
\usepackage{bm}
\usepackage{amssymb}
\usepackage{amsmath}
\usepackage{wasysym}
\usepackage{mathrsfs} 

\usepackage{graphicx}
\usepackage[version=3]{mhchem}
\usepackage{siunitx}
\usepackage{comment}

\begin{document}

\title{Advanced data analysis procedure for hard x-ray resonant magnetic reflectivity\\ discussed for Pt thin film samples of various complexity}
\author{Jan Krieft}
\email{Electronic mail: jkrieft@physik.uni-bielefeld.de}
\affiliation{Center for Spinelectronic Materials and Devices, Department of Physics, Bielefeld University, Universit\"atsstra{\ss}e 25, 33615 Bielefeld, Germany}
\author{Dominik Graulich}
\affiliation{Center for Spinelectronic Materials and Devices, Department of Physics, Bielefeld University, Universit\"atsstra{\ss}e 25, 33615 Bielefeld, Germany}
\author{Anastasiia Moskaltsova}
\affiliation{Center for Spinelectronic Materials and Devices, Department of Physics, Bielefeld University, Universit\"atsstra{\ss}e 25, 33615 Bielefeld, Germany}
\author{Laurence Bouchenoire}
\affiliation{XMaS, The UK-CRG, European Synchrotron Radiation Facility (ESRF), 71 Avenue des Martyrs, 38043 Grenoble CEDEX 9, France}
\affiliation{Department of Physics, University of Liverpool, Oxford Street, Liverpool L69 7ZE, UK}
\author{Sonia Francoual}
\affiliation{Deutsches Elektronen-Synchrotron DESY, Notkestra{\ss}e 85, 22607 Hamburg, Germany}
\author{Timo Kuschel}
\affiliation{Center for Spinelectronic Materials and Devices, Department of Physics, Bielefeld University, Universit\"atsstra{\ss}e 25, 33615 Bielefeld, Germany}

\date{\today}


\begin{abstract}

X-ray resonant magnetic reflectivity (XRMR) is a powerful method to determine the optical, structural and magnetic depth profiles of a variety of thin films. Here, we investigate samples of different complexity all measured at the Pt $\text{L}_3$ absorption edge to determine the optimal procedure for the analysis of the experimental XRMR curves, especially for nontrivial bi- and multilayer samples that include differently bonded Pt from layer to layer. The software tool \textsc{ReMagX} is used to fit these data and model the magnetooptic depth profiles based on a highly adaptable layer stack which is modified to be a more precise and physically consistent representation of the real multilayer system. Various fitting algorithms, iterative optimization approaches and a detailed analysis of the asymmetry ratio features as well as $\chi^2$ (goodness of fit) landscapes are utilized to improve the agreement between measurements and simulations. We present a step-by-step analysis procedure tailored to the Pt thin film systems to take advantage of the excellent magnetic sensitivity and depth resolution of XRMR.
\end{abstract}

\maketitle

\section{Introduction}

Since the famous experiment by Lawrence and William Henry Bragg \cite{bragg1913reflection}, the analysis of condensed matter by x-ray radiation is one of the most important characterization techniques for any kind of solid states. Here, x-ray reflectivity (XRR) is a well-established powerful tool to determine layer thickness, analyze the density depth profile, and characterize the quality of interfaces \cite{holy1993x, tolan1999x, daillant2008x}. 

Nowadays, XRR is usually evaluated by a simulation and fit of the experimental data utilizing the recursive Parratt formalism \cite{parratt1954surface}. In this process, a multitude of structural and optical parameters have to be considered confronting the fitting algorithms with a wide parameter space. Therefore, the choice of an appropriate fitting algorithm is a nontrivial process since various non-global optimization algorithms only converge to local optima \cite{dane1998application}.

The modeling becomes even more complicated when we combine reflectivity techniques with the 
energy tuneability and polarization properties of synchrotron x-ray sources.
When tuned to the absorption edge energies, 
XRR is sensitive to the magnetic moments of the  selected element
due to the x-ray magnetic circular (XMCD) and linear dichroism influencing the intensity of the reflected and absorbed light \cite{van1986experimental, schutz1987absorption, chen1990soft, van1991strong, thole1992x, carra1993x, stohr1995x, stohr1999exploring}. This expands the parameter space into the magnetooptic regime, considering the atomic scattering as a function of the magnetic properties.

Combining XRR with XMCD creates a technique that unites the structural and optical depth profiles of the reflectivity experiment with the magnetic information of a given element, thus obtaining a highly accurate magnetooptic depth profile. X-ray resonant magnetic reflectivity (XRMR) measures the change of the specular reflection intensity when the direction of either the circular polarization or applied external magnetic field is reversed in order to obtain the spin depth profile of a material at a specific absorption edge \cite{tonnerre1998soft,geissler2001pt,lee2003x,roy2007evidence,macke2014magnetic,kuschel2015static,PhysRevB.93.214440}.

While XMCD only gives the mean polarization of a thin film, XRMR combines this magnetic information with conventional XRR depth profiling and therefore provides additional spatial resolution which is very high on the order of \SI{1}{\angstrom} \cite{seve1999profile,kuschelstatic}.
This gain in magnetic information comes at the expense of a more complex analysis since we add the magnetic dimension to the probed parameter space. For particularly complex systems, additional structural information, e.g., surface roughness may be necessary to accurately simulate a multilayer system, especially given that the roughness model is indistinguishable between smooth chemically diffuse or rough diffusion-free interfaces. The structural accuracy directly influences the modeling of the magnetic depth profile.
However, there is no standard procedure to analyze XRMR asymmetry ratios even though determining the magnetic depth profile of a thin film is essential for a deeper understanding of spin transport phenomena. These include for instance anisotropic magnetoresistance contributions in spin Hall magnetoresistance measurements \cite{althammer2013quantitative} or anomalous Nernst effects in spin Seebeck experiments \cite{huang2012transport, bougiatioti2017quantitative}.

Here, we present the spin polarization depth profile of Pt in different types of thin films investigated by interface-sensitive XRMR in the hard x-ray regime.
We use \textsc{ReMagX}, a software developed specifically to analyze XRMR curves \cite{ReMagX}, in order to fit specular reflectivity and asymmetry ratio curves and subsequently determine the magnetooptic depth profile. 
We show the steps to achieve a real best-fit 
for the XRR and XRMR scans of
three specific systems where Pt is integrated in the sample in an increasingly complex manner from simple bilayer over multi-layered stacks to Heusler compounds that include Pt.

First of all, 
we study a Pt/Fe//MgO bilayer system to illustrate the basic principles of the XRMR analysis describing the magnetic proximity effect (MPE) in Pt and highlight 
the pros and cons of the different modeling modes.
The detailed analysis of a second system, in which Pt is an integral part of a complex multilayer made of a single Pt layer adjacent to a Co layer that induces MPE, is discussed in Appendix C \cite{rowan2017interfacial,mukhopadhyay2019,moskaltsova2019}.
The third example is a PtMnSb thin film, a half-metallic alloy, which is an interesting candidate for studies of intrinsic spin-orbit torques \cite{krieft2017co}, with and without a single Pt layer on top.

Besides versatile fitting algorithms and optimization approaches to improve convergence between measured and simulated reflectivity data, we employ an advanced element specific fitting mode to simulate separate density and magnetic depth profiles. Finally, we outline a procedure to analyze features of the XRMR asymmetry ratios in great detail. This process turns out to be crucial for the precise identification of MPE or a magnetic dead layer of Pt based multilayer thin films and comparable systems.
In the following, we discuss common limitations
of the prevalent analysis procedures and expand the standard approach
to optimize the results of our various Pt $\text{L}_3$ XRMR measurements.

\section{Applicable steps and limitations in the refinement of the XRMR analysis}

The studies discussed in the following are examples of the current XRMR literature presenting possible solutions to improve the significance of the obtained results with regard to the study of the Pt specific magnetic depth profiles in bi- and multilayer systems. For instance,
many simulations of the specular reflectivity and asymmetry ratio can be found in literature which do not fit the data within the constraints set by the model or yield ambiguous results. The discrepancies are occasionally attributed to very noisy data in critical regions of the simulation \cite{geissler2002interplay}.

There are various approaches to challenge these methodical limitations usually resorting to additional degrees of freedom in the structural and magnetooptic simulation process or gathering more structural and magnetic information on the system. 
The biggest challenge is to find a modeling method to extract the magnetic depth profile unambiguously, which has been addressed in manifold ways usually explicitly tailored for specific samples so far.

In general, very few constraints or a vague model allow for the simulation of a perfect asymmetry ratio. However, the interpretation of the simulated magnetooptic parameters is often inconsistent and not convincing.
Depending on the system being investigated and the x-ray photon energy, different methods are thus utilized to counter those limitations and to exploit the excellent sensitivity of the XRMR technique to probe magnetic phenomena. 

Complex interfacial depth profiles are often accounted for by multi-slicing layers, a method which facilitates a perfect agreement between experiment and simulation at the expense of numerous free parameters to model interface transitions. Those simulated depth profiles should thus be interpreted very carefully regarding diffusion, hybridization, exchange interactions or other interface effects. As shown by Awaji \textit{et al.} \cite{awaji2007soft}, multiple depth profiles of magnetism have to be evaluated to rule out the implausible variations and to show the most probable solution. Another approach is to revert back to theoretical calculations to refine the specifications of the simulated magnetic depth profiles for an unambiguous result \cite{ederer2002theory}.

A reasonable first step to refine the fitting process is to determine the sample structure by fitting reflectivity measured at energies far from the resonance to be able to derive a magnetic depth profile from a fit of resonant data with constant structural parameters later on.
This supports the XRMR studies on thin films reported in the hard x-ray \cite{PhysRevB.93.214440} as well as in the soft x-ray regime \cite{bertinshaw2014element}.
Where applicable, the magnetization can be split and confined to separate interface layers \cite{blackburn2008pinned} similarly to the structural part. Thus, it can be modeled by an inhomogeneous distribution of magnetization throughout the magnetic layer via additional sublayers \cite{przybylski2012non}. In some studies, primarily the charge peaks (resonant reflections) or only the main features of the asymmetry ratio are taken into account \cite{haskel2001enhanced,gibert2016interlayer} to enable an adequate simulation.
 
Even a combination of these modeling steps does not guarantee a successful simulation of the magnetic depth profile since the information derived from simpler models may be insufficient to reproduce the observed magnetic reflectivity \cite{hosoito2014charge}. Performing polarized neutron reflectivity experiments \cite{felcher1987polarized,majkrzak1996neutron} can refine the simulated complex magnetization depth profiles in heterostructures \cite{roy2005depth,bjorck2009segregation} 
and provide results which are not even supported by one of the techniques separately \cite{kravtsov2009complementary}.
However, neutron reflectivity is only sensitive to the total magnetic moment and cannot distinguish between the magnetic moments of the different elements.
Furthermore, the yield of magnetic information by neutron reflectivity takes much more time, especially for small magnetic moments as in the case of interface magnetism or MPE.

As shown by Zafar \textit{et al.}, multiple measurements and simulations on one system should be performed to check how robust the simulated magnetic depth profile actually is, especially when XRMR is used to detect very small moments or tiny spin polarized layers \cite{zafar2011cr}.
This variety of complex approaches used to compensate for the limitations of the standard XRMR analysis procedures underlines the need for a consistent and reliable method, e.g., to study MPE samples of increasing complexity.
In the following, we establish a step-by-step procedure to obtain robust results of the Pt specific magnetic depth profiles in bi- and multilayer thin films.

\section{EXPERIMENTAL AND THEORETICAL DETAILS}

In our work, we focus on XRMR measurements performed on systems with increasing complexity but at the same absorption edge to facilitate the comparison.
The resonant XRMR scans were measured at a fixed photon energy slightly below the whiteline energy of the Pt $\text{L}_3$ absorption edge at the the maximum of the magnetic dichroism \cite{schutz1990spin, geissler2001pt, kuschel2015static, kuschelstatic}.
We prepared three different multilayers on $10 \times \SI{10}{mm^2}$ \ce{MgO}(001) and \ce{SiO_x}/\ce{Si}(001) substrates by magnetron sputter deposition in $\ce{Ar}^+$ atmosphere in the range of
\SI{3E-3}{mbar} in two different sputter deposition systems at Bielefeld University. 
The epitaxial Pt/Fe//MgO(001) bilayer and \ce{TaO_x}/\allowbreak\ce{MgO}/\allowbreak\ce{Ta}/\allowbreak\ce{Co}/\allowbreak\ce{Pt}//\allowbreak\ce{SiO_x}/\ce{Si}(001) multilayer
were deposited at room temperature. In contrast, the \ce{PtMnSb} thin films were prepared at high temperatures to achieve optimal growth without utilizing any seed layer \cite{krieft2017co} and were \textit{in situ} capped with either \ce{AlO_x}/\ce{MgO} or \ce{Pt}. 
The theoretical and experimental details regarding the XRMR measurements and beamline specifications can be found in Appendix A.

The goal of the magnetooptic analysis is to accurately extract the structural and magnetic depth profiles.
Therefore, the layer thickness and roughness are usually defined as free fitting parameters within realistic intervals while literature values are used to keep the optical constants fixed during the first step of the optimization process.
In case of the Pt/Fe bilayer, the off-resonant XRR measurements \SI{100}{eV} below the \ce{Pt} absorption edge were used to determine the correct structural properties which were in turn used to derive the optical constants at the resonance energy. Here, each element is assumed to have a homogeneous layer density based on the epitaxial growth \cite{krieft2017co} as well as a finite interface roughness which are determined far from the absorption edge.
This procedure detailed by Klewe \textit{et al.} \cite{PhysRevB.93.214440} is only applicable for simple and well-defined systems that are easy to model with high accuracy (e.g. bilayers with smooth interfaces).

For structurally more sophisticated systems or systems with a distinct level of inter-diffusion, interface roughness, oxidation or a general sample inhomogeneity, the structural parameters are not fully transferable. 
Based on experience, the thin film model is usually limited to the most reasonable parameters. However, these can be insufficient when dealing with structurally complex systems exhibiting various minor inhomogeneities which are not independently parameterized following the principle of Occam's razor. Here, the fitted parameters have to account for the inevitable shortcomings of the applied model. Generally speaking, all sample complexity beyond the model capability is compensated to a certain degree by minor adjustments of the structural and optical parameters. Since these best-fit approximations are not necessarily independent of the photon energy, the transferability of the structural parameters is not guaranteed.
In that case, the structural parameters should be first extracted by modeling the resonant XRR intensity $I$ and then used as input parameters in the XRMR analysis of the asymmetry ratio $\Delta I$. The exact process of calculating this ratio in the XRMR measurement is detailed in Appendix A.

The non-magnetic reflectivity is simulated as a function of the scattering vector $q = 4\pi/\lambda \sin(\theta)$ using the recursive Parratt algorithm \cite{parratt1954surface} and a N\'{e}vot-Croce \cite{nevot1980caracterisation} roughness model. 
This approach is fine to determine the structural and optical depth profiles, but not sufficient to obtain a fully accurate model of the asymmetry ratio. Therefore, we use a full matrix-based magnetooptic representation. The asymmetry ratio is simulated relying on a Zak matrix \cite{zak1990universal} formalism which simulates the roughness by adaptively slicing the interface into a series of segments.
As a result, the sample is divided into thin layers where each section has optical and magnetooptic properties corresponding to the density and magnetic depth profile spread vertically through the layer stack. A step-by-step guide to the basics of analyzing reflectivity measurements is presented in the review by Macke and Goering on magnetic reflectivity of heterostructures \cite{macke2014magnetic}.

The quality of the asymmetry ratio simulation is always determined by the sum of the squared error 
\begin{equation}
\chi^2 = \sum_i \left(A_{i,\text{meas.}} - A_{i,\text{sim.}} \right)^2
\end{equation}
of every asymmetry ratio data point $A_i$. The goodness of fit of the specular reflectivity is defined as
\begin{equation}
\chi^2 = \sum_i \left(\log I_{i,\text{meas.}} - \log I_{i,\text{sim.}} \right)^2,
\end{equation}
since the detected intensity $I_i$ decays by orders of magnitude as a function of $q$.
However, this method of least squares based on simple optimization is not sufficient to define a good fit.
Due to the often complex sample structure and variety of free parameters, $\chi^2$ usually has a multi-dimensional landscape with numerous local minima impeding simple downhill algorithms \cite{lagarias1998convergence}.
For more complex thin films with numerous structural parameters a second more sophisticated optimization algorithm is required.
Hence, the asymmetry ratio simulation and its corresponding $\chi^2$ landscape have to be analyzed in great detail to obtain reliable solutions for non-trivial magnetic depth profiles.
It is especially important to set reasonable constraints to the fitting parameters and to use the most probable starting configuration determined from the sample growth as well as a procedure sampling the complete configured parameter landscape like a heuristic algorithm. Therefore, aside from the standard Simplex optimization algorithm, an evolution approach based on a generic algorithm fitting routine \cite{tiilikainen2007genetic,tiilikainen2006nonlinear,ulyanenkov2005extended} was implemented in \textsc{ReMagX}.

\begin{figure*}[tb!]
\centering
\includegraphics[width=\linewidth]{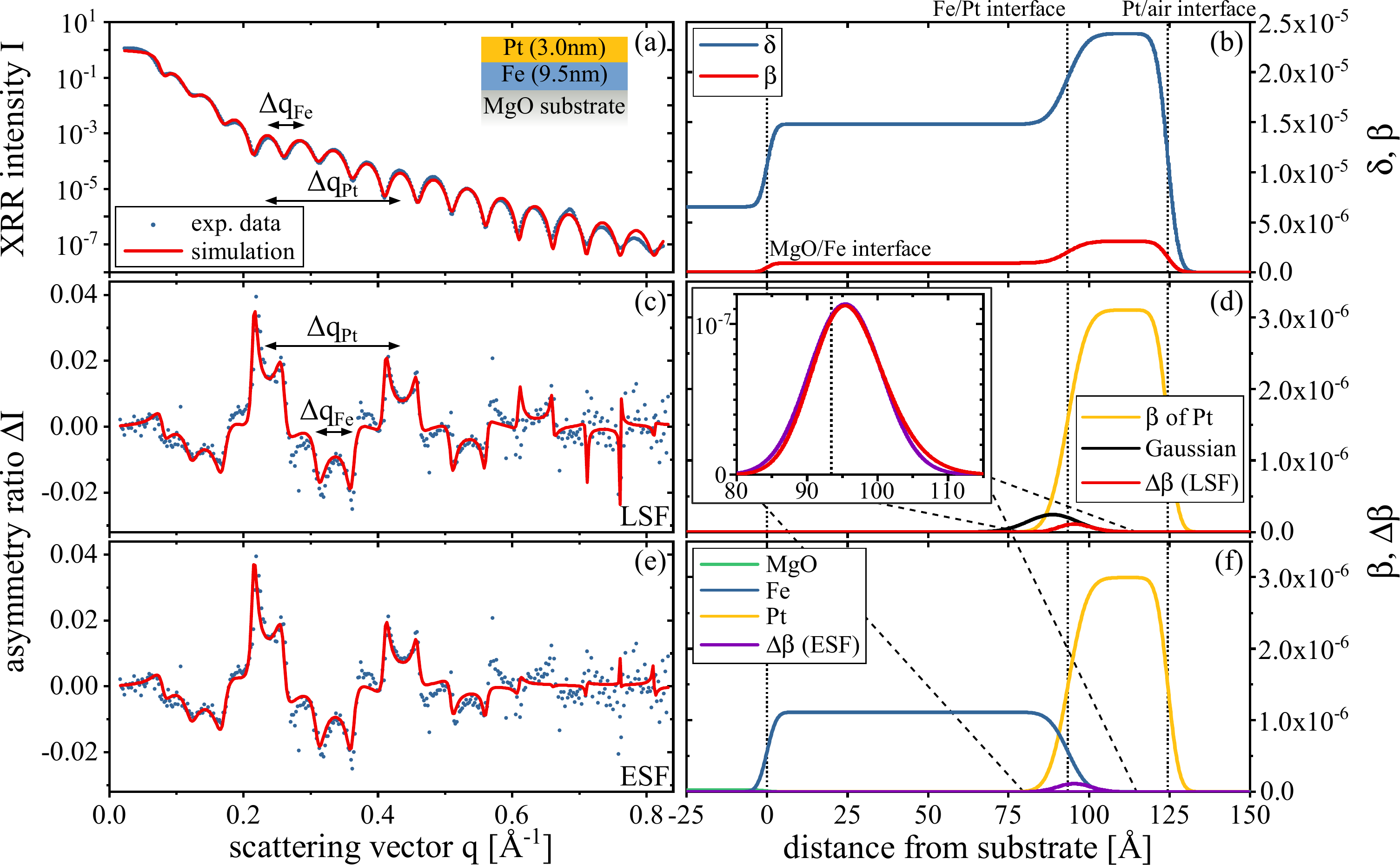}
\caption{(a) XRR measurement for \ce{Pt}(\SI{3.0}{nm})/\ce{Fe}(\SI{9.5}{nm})//\ce{MgO} (at \SI{11567.5}{eV}). (b) Optical $\delta$ and $\beta$ depth profiles used for the asymmetry ratio simulation. (c) Corresponding asymmetry ratio $\Delta I(q)$ and simulated data using the \textsc{ReMagX} LSF mode. (d)~Optical~$\beta$ and magnetooptic $\Delta\beta$ depth profile obtained with the LSF simulation. The $\Delta\beta$ depth profile is obtained from the displayed Gaussian function convolved with the $\beta$ depth profile and, thus, with the roughness of the Pt/Fe interface. (e)~Equivalent simulation of the asymmetry ratio $\Delta I(q)$ using the \textsc{ReMagX} ESF mode. (f) Optical $\beta$ depth profile for each element and the magnetooptic $\Delta\beta$ depth profile corresponding to the asymmetry ratio simulation in (e). Inset in (d): Close-up comparison of the $\Delta\beta$ depth profile obtained from LSF and ESF in \textsc{ReMagX}.}
\label{fig:1}
\end{figure*}

The procedure for modeling reflectivity curves of thin films is usually based on layers of different materials or compounds with a homogeneous density. Therefore, the modeling of thin films utilizes a specific list of the various layered materials or compounds, where thickness, roughness of the interfaces as well as the dispersion and absorption coefficients $\delta$ and $\beta$ are the defining parameters.
Beside this layer specific fitting (LSF) mode, \textsc{ReMagX} supports an advanced element specific fitting (ESF) mode to simulate separate density and magnetic depth profiles.
In this work, we combine the ESF mode with a detailed analysis
of the asymmetry ratio features and $\chi^2$ landscape to determine a global best-fit of the XRR and XRMR data
and create a consistent representation of the structural and magnetic depth profiles of our bi- and multilayer thin films.

Element specific simulations are based on separate depth profiles for each element with specific scattering factors \cite{hannon1988x}. This method enables us to model entirely different depth profiles for each element of a thin film or heterostructure. It can also be extended to describe a depth profile based on an exclusive magnetic scattering factor (no non-magnetic scattering contribution) and, thus, allows us to simulate a layer independent magnetization. Therefore, separate layer depth profiles are used for each element in the ESF mode with specific scattering factors $f_1$ and $f_2$. Here, the magnetic depth profile is modeled using a separate dummy element having the scattering factors set to zero and the magnetic scattering factor $f_\text{m} = f_{1\text{m}} + i f_{2\text{m}}$ providing the magnetooptic properties. 
For comparison, the asymmetry ratio is also fitted in the classical LSF mode, simulating the magnetooptic $\Delta\delta$ and $\Delta\beta$ depth profiles and specifying the magnetization dependent changes in the optical parameters of the refractive index.

When it comes to analyzing reflectivity data obtained for complex multilayers with structures possessing imperfect interfaces, significant contamination or oxidation, even this advanced ESF technique is still no guarantee for a successful simulation of the asymmetry ratio. Therefore, we suggest an extended analysis approach, including $\chi^2$ mapping, to avoid discrepancies within the results 
and to reduce the need for further measurements, which is a common problem in diverse XRMR studies of comparable thin film systems.

\section{RESULTS AND DISCUSSION}

In the following, we outline a systematic method to simulate XRMR asymmetry ratio data and achieve reliable XRMR simulations for our increasingly complex Pt based structures without the implementation of any additional layers, unrestricted fit parameters or supplementary measurements.

\subsection{I. Standard Pt/Fe bilayer}

In order to visualize the difference between the layer and independent element based simulation methods and to point out the strength of the ESF mode of \textsc{ReMagX}, we present a direct comparison of both simulations in Fig.~\ref{fig:1}. Here, we show the XRR intensity $I$ and corresponding asymmetry ratio $\Delta I$ of a \ce{Pt}/\ce{Fe} bilayer on \ce{MgO} measured at the \ce{Pt} $\text{L}_3$ edge. This is a prominent example of an MPE in a bilayer \cite{kuschel2015static}, for which an unambiguous asymmetry ratio is observed, and confirms the presence of spin polarization at the Pt/FM interface.
The specular reflectivity curve of this sample is shown in Fig.~\ref{fig:1}(a) along with the corresponding best-fit simulation.
They show Kiessig fringes \cite{kiessig1931untersuchungen, kiessig1931interferenz} with the expected oscillation length $\Delta q_\text{Fe}$ and faint overlaying additional Kiessig fringes with the oscillation length $\Delta q_\text{Pt}$. These determine the \ce{Pt} thickness to be \SI{3.0}{nm} and the \ce{Fe} thickness to be \SI{9.5}{nm} with a roughness of \SI{0.3}{nm} and \SI{0.5}{nm}, respectively, and a \ce{MgO} substrate roughness of \SI{0.2}{nm}.
Figures~\ref{fig:1}(c) and (e) present the magnetic asymmetry ratio with an amplitude of up to \SI{4}{\percent} and corresponding fits obtained by the different layer and element specific modes of \textsc{ReMagX}.

\begin{figure}[tb!]
\centering
\includegraphics[width=\linewidth]{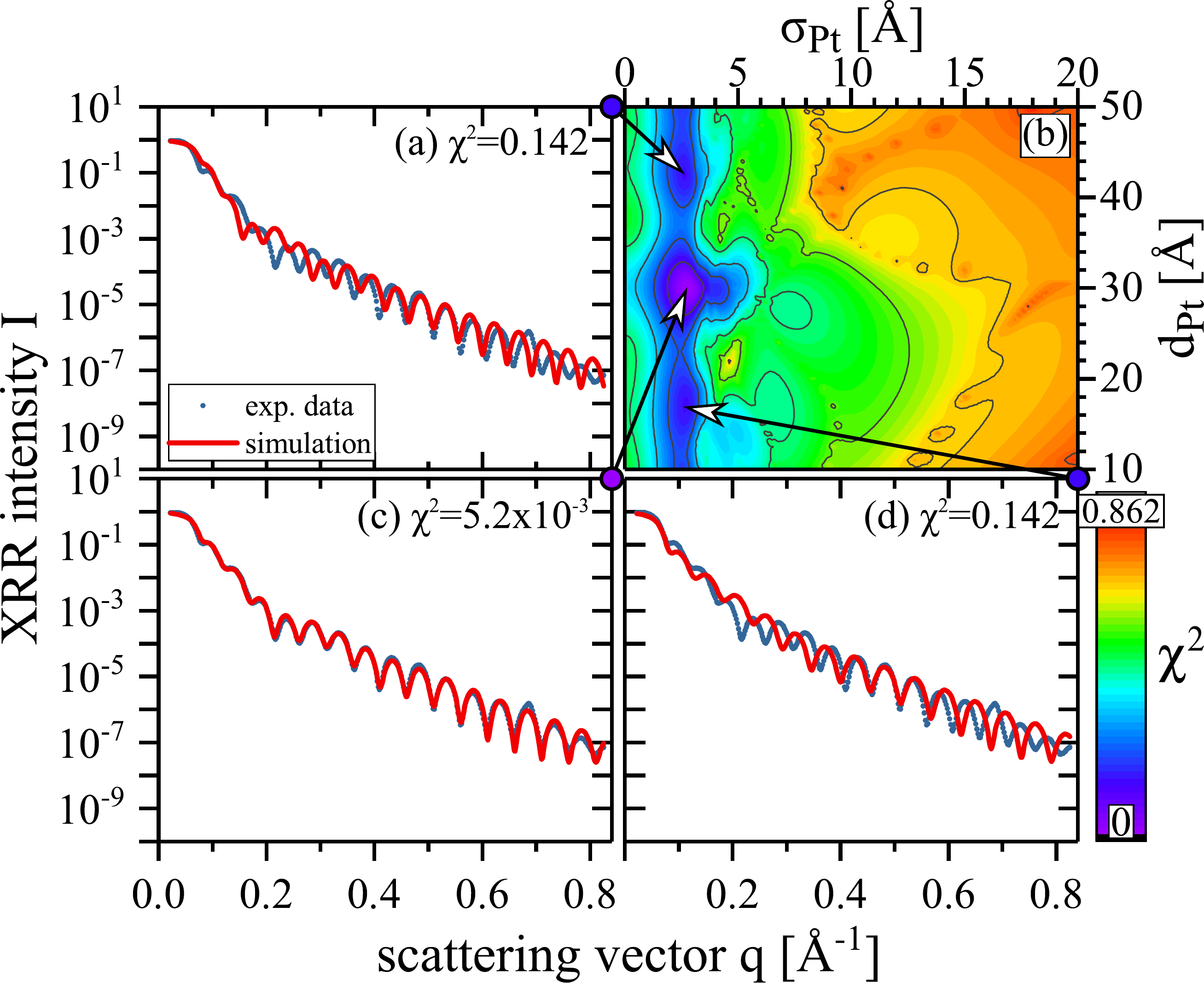}
\caption{XRR measurement of \ce{Pt}(\SI{3.0}{nm})/\ce{Fe}(\SI{9.5}{nm})//\ce{MgO} and various simulations of the XRR intensity $I$ for different thicknesses $d$ and roughnesses $\sigma$ of the Pt layer. (b) 2D mapping plot of $\chi^2$ depending on $d$ and $\sigma$ of the Pt layer. (c)~shows the optimal simulation, while (a) and (d) represent local minima in the $\chi^2$ landscape.}
\label{fig:2}
\end{figure} 

Figure~\ref{fig:1}(b) shows the optical depth profiles for the real $\delta$ and imaginary parts $\beta$ derived from the reflectivity simulation. This is fundamental for the simulation of the magnetic depth profile using the LSF mode, where the magnetooptic depth profile is derived from a convolution of a Gaussian function with the optical depth profile. Thus, we derive the $\Delta\delta$ and $\Delta\beta$ depth profiles at the Pt/Fe interface based on the optical depth profile of the atoms at the chosen absorption edge,
in this case \ce{Pt} \cite{PhysRevB.93.214440}. This is illustrated in Fig.~\ref{fig:1}(d), where the best-fit places the Gaussian function far into the \ce{Fe} layer, yet the convolution with the interface roughness of \ce{Pt} yields the reasonable $\Delta\beta$ depth profile close to the Pt/Fe interface as illustrated in the inset of Fig.~\ref{fig:1}(d).
Notably, the full width at half maximum (FWHM) of the Gaussian function is much larger than the FWHM of the resulting magnetic depth profile (compare black and red curve in Fig.~\ref{fig:1}(d)).
The corresponding fit of the asymmetry ratio perfectly simulates all main features despite an increasing noise level for higher scattering vectors $q$. 

We repeated the simulation using the ESF mode with separate element specific density depth profiles.
Here, we obtained an equivalent simulation 
and confirmed that this mode is as precise as the conventional LSF approach when modeling simple magnetic layers (see Fig.~\ref{fig:1}(f)), yet more flexible in the simulation of realistic magnetic depth profiles especially of more sophisticated magnetic systems. Since the magnetooptic parameter $\Delta\delta$ is close to zero at the absorption edge \cite{kuschel2015static}, we focus on the parameter $\Delta\beta$. This parameter is now modeled as an independent layer with its own thickness and interface roughness. So it has to be constricted to the sample structure to model a realistic spin polarized layer at the interface. In the ESF mode, this is done by assuming that the magnetic depth profile adapts the structural interface roughness between the non-magnetic metal and the ferromagnet layer, which is therefore significantly affecting the magnetic coupling \cite{PhysRevB.93.214440, kim2016asymmetric}. 

\begin{figure}[tb!]
\centering
\includegraphics[width=\linewidth]{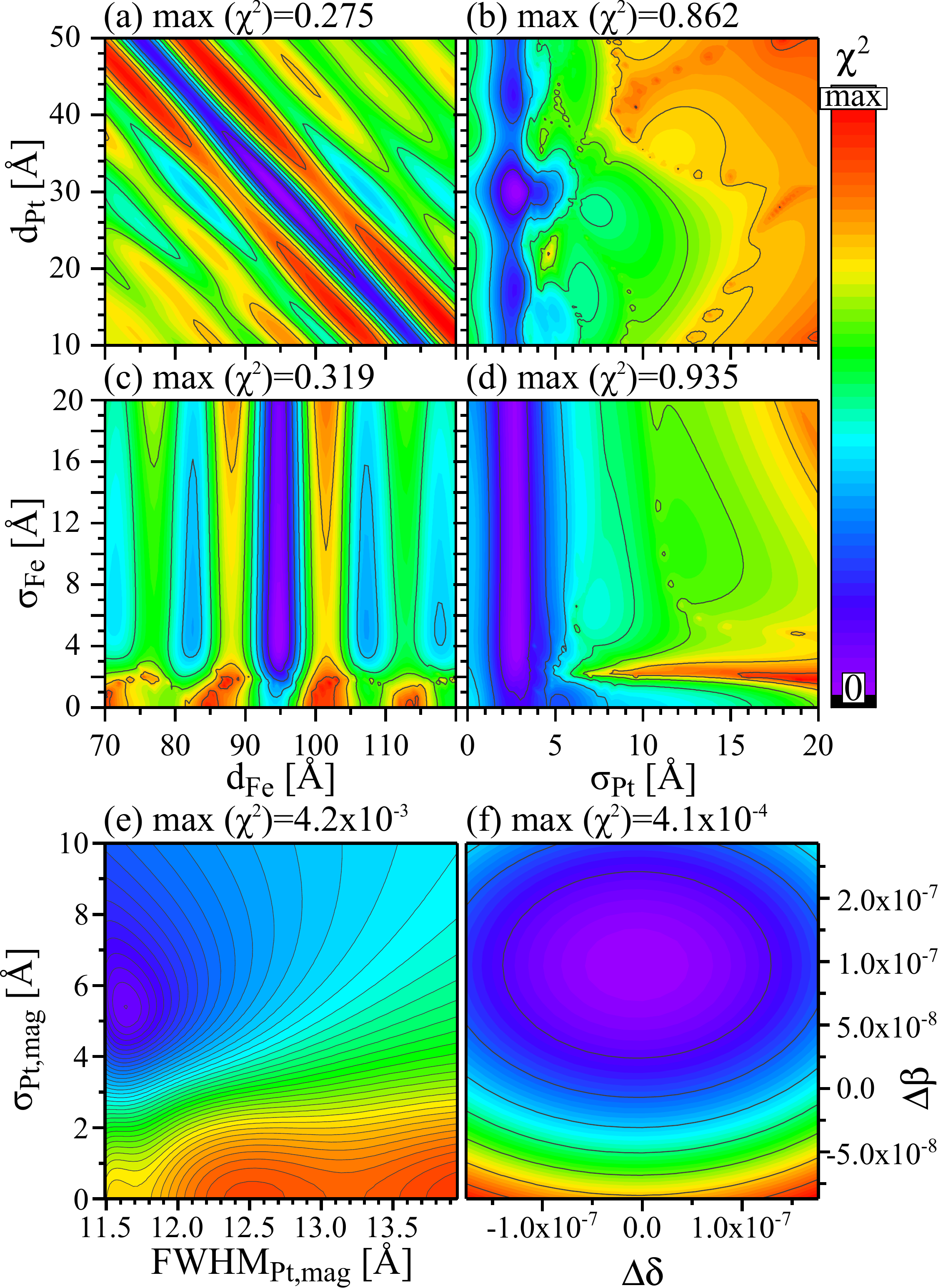}
\caption{Selected 2D maps of the $\chi^2$ value for the XRR fits of \ce{Pt}(\SI{3.0}{nm})/\ce{Fe}(\SI{9.5}{nm})//\ce{MgO}. (a) $\chi^2$ map plot varying the thickness of the \ce{Fe} layer $d_\text{Fe}$ and the thickness of the \ce{Pt} layer $d_\text{Pt}$. (b) $\chi^2$ map plot varying the roughness $\sigma_\text{Pt}$ and $d_\text{Pt}$. (c) $\chi^2$ map plot varying $d_\text{Fe}$ and the roughness of the \ce{Fe} layer $\sigma_\text{Fe}$. (d) $\chi^2$ map plot varying $\sigma_\text{Pt}$ and $\sigma_\text{Fe}$.
(e)~2D plot for the magnetooptic fits of the XRMR asymmetry ratios. The FWHM of the magnetic depth profile $\text{FWHM}_\text{Pt,mag}$ is varied as well as the roughness $\sigma_\text{Pt,mag}$. (f) Corresponding 2D $\chi^2$ plot of the magnetooptic parameters $\Delta\delta$ vs. $\Delta\beta$.}
\label{fig:3}
\end{figure}  

Simulating a magnetic depth profile based on a discrete layer adapting the roughness of the \ce{Pt} density depth profile results in a magnetooptic $\Delta\beta$ depth profile very similar to that obtained using the LSF mode by convolving the Gaussian function with the optical depth profile.
Both results are directly compared in the inset of Fig.~\ref{fig:1}(d), revealing a minor increase in $\Delta\beta$ on the right flank of the corresponding depth profile ($100$ - \SI{115}{\angstrom} from the substrate) derived in the LSF mode in relation to the ESF solution. Overall, the ESF simulation of the asymmetry ratio only shows slight variations from the LSF result constituting an equivalent best-fit quality.
Hence, one should consider utilizing the ESF mode to benefit from the versatility in modeling complex compounds and multilayer structures as well as independent magnetic depth profiles.

We can calculate and analyze the multi-dimensional $\chi^2$  landscape to optimize our fitting routine and determine a true global best-fit. Since this is a problem scaling with the number of free parameters in our simulation, we focus on a simple fit of the XRR intensity in the ESF mode and two specific parameters for illustration. In Fig.~\ref{fig:2}, the simulation of the \ce{Pt}/\ce{Fe} bilayer reflectivity is mapped onto the $\chi^2$ landscape determined as a function of the thickness $d$ and roughness $\sigma$ of the \ce{Pt} layer. The XRR best-fit at the global minimum is compared to two adjacent fits in local minima clearly showing a deviation from the experimental Kiessig fringes 
because of a variation in thickness. Due to the nature of the oscillatory pattern, we see multiple local minima at fixed multiples of $d_\text{Pt}$. In general, higher values of \ce{Pt} roughness $\sigma_\text{Pt}$ are accompanied by increasing $\chi^2$ effective errors. A downhill algorithm operating on this simple map could easily fit the reflectivity if all other parameters are already determined. Naturally, the multi-dimensional $\chi^2$ landscape is highly complex when we vary the parameters outlined above to simulate a real multilayer system.

Selected other 2D maps within the same multi-dimensional landscape of $\chi^2$ are presented in Fig.~\ref{fig:3}.
These maps serve to illustrate the principles of a detailed analysis of the $\chi^2$ landscape. Although the global minimum is pronounced in every displayable combination, we see many local minima demanding heuristic algorithms.
The values of $\chi^2$ in Fig.~\ref{fig:3}(a) are determined by varying the \ce{Fe} and \ce{Pt} thickness parameters $d_\text{Fe}$ and $d_\text{Pt}$ which induce a combined oscillatory character. 
This results in a striped oscillatory pattern rotated by \SI{45}{\degree} relative to the coordinate axes. 
Here, the pattern is defined by multiples of $d_\text{Fe}$ and $d_\text{Pt}$ evenly arranged with regard to the best-fit parameters in the center of the map (see Appendix B).

In contrast, the maps generated by varying the thickness in one dimension and the corresponding roughness in the other show an interference like pattern.
The map displayed in Fig.~\ref{fig:3}(b) was analyzed in Fig.~\ref{fig:2} and is included here for the sake of completeness.
Figure~\ref{fig:3}(c) illustrates the map as a function of the \ce{Fe} layer thickness $d_\text{Fe}$ and the \ce{Fe} roughness $\sigma_\text{Fe}$ while Fig.~\ref{fig:3}(d) is a function of both roughnesses $\sigma_\text{Pt}$ and $\sigma_\text{Fe}$.
The run through the $d_\text{Fe}$ dimension of the $\chi^2$ map reveals the established oscillatory pattern while higher values of \ce{Pt} roughness $\sigma_\text{Pt}$ again lead to increasing $\chi^2$ values.
In both maps of Figs.~\ref{fig:3}(c) and (d), the roughness of the upper \ce{Fe} interface $\sigma_\text{Fe}$ yields particularly broad minima along this parameter dimension which indicates a higher level of uncertainty regarding this value in the simulation.
Therefore, $\sigma_\text{Fe}$ must be confirmed by secondary measurements or a detailed check of $\chi^2$ to restrict this parameter to realistic values. 
This example illustrates how important a detailed analysis of the $\chi^2$ landscape is to determine the real best-fit. 

Moreover, the simulation of asymmetry ratios can be equally sophisticated.
The $\chi^2$ maps defined by the simulation of the asymmetry ratio and thus the magnetooptic depth profiles are presented at the bottom of Fig.~\ref{fig:3}. 
The $\chi^2$ landscape of map Fig.~\ref{fig:3}(e) is generated by a variation of the magnetic depth profile roughness $\sigma_\text{Pt,mag}$ in relation to the FWHM of the magnetic depth profile $\text{FWHM}_\text{Pt,mag}$. This effective thickness of the magnetic Pt depends on the thickness $d_\text{Pt,mag}$ of the initially used magnetic layer that becomes the magnetic depth profile $\Delta\beta$ of Fig.~\ref{fig:1}(f) when the magnetic roughness $\sigma_\text{Pt,mag}$ is taken into account. Here, the dependence is $\text{FWHM}_\text{Pt,mag} = \SI{0.0198}{nm^{-1}} \cdot d^2_\text{Pt,mag} + 0.0476 \cdot d_\text{Pt,mag} + \SI{11.5}{nm}$ for \ce{Pt}(\SI{3.0}{nm})/\ce{Fe}(\SI{9.5}{nm})//\ce{MgO}.
The map shows a distinct global minimum and no local minima which allows us to obtain an optimal solution in this simple case by utilizing a basic downhill fitting algorithm. This approach is equally applicable when the magnetooptic parameters $\Delta\delta$ and $\Delta\beta$ are determined in the asymmetry ratio simulation since the 2D $\chi^2$ map Fig.~\ref{fig:3}(f) illustrates a global minimum best described as a $\chi^2$ well.
In general, it is particularly important to check multiple $\chi^2$ maps, as presented in Fig.~\ref{fig:3}, when there are local minima close to the global best-fit. These minima usually represent good fits yet semi-optimal solutions which can impede optimal simulations of the asymmetry ratio established on the corresponding structural parameters.
 
A similar analysis of a structurally more complex system still involving an interface of \ce{Pt} adjacent to a $3d$ ferromagnet is presented in Appendix C. This sample is particularly challenging due to a partially inhomogeneous multilayer structure and intermixed oxide based capping which has a significant influence on the reflectivity curve. A combination of fitting algorithms and an in-depth examination of the $\chi^2$ maps is presented in order to identify the global best-fit minimum and ultimately the MPE based magnetic depth profile within the buried Pt layer \cite{moskaltsova2019}.

\begin{figure}[!tb]
\centering
\includegraphics[width=\linewidth]{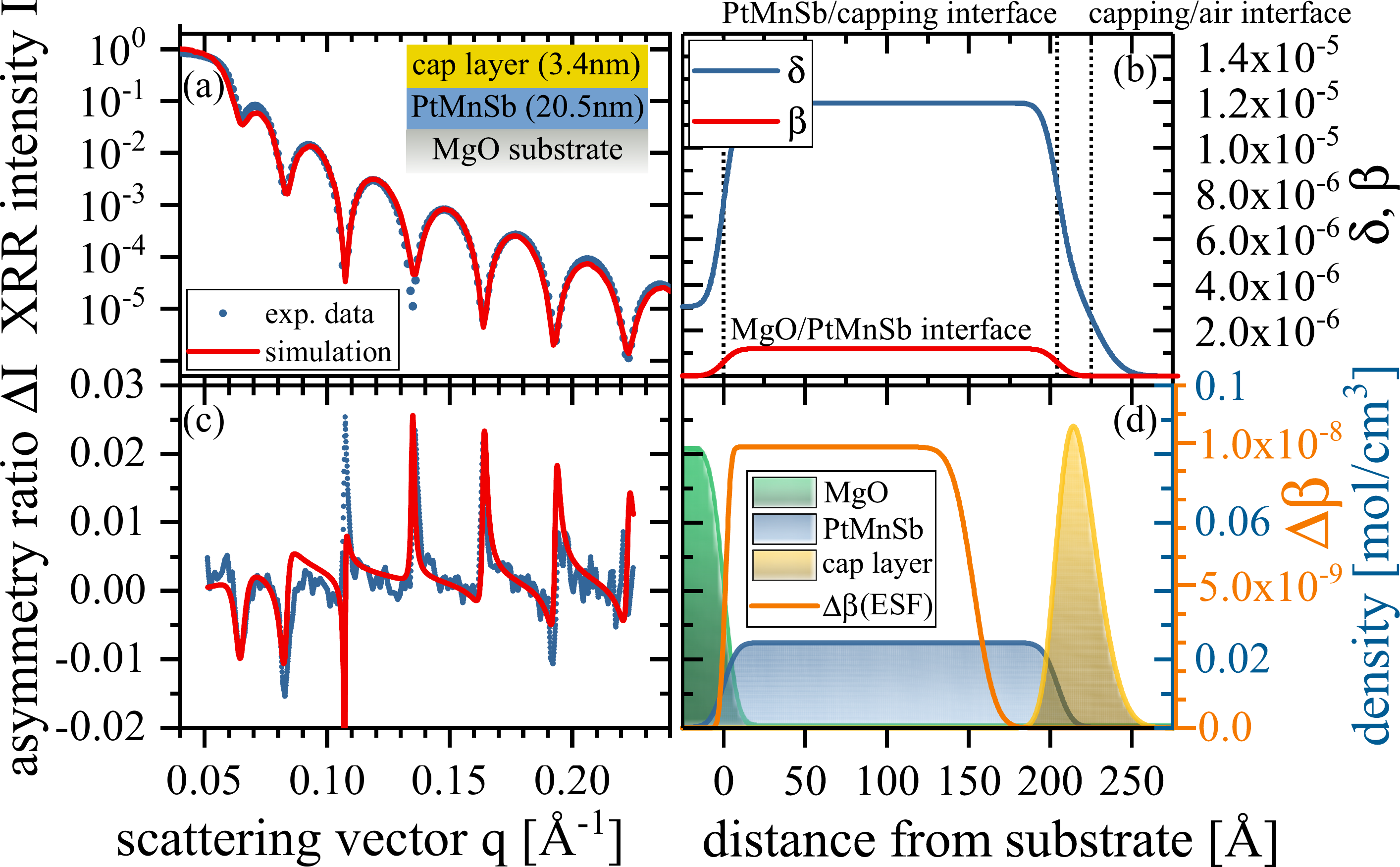}
\caption{(a) XRR measurements of \ce{PtMnSb}(\SI{20.5}{nm}) with \ce{AlO_x}/\ce{MgO} cap layer. (b)~Magnetooptic $\delta$ and $\beta$ depth profile used in the simulation. (c)~Asymmetry ratio $\Delta I$ and the corresponding simulation generated with the ESF mode. (d)~XRR density depth profile together with the magnetooptic $\Delta \beta$ depth profile used in the asymmetry ratio simulation in (c).}
\label{fig:6}
\end{figure}

\subsection{II. Half-Heusler compound PtMnSb with and without additional Pt layer}

\begin{figure*}[htb!]
\centering
\includegraphics[width=\linewidth]{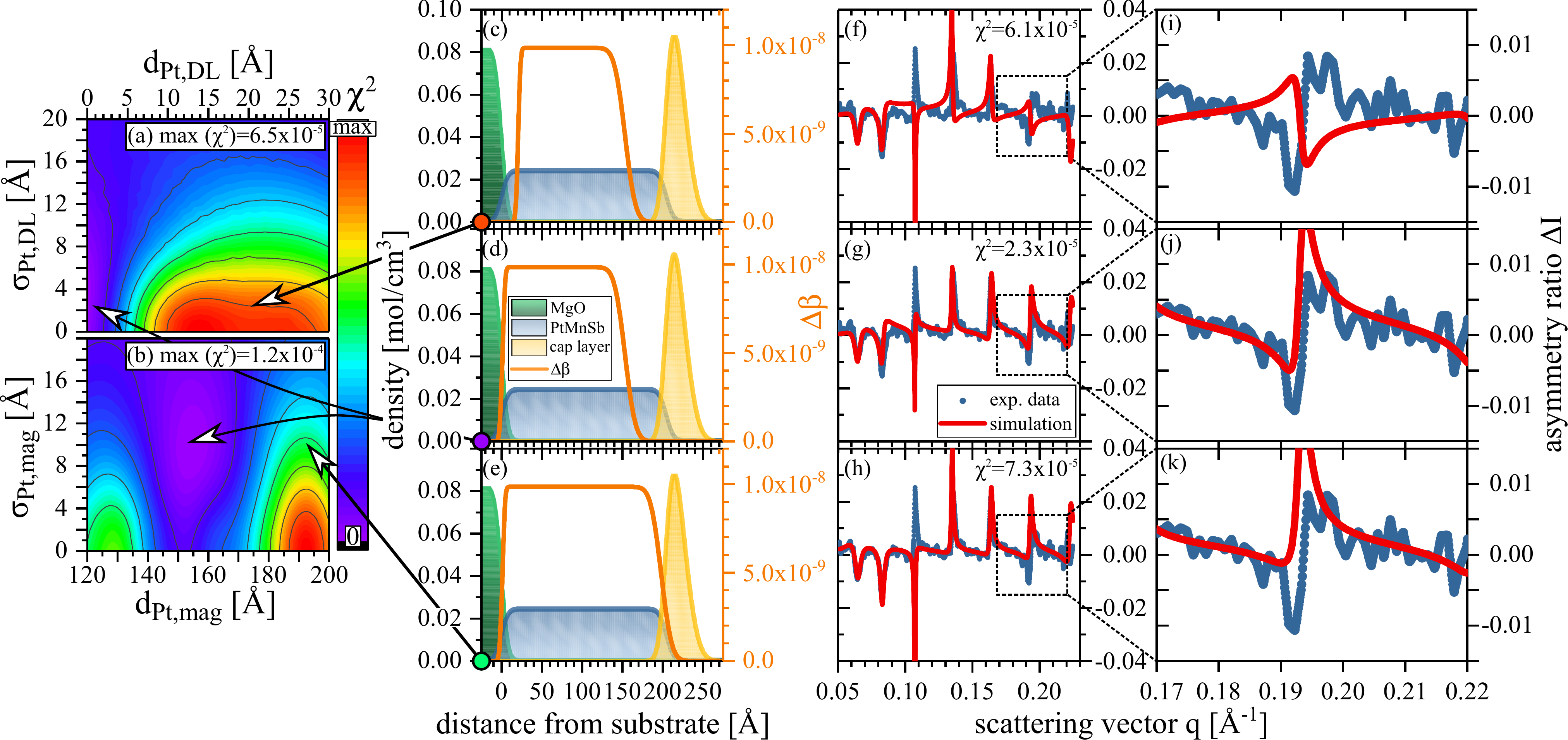}
\caption{(a) and (b) show sections of the general 2D $\chi^2$ landscape of \ce{PtMnSb}(\SI{20.5}{nm}) with a \ce{AlO_x}/\ce{MgO} cap layer. In (a) the thickness of the magnetic dead layer $d_\text{Pt,DL}$ is plotted against the roughness of the lower magnetic interface $\sigma_\text{Pt,DL}$, (b) shows the thickness of the magnetic layer $d_\text{Pt,mag}$ vs. the roughness of the upper magnetic interface $\sigma_\text{Pt,mag}$. We chose two distinct spots of the 2D maps to compare to the corresponding asymmetry ratio simulations of the optimal fit (g). (c) - (e) show the combined density and magnetooptic $\Delta\beta$ depth profiles of (d) the optimal fit, (c) an asymmetry ratio fit with a magnetic dead layer of \SI{20}{\angstrom} at the substrate interface and (e) an extended magnetooptic depth profile by \SI{40}{\angstrom} at the upper  interface. The asymmetry ratio $\Delta I(q)$ and simulated data (f) - (h) are drawn to a larger scale (i) - (k) to illustrate crucial parts of the asymmetry ratio simulation.}
\label{fig:7}
\end{figure*}

When we analyze the structurally and magnetically more complex \ce{PtMnSb} bilayer system, the need for a detailed XRR and XRMR analysis becomes obvious, in particular for the investigation of the magnetic depth profile in an \ce{AlO_x}/\ce{MgO} capped system. Figure~\ref{fig:6} summarizes the fit results of the reflectivity and asymmetry ratio scans and the corresponding optical, density and magnetooptic depth profiles.
Figure~\ref{fig:6}(a) shows an almost perfect XRR fit of the experimental data with only slight deviation for scattering vectors $q > \SI{0.2}{\angstrom^{-1}}$. The optical depth profile presented in Fig.~\ref{fig:6}(b) is dominated by the homogeneously grown half-Heusler alloy. The oxidized capping layer primarily shapes the decline of $\delta$ and $\beta$ at the air interface since the intermixed light elements exhibit optical parameters roughly one order of magnitude lower than \ce{PtMnSb} at the \ce{Pt} absorption edge. 

The fitted asymmetry ratio is presented in Fig.~\ref{fig:6}(c). Here, every main feature of this curve is reproduced within the best-fit. The only major deviation can be found at the third peak of the asymmetry ratio. However, the nature of the deviation is highly important to determine the quality of the fit which is discussed in detail later one. 
This asymmetry ratio feature is in principle part of the simulated curve at an identical scattering vector merely with a smaller amplitude. The density depth profile obtained from the XRR analysis and the magnetooptic $\Delta\beta$ depth profile calculated within the XRMR simulation are combined in Fig.~\ref{fig:6}(d).
This magnetic depth profile is the best fit result representing a uniformly magnetized half-Heusler layer with an unpolarized layer at the upper interface probably due to oxidization.

Based on this simulation result, we analyze the $\chi^2$ landscape of the magnetooptic depth profile parameters of the lower and upper \ce{PtMnSb} interface. The Figs.~\ref{fig:7}(d), (g) and (j), placed in the middle row, represent the global minimum of the $\chi^2$ maps shown in Figs.~\ref{fig:7}(a) and (b). In the top row, we present the simulation of a similar magnetic depth profile with a modification of the lower boundary introducing a \SI{20}{\angstrom} non-magnetic Pt layer at the interface (see Fig.~\ref{fig:7}(c)). By contrast, we show the results of an extension of the magnetic depth profile of \SI{40}{\angstrom} at the upper boundary to match the \ce{PtMnSb} interface in the bottom row of Fig.~\ref{fig:7}. Comparing the corresponding asymmetry ratio simulations in Figs.~\ref{fig:7}(f) - (h), we see that all three 
display the main peaks. In Fig.~\ref{fig:7}(f), deviations of the fitted curves are apparent 
between the main asymmetry ratio peaks while the significantly better fits shown in Figs.~\ref{fig:7}(g) and (h) are very similar although based on substantially different upper magnetooptic boundary positions.

Here, we focus on minor details of the fit to draw conclusions from this XRMR analysis regarding the Pt magnetization depth profile. 
The displayed difference in the magnetic depth profiles of Pt at the bottom interface (compare Figs.~\ref{fig:7}(c) and (d)) most significantly changes the slope of the asymmetry ratio features as shown in Figs.~\ref{fig:7}(i) and (j). An even closer look at the simulated asymmetry ratio is necessary to investigate the magnetic Pt depth profile at the upper interface (compare Figs.~\ref{fig:7}(d) and (e)). The asymmetry ratio simulation of the best-fit depth profile and the extended magnetic depth profile (see Figs.~\ref{fig:7}(g) and (h)) are hard to 
differentiate by eye and closely follow the measured dataset with only marginal differences in the third and sixth feature of the scan. However, this is exactly the part of the asymmetry ratio simulation we have to take into account very carefully since an adequate fit here is crucial for the identification of the real magnetic Pt depth profile. When we compare the enlarged graphic parts Figs.~\ref{fig:7}(j) and (k) it appears that the first dip of the feature is not exactly modeled and the upwards counterpart is exaggerated in the case of the extended magnetooptic depth profile. This small difference, conveniently but not necessarily accompanied by a lower value of $\chi^2$, serves to identify the most probable magnetic Pt depth profile which includes an unpolarized Pt layer at the oxide interface. 

The detailed XRMR analysis based on the variation of the magnetooptic depth profile of an identical layer of \SI{20.5}{nm} \ce{PtMnSb} capped with \SI{3.4}{nm} \ce{Pt} instead of \ce{AlO_x}/\ce{MgO} is presented in Fig.~\ref{fig:13} and discussed in Appendix D. 
In accordance with the results presented here, we are able to eliminate the possibility of a pronounced magnetic dead layer at the substrate interface of the Pt/PtMnSb bilayer based on a detailed analysis of the main periodic features of the asymmetry ratio and the corresponding $\chi^2$ maps. The potential extension of the magnetic depth profile into the \ce{Pt} layer is considered as well, however the simulated asymmetry ratio based on an extended magnetic depth profile involves an inconspicuous yet important deviation between data and fit. The best-fit solution represents a magnetic depth profile which is closely resembling the determined structural half-Heusler depth profile yet identifying a few angstroms of unpolarized Pt at both interfaces.

\subsection{III. General XRMR analysis recipe procedure}

\begin{figure}[htb!]
\centering
\includegraphics[width=\linewidth]{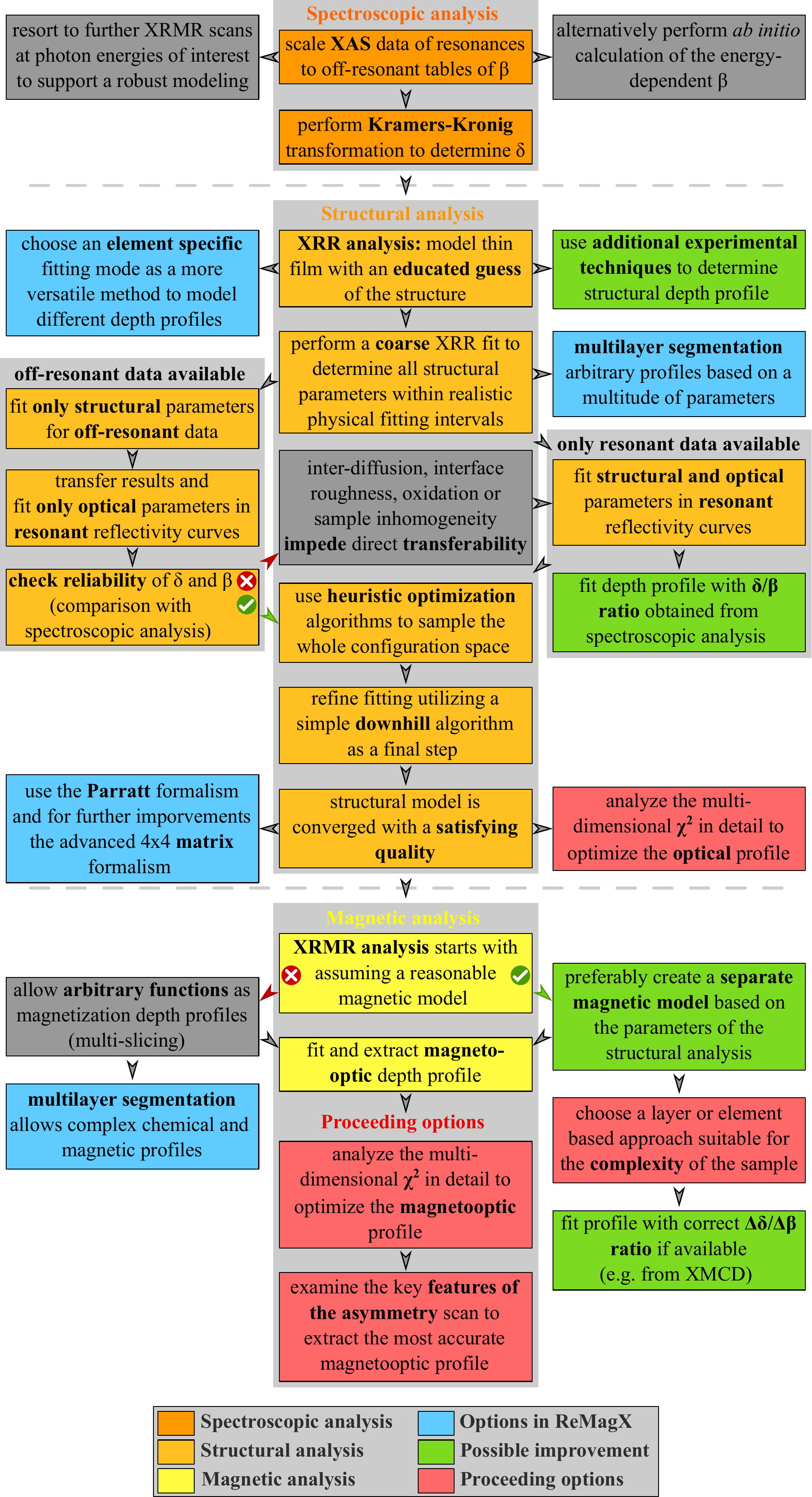}
\caption{Recipe procedure for the determination of a robust XRMR asymmetry ratio simulation for our bi- and multilayer systems. The central blocks of the spectroscopic, structural and magnetic analysis correspond to the general steps suggested by Macke and Goering \cite{macke2014magnetic}. The structural part includes the more accurate evaluation procedure using off-resonant scattering data presented by Klewe \textit{et al.} \cite{PhysRevB.93.214440} and additionally takes into account the possibility of an impeded transferability of the structural parameters.
The additional boxes demonstrate the alternative approaches and more complex analysis methods discussed here. Proceeding options regarding $\chi^2$ landscape and asymmetry ratio feature analysis are appended in red while optional methodical alternatives are presented in grey and improvements in green, respectively. Beneficial fitting options implemented in \textsc{ReMagX} are shown in blue. 
\quad\newline
}
\label{fig:8}
\end{figure}

The XRMR analysis presented here for samples at the Pt $\text{L}_3$ edge is summarized in a process flow diagram. Due to its wide applicability for structurally similar bi- and multilayer thin films, a guideline for the XRMR evaluation is created which includes all options for an improved fit convergence discussed in this work. This recipe procedure for the determination of a robust XRMR asymmetry ratio simulation is summarized in the flowchart presented in Fig.~\ref{fig:8}. The core is based on the guide for analyzing XRMR measurements presented by Macke and Goering \cite{macke2014magnetic} extended by the conclusions of our work. The whole procedure can be divided into three sections: the spectroscopic, structural and magnetic analysis.

Within the spectroscopic analysis, the first step is to scale the resonant x-ray absorption spectrum (XAS) to the tabulated off-resonant values of the absorptive optical parameter $\beta$ \cite{henke1993x, chantler2000detailed}, which enables us to derive the corresponding dispersive part $\delta$ of the index of refraction via a Kramers-Kronig transformation.
Hence, this step requires a measurement of the material absorption. An alternative is to perform \textit{ab initio} calculations of the energy-dependent absorption which is increasingly difficult for complex materials when the Coulomb interaction is not negligible. 

The central part of this procedure is the structural XRR analysis in order to obtain the correct thickness, roughness as well as optical parameters. Here, off-resonant XRR data is useful to eliminate any influence of the absorption edges. However, the reliability of the obtained optical fitting parameters must be checked since a direct transferability is only valid within the strict constrains and complexity level of the employed model. 
Additional information on the composition and material growth is beneficial in assessing the obtained parameters. In particular the optical constants should be checked for their reliability in order to evaluate the applicability of the model.

When the structural model has been determined with adequate quality, the obtained information is used as the structural basis on which a reasonable and parameterized magnetic model can be developed for the XRMR analysis. The thereby obtained magnetooptic depth profiles are subsequently optimized based on a detailed analysis of the $\chi^2$ landscapes and asymmetry ratio features until satisfactory results are achieved.
Where possible, measurements at further photon energies of interest should be performed, specified with the help of the spectroscopic analysis, in order to separate out different models. For instance, XRMR scans at the $\text{L}_2$ edge or the inflexion point of the dichroic response are useful to create a robust model of the magnetic depth profile.

The applicable improvements (green) and advanced proceeding options (red) are discussed here and by Klewe \textit{et al.} \cite{PhysRevB.93.214440}. Options available in the program \textsc{ReMagX} are presented in blue.
The basic steps of the conventional approach, outlined in the center, are sufficient for most simple systems such as bilayer samples. 
The multi-slicing method is accompanied by the introduction of various free parameters often leading to complex depth profiles, but less physically feasible results. For more complex heterostructures and multilayer systems, the extended steps beyond standard procedures are recommended for a detailed analysis.

\section{CONCLUSION}

In conclusion, we have investigated various samples of increasing complexity by means of XRMR at the Pt $\text{L}_3$ edge 
and determined a robust procedure for the analysis of the measured XRMR curves. 
We have shown that methods based on simple downhill algorithms used to simulate the XRMR measurements are reaching their limit when investigating magnetic profiles with a complex landscape of $\chi^2$ and proposed methods to 
reliably obtain qualitative and quantitative fits without the need for additional measurements or exploiting further unrealistic degrees of freedom in the simulation. Furthermore, we utilized the software tool \textsc{ReMagX} to fit the experimental 
curves and model magnetooptic depth profiles based on different fitting algorithms, iterative optimization approaches and a detailed analysis of the asymmetry ratio features as well as $\chi^2$ landscapes in order to improve the agreement between the XRMR data and simulation.

We have investigated a standard Pt/Fe bilayer to discuss the difference between layer and independent element based simulation methods. The latter is preferable for the independent modeling of nontrivial magnetic depth profiles.
The supporting detailed analysis of the complex multi-dimensional $\chi^2$ landscape has been introduced as a fundamental tool to determine accurate structural and magnetooptic depth profiles. Additionally, the detailed study of the dominating asymmetry ratio features is introduced as a potent method to determine and evaluate supposed best-fit solutions. Both approaches are especially important to obtain physically consistent structural and magnetic information of complex sample structures. 

Crucial parts of the asymmetry ratio for various magnetic depth profiles in bilayers based on the half-Heusler compound \ce{PtMnSb}
have been studied to visualize the significance of even minimal deviations in an asymmetry ratio simulation and underline the importance of a close examination of the key features to identify the most probable solution. Here, the influence of an oxide capping on the magnetic depth profile has been studied in relation to a Pt capped twin sample, discussed in Appendix D. The XRMR study of a multilayer including a Co/Pt interface, which has been investigated to refine our analytic approach for structurally and magnetically complex systems, is presented in Appendix C. By developing a recipe procedure for the determination of a robust XRMR asymmetry ratio simulation, in particular applicable for Pt based thin film samples as well as adaptable for a wide range of similar systems, we are able to use the excellent magnetic sensitivity and depth resolution of XRMR to full capacity.

\section*{ACKNOWLEDGMENTS}

We acknowledge DESY (Hamburg, Germany), a member of the Helmholtz Association HGF, and the European Synchrotron Radiation Facility (ESRF) (Grenoble, France) for the provision of experimental facilities. Parts of this research
were carried out at PETRA III. We would like to thank David Reuther and Philipp Glaevecke at DESY and the beamline BM28 staff at the ESRF for technical support as well as Tobias Pohlmann, Panagiota Bougiatioti and Johannes Mendil for assistance during the beamtimes. We also thank Sebastian Macke for providing software support of the fitting tool \textsc{ReMagX} and G\"unter Reiss for making available the laboratory equipment in Bielefeld.

\section*{Appendix A. Supplementary experimental and theoretical details}

The asymmetry ratio of the reflectivity is measured at the \ce{Pt} $\text{L}_3$ absorption edge using circularly polarized x-rays at room temperature. The magnetic contrast is achieved in two ways either by flipping a saturating external magnetic field and, thus, the magnetization of the thin film system or by switching the helicity of the circularly polarized light. Each method independently results in slightly different x-ray reflectivity curves when varying the magnetization direction relative to the x-ray polarization due to a change in the optical constants of the spin polarized material based on the refractive
index $n = 1 -\delta + i\beta$. Thus, the dispersion and absorption coefficients $\delta$ and $\beta$ are modified 
by the fraction $\Delta\delta$ and $\Delta\beta$ defined by the magnetooptic index $\Delta n = \Delta\delta-i \Delta\beta$ of the circular dichroism for different magnetization directions. In a geometry, in which the x-ray light is propagating parallel to the magnetization, the complex refractive index is generally expressed by $n_\pm =  1 -\left(\delta \mp \Delta\delta\right) + i\left(\beta \mp \Delta\beta\right)$.

In case of magnetization as well as helicity switching the asymmetry ratio is detected for both x-ray polarizations or  magnetization orientations, respectively.
For instance, in case of helicity switching, the intensity is collected for right $I_+$ and left $I_-$ circular polarization for each angle of incidence~$\theta$ and for the sample magnetization parallel to the scattering vector. The magnetic asymmetry ratio \mbox{$\Delta I = (I_+ - I_-)/(I_+ + I_-)$} can thus be calculated at each angle. The same measurement can also be repeated with the sample magnetization being reversed. The new asymmetry ratio should thus be equal but opposite in sign. The final asymmetry ratio can thus be extracted by calculating the half difference between the two measurements $A_c = 1/2 \cdot \left(A_+-A_-\right)$. This method allows to subtract non-magnetooptic effects.

The Pt/Fe bilayer and \ce{TaO_x}/\allowbreak\ce{MgO}/\allowbreak\ce{Ta}/\allowbreak\ce{Co}/\allowbreak\ce{Pt}//\ce{SiO_x} multilayer sample were measured at the P09 beamline at PETRA III (Hamburg, Germany). 
A single diamond quarter-wave plate was used to produce \SI[separate-uncertainty = true]{99\pm1}{\percent} degree of circular polarization. An external magnetic field of \SI{\pm90}{mT} applied parallel to the sample surface in the scattering plane was produced by a four-coil electromagnet. The reflectivity curves were collected in a conventional $\theta - 2\theta$ scattering geometry.

The \ce{PtMnSb} thin films were investigated at the XMaS beamline BM28 \cite{brown2001xmas} at ESRF (Grenoble, France) in the same scattering geometry at room temperature using circularly polarized x-rays.
A $\pm\SI{200}{mT}$ field was applied in the scattering plane parallel to the sample surface. The incoming x-ray photons were converted to \SI[separate-uncertainty = true]{88(1)}{\percent} of circular polarization \cite{bouchenoire2003} by means of a 
diamond phase-plate. To improve the signal to noise ratio, the helicity was reversed at \SI{11.5}{Hz} by a piezo driven device \cite{bouchenoire2007development}. The measurements were carried out at the Pt $\text{L}_3$ absorption edge (\SI{11568}{eV}) and off-resonance (\SI{11468}{eV}). These energies can be slightly different from our prior XRMR experiments \cite{kuschel2015static, PhysRevB.93.214440, kuschelstatic, bougiatioti2017quantitative, bougiatioti2018impact} depending on the current energy calibration of the beamline.

\section*{Appendix B. Analysis of the $\chi^2$ landscape for the $\textsc{\ce{Pt}/\ce{Fe}}$ bilayer}

\setcounter{figure}{0} \renewcommand{\thefigure}{A.\arabic{figure}} 

\begin{figure}[t!]
\centering
\includegraphics[width=1\linewidth]{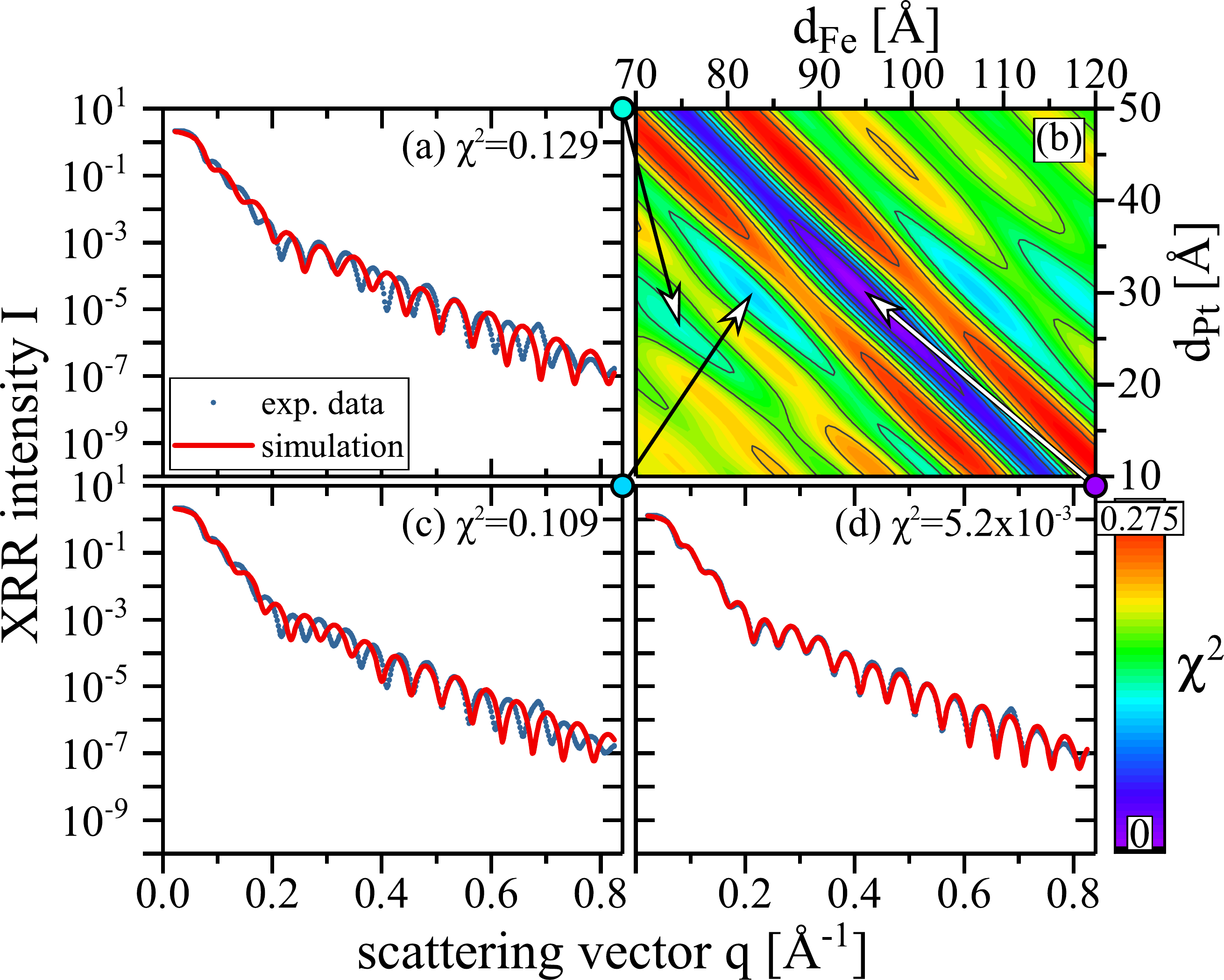}
\caption{XRR measurements of \ce{Pt}(\SI{3.0}{nm})/\allowbreak\ce{Fe}(\SI{9.5}{nm})//\allowbreak\ce{MgO} and various simulations of the XRR intensity $I$ at specific locations within the (b) 2D mapping plot of $\chi^2$ as a function of the thickness of the Pt $d_\text{Pt}$ and \ce{Fe} layer $d_\text{Fe}$. (d)~shows the optimal simulation, while (a) and (c) represent local minima in the $\chi^2$ landscape.}
\label{fig:9}
\end{figure} 

\begin{figure*}[tb!]
\centering
\includegraphics[width=\linewidth]{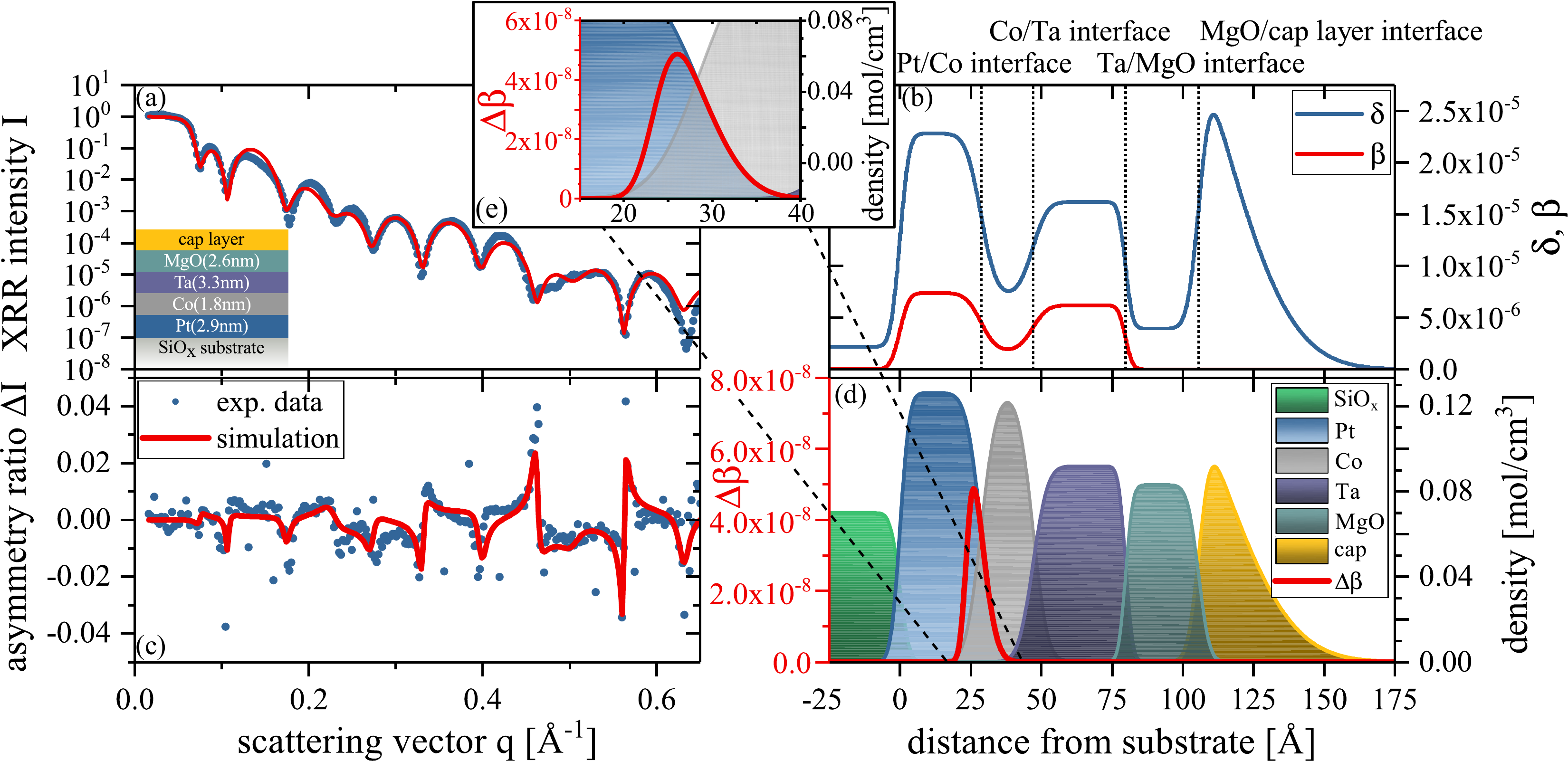}
\caption{(a) XRR and (c) XRMR measurement of a \ce{TaO_x}/\ce{MgO}(\SI{2.6}{nm})/\allowbreak\ce{Ta}(\SI{3.3}{nm})/\allowbreak\ce{Co}(\SI{1.8}{nm})/\allowbreak\ce{Pt}(\SI{2.9}{nm})//\allowbreak\ce{SiO_x} multilayer. (b) shows the optical depth profile at the $\text{L}_3$ edge of \ce{Pt} leading to the simulation in (a). (d) The element specific density and the magnetooptic depth profile used for the asymmetry ratio simulation in (c) obtained using the ESF mode. (e) Close-up focusing on the magnetooptic depth profile of $\Delta\beta$ which is bound to the depth profile of the Pt density at the right side of the $\Delta\beta$ depth profile.}
\label{fig:4}
\end{figure*}

The $\chi^2$ landscape spanned by the thickness of the \ce{Pt} layer $d_\text{Pt}$ and the \ce{Fe} layer $d_\text{Fe}$ is shown in Fig.~\ref{fig:9} together with three reflectivity simulations at particular $\chi^2$ minima of the parameter space.
The global minimum of the best-fit is compared to two adjacent local minima on this $\chi^2$ map which is part of the selected 2D landscapes presented in Fig.~\ref{fig:3}. The two fits representing the local minima are clearly showing a deviation from the Kiessig oscillations due to a variation in thickness. The global minimum is very clearly located in the center of the map. For the same value of $d_\text{Pt}$, we see further repeats of local minima with a regular change in $d_\text{Fe}$. The reflectivity simulation of the best-fit is presented in Fig.~\ref{fig:9}(d) while the fits of the neighboring minima are plotted in Figs.~\ref{fig:9}(a) and (c). Obviously, the reflectivity simulations on the left side of Figs.~\ref{fig:9}(a) and (c) are far from ideal also indicated by a comparatively large $\chi^2$ error value. It becomes apparent that those fits are partially inaccurate due to the misrepresented \ce{Fe} thickness which manifests in a displacement of the Kiessig oscillation and will ultimately preclude simulations of the asymmetry ratio constituted on those structural parameters. 

\section*{Appendix C. Structurally complex multilayer including a $\textsc{\ce{Co}/\ce{Pt}}$ interface}

The analysis of a structurally complex multilayer system involving an interface of \ce{Pt} adjacent to a $3d$ ferromagnet is presented in Fig.~\ref{fig:4}.
Depending on the method of preparation and capping, we expect  density depth profiles with varying layer roughness and start the simulation process with little information on the diffusive or oxidative condition.
Next to the XRR intensity plot with the best-fit simulation shown in Fig.~\ref{fig:4}(a), the corresponding optical depth profile is presented in Fig.~\ref{fig:4}(b) illustrating the interface positions and roughness parameters for each transition between the elements or compounds in this ESF simulation. The interface positions and the roughnesses of adjacent materials are coupled to each other similar to the simulation in the LSF mode. 
For the partially oxidized \ce{TaO_x} capping structure, the description of layers consisting of individual elements is not necessary. 
It is expedient to define a single compound for the capping layer and thus reduce the number of free parameters in the XRR simulation. Here, the structural parameters are not transferable without modifying the off-resonant XRR fit results to the fitting simulation of the resonant XRR due to the increased sample inhomogeneity and complex oxide capping which significantly alters the reflectivity curve. The reliability of the optical constants has to be checked in order to guarantee that structural and optical parameters are fitted within realistic physical intervals based on the resonant reflectivity curve.

Optimizing the fit of the XRR intensity with a combination of an evolutionary algorithm and a simplex fitting approach allows us to determine the best parameters. 
Due to the challenging modeling of the capping layer, the XRR fit is reasonably accurate yet no perfect agreement with the obtained curve is reached in particular for higher scattering vectors $q$. 
As a consequence, the simulated asymmetry ratio has to be carefully tested to determine the influence of these inaccuracies since the magnetooptic depth profile is directly connected to the obtained structural parameters.
The experimental asymmetry ratio and best simulation are plotted in Fig.~\ref{fig:4}(c) where every major characteristic of the presented curve is reasonably featured by the simulation
although the measured ratio is on average below \SI{2}{\percent}.
Figure~\ref{fig:4}(e) shows a close-up of the \ce{Pt} magnetic depth profile of this sample fixed to the right flank of the \ce{Pt} density depth profile. Thus, the \ce{Pt} magnetic depth profile follows the \ce{Pt} density depth profile at the interface and drops towards the substrate following the depth profile of a realistic magnetooptic depth profile (compare Fig.~\ref{fig:1}(d) and inset).

\begin{figure}[!tb]
\centering
\includegraphics[width=\linewidth]{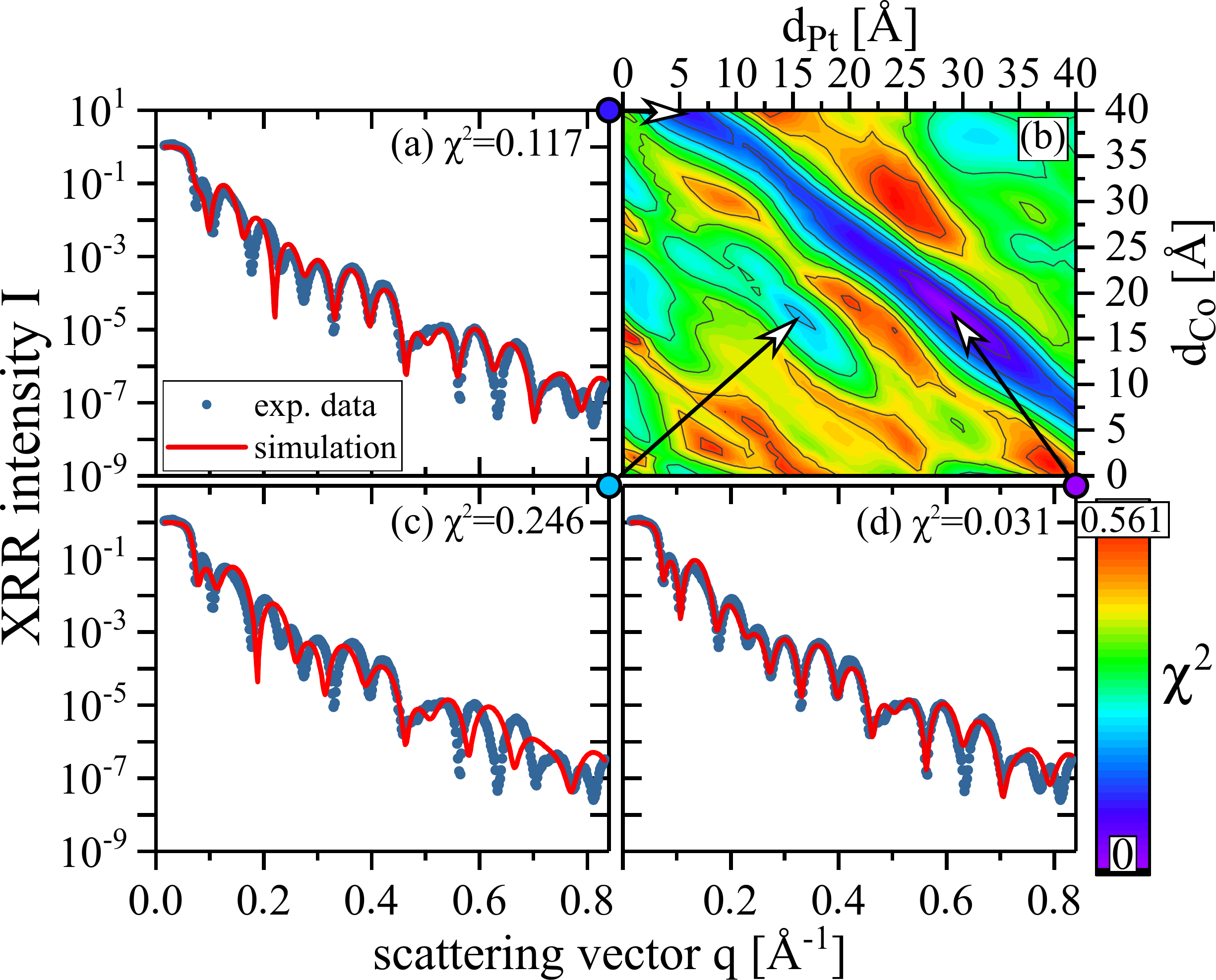}
\caption{XRR intensity $I$ and three exemplary simulations for \ce{TaO_x}/\ce{MgO}(\SI{2.6}{nm})/\ce{Ta}(\SI{3.3}{nm})/\ce{Co}(\SI{1.8}{nm})/ \ce{Pt}(\SI{2.9}{nm})//\ce{SiO_x} at selected locations in the parameter space visualized as (b) a 2D mapping plot of $\chi^2$ depending on the thickness of \ce{Pt} $d_\text{Pt}$ and thickness $d_\text{Co}$ of the \ce{Co} layer. (d) shows the optimal simulation, while (a) and (c) represent local minima in the $\chi^2$ landscape.}
\label{fig:5}
\end{figure} 

The importance of accurately modeling every feature of a reflectivity or asymmetry ratio becomes apparent when we examine the $\chi^2$ map of this heterostructure spanned by the thicknesses $d_\text{Co}$ and $d_\text{Pt}$ shown in Fig.~\ref{fig:5}. In principle this map is comparable to the one presented in Fig.~\ref{fig:3}(a) since it is spanned by a variation of two thickness parameters thus establishing an oscillatory pattern of local minima. Here, we can clearly identify the global minimum broadened along the map diagonal 
as observed for the Pt/Fe bilayer.
However, due to the irregular nature of the Kiessig fringes of this complex multilayer, the oscillatory pattern is diluted resulting in a landscape of various overlapping and a few sharp local minima. The XRR intensity simulations in Figs.~\ref{fig:5}(a) and (c) show two distinct fits using the parameters of local minima. The first one illustrates a sharp local minimum being close to the global minimum (Fig.~\ref{fig:5}(d)) which represents the best-fit established in Fig.~\ref{fig:4}. The difference in the fit quality between the local minima is dramatic although the parameters of the worse fit are similar to the best-fit except for $d_\text{Pt}$ while the objectively better fit is far from the global minimum in both mapped parameters. The reflectivity simulations in a local (Fig.~\ref{fig:5}(a)) and the global minimum (Fig.~\ref{fig:5}(d)) differ in just fitting one Kiessig fringe at the start of the reflectivity curve. However, the simulation is a good coarse fit which cannot be optimized by established fitting algorithms in this local minimum and would ultimately render the subsequent asymmetry ratio simulation impossible.

If we want to take advantage of the excellent sensitivity of XRMR in detecting magnetic depth profiles and variations of those in a series of samples, we have to analyze the features of our reflectivity and asymmetry ratio data in detail.
This is ultimately more important than a pure minimization of $\chi^2$ since a straight line can qualify as a stable fit through an oscillation in extreme cases.

\begin{figure}[t!]
\centering
\includegraphics[width=1\linewidth]{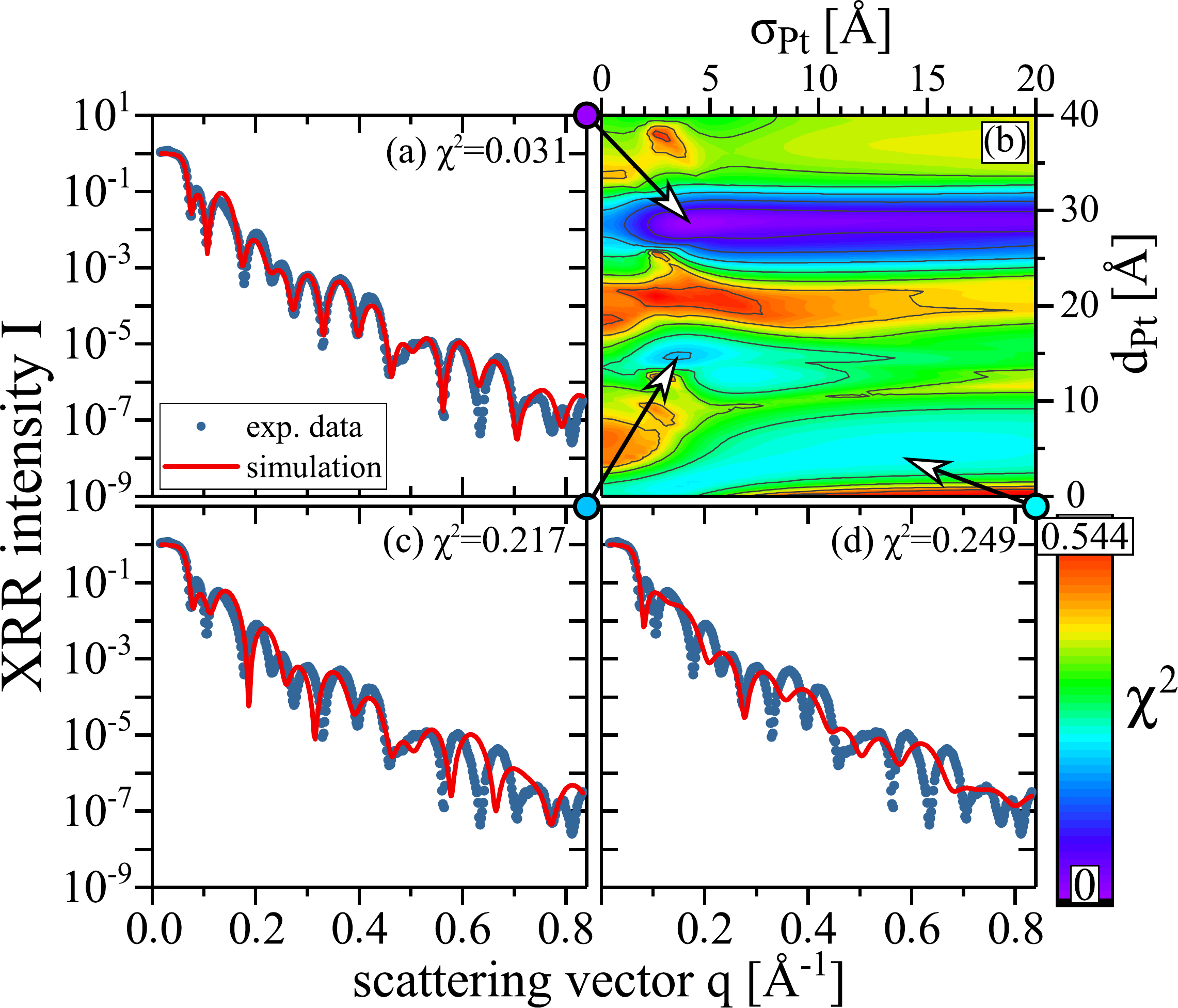}
\caption{XRR intensity $I$ and various simulations for \ce{TaO_x}/\allowbreak\ce{MgO}(\SI{2.6}{nm})/\allowbreak\ce{Ta}(\SI{3.3}{nm})/\allowbreak\ce{Co}(\SI{1.8}{nm})/\allowbreak\ce{Pt}(\SI{2.9}{nm})//\allowbreak\ce{SiO_x} at selected locations within (b) the 2D mapping plot of $\chi^2$ depending on the thickness $d$ and roughness $\sigma$ of the \ce{Pt} layer. (a)~shows the optimal simulation, while (c) and (d) represent local minima in the $\chi^2$ landscape.}
\label{fig:10}
\end{figure}

In contrast to the mere variation of the layer thickness, Fig.~\ref{fig:10} shows a second specific set of XRR simulations for the \ce{TaO_x}/\allowbreak\ce{MgO}(\SI{2.6}{nm})/\allowbreak\ce{Ta}(\SI{3.3}{nm})/\allowbreak\ce{Co}(\SI{1.8}{nm})/\allowbreak\ce{Pt}(\SI{2.9}{nm})//\allowbreak\ce{SiO_x} multilayer stack and the corresponding $\chi^2$ landscape spanned by the parameters $d_\text{Pt}$ and the upper roughness of the \ce{Pt} layer $\sigma_\text{Pt}$ at the \ce{Co} interface. 
The fit shown in Fig.~\ref{fig:10}(c), representing a local minimum at half the \ce{Pt} thickness of the best simulation, coarsely matches most Kiessig fringes in the first part of the plot yet it lacks accuracy at higher values of the scattering vector $q$. Figure~\ref{fig:10}(d) shows a fit in a shallow local minimum. However, the quality of this fit is hardly better than a straight line through the middle of the oscillations although it is a stable minimum in terms of $\chi^2$ inescapable by a standard simplex algorithm \cite{olsson1975nelder}. Furthermore, the $\chi^2$ map is highly anisotropic. While the thickness $d_\text{Pt}$ is defined precisely by the global minimum which is steeply sloping along this axis within the map, $\chi^2$ is increasing only marginally when leaving the global minimum along the axis defined by the parameter $\sigma_\text{Pt}$. In order to take advantage of the outstanding sensitivity of XRMR to magnetic moments at interfaces, the roughness should be defined precisely in the reflectivity simulation. Therefore, the best-fit value for the roughness has to be identified as an unambiguous global minimum on the $\chi^2$ landscape and if possible confirmed by other methods such as transmission electron microscopy.

\begin{figure}[t!]
\centering
\includegraphics[width=1\linewidth]{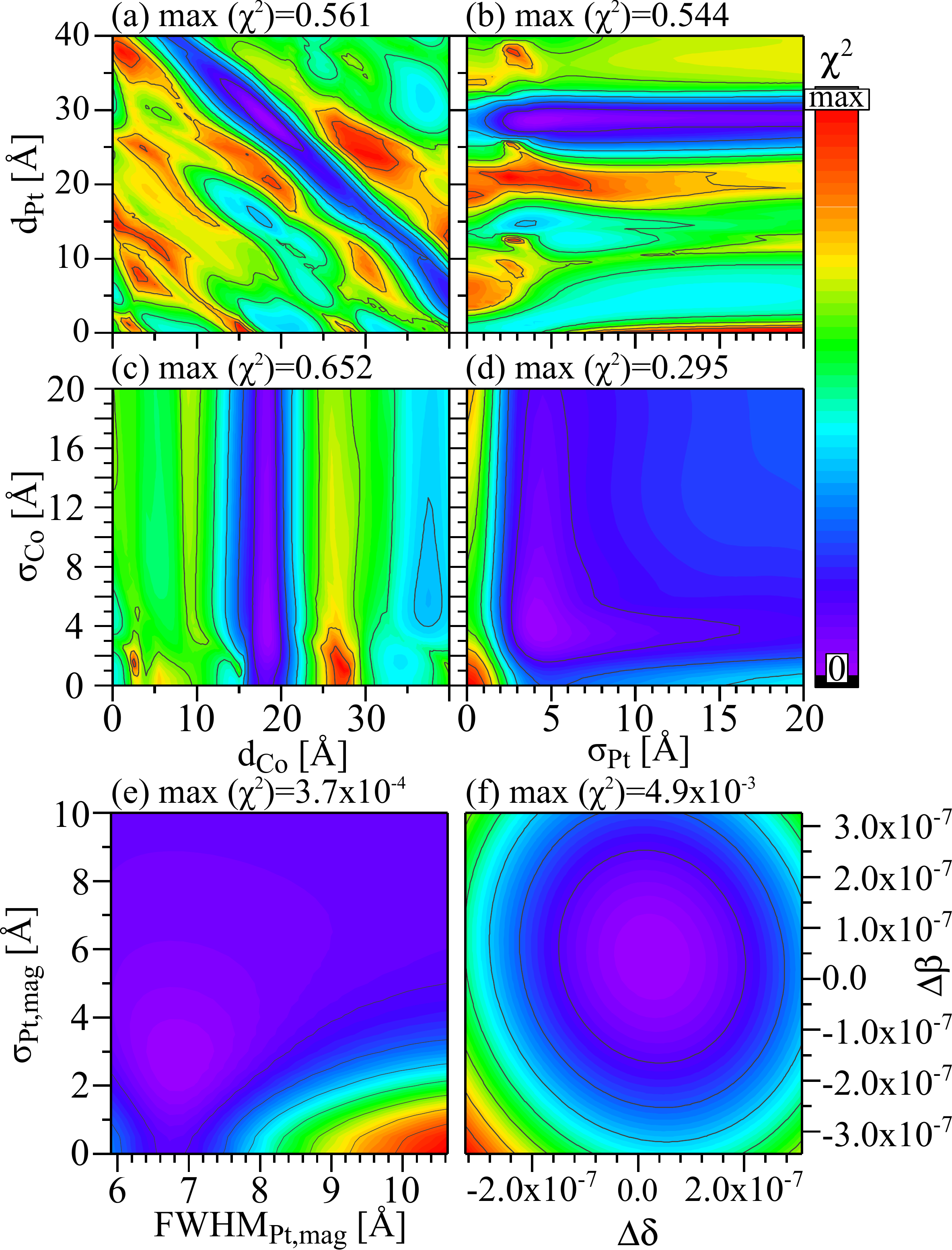}
\caption{Selected 2D landscapes of the $\chi^2$ value for \ce{TaO_x}/\allowbreak\ce{MgO}(\SI{2.6}{nm})/\allowbreak\ce{Ta}(\SI{3.3}{nm})/\allowbreak\ce{Co}(\SI{1.8}{nm})/\allowbreak\ce{Pt}(\SI{2.9}{nm})//\allowbreak\ce{SiO_x}. (a) $\chi^2$ map plot of the thickness of the \ce{Co} layer $d_\text{Co}$ vs. the thickness of the \ce{Pt} layer $d_\text{Pt}$. (b) $\chi^2$ map plot of the roughness of the \ce{Pt} layer $\sigma_\text{Pt}$ vs. $d_\text{Pt}$. (c) $\chi^2$ map plot of $d_\text{Co}$ vs. the roughness of the \ce{Co} layer $\sigma_\text{Co}$. (d) $\chi^2$ map plot of $\sigma_\text{Pt}$ vs. $\sigma_\text{Co}$.
(e) 2D plot of the magnetooptic simulation. The FWHM of the magnetic depth profile $\text{FWHM}_\text{Pt,mag}$ is plotted vs. the roughness $\sigma_\text{Pt,mag}$. (f) Corresponding 2D $\chi^2$ plot of the magnetooptic parameters $\Delta\delta$ vs. $\Delta\beta$.}
\label{fig:11}
\end{figure}

\begin{figure}[t!]
\centering
\includegraphics[width=1\linewidth]{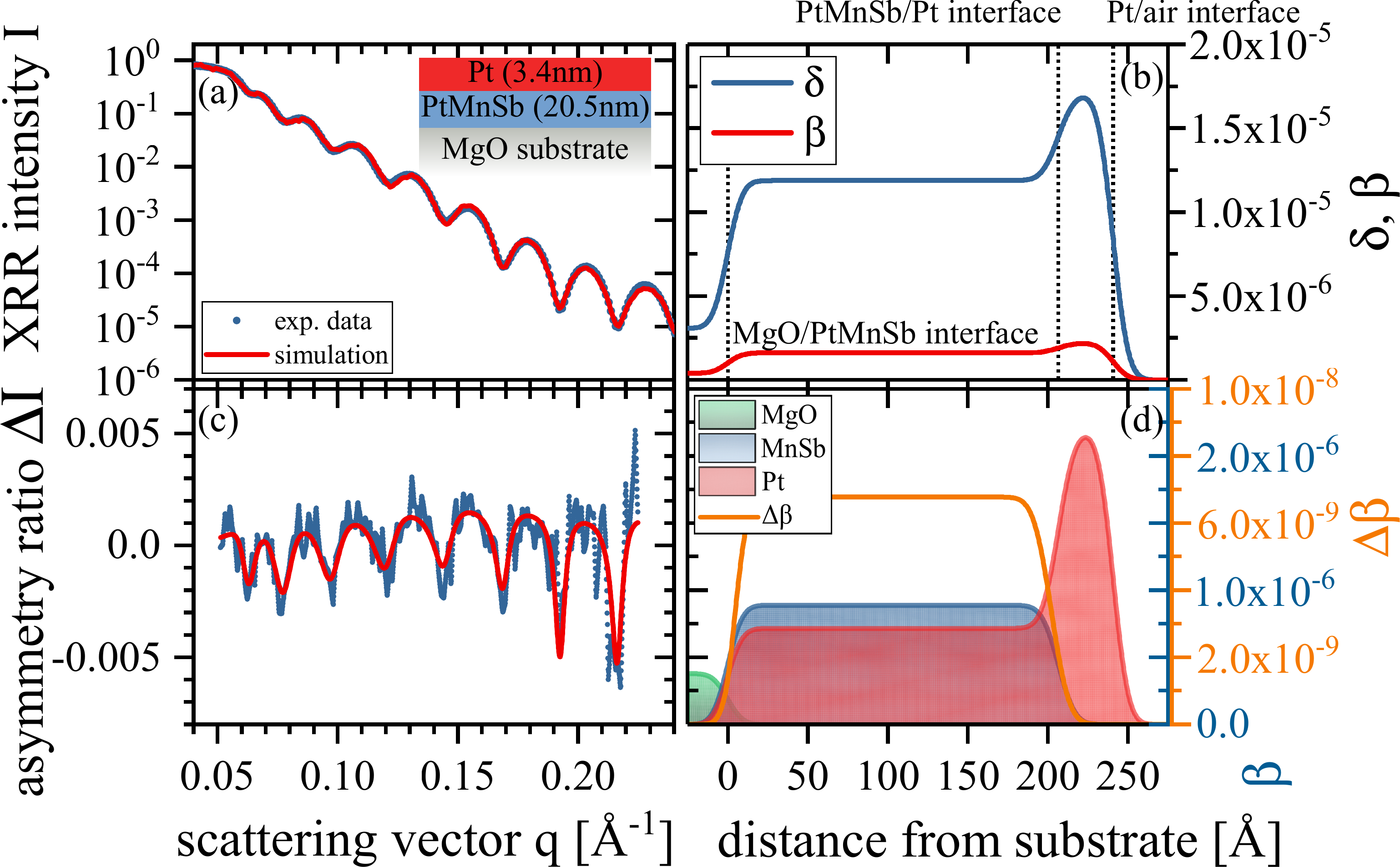}
\caption{(a)~XRR measurement of the \ce{Pt}(\SI{3.4}{nm})/\allowbreak\ce{PtMnSb}(\SI{20.5}{nm})//\allowbreak\ce{MgO} bilayer and (b) magnetooptic $\delta$ and $\beta$ depth profile used in the simulation. (c)~shows the asymmetry ratio $\Delta I$ and the corresponding simulation generated in the ESF mode. The XRR density depth profile is visualized in (d) and matched to the magnetooptic $\Delta\beta$ depth profile used in the asymmetry ratio simulation (c).}
\label{fig:12}
\end{figure} 

Sections of the complex multi-dimensional $\chi^2$ landscape for \ce{TaO_x}/\allowbreak\ce{MgO}(\SI{2.6}{nm})/\allowbreak\ce{Ta}(\SI{3.3}{nm})/\allowbreak\ce{Co}(\SI{1.8}{nm})/\allowbreak\ce{Pt}(\SI{2.9}{nm})//\allowbreak\ce{SiO_x} are
presented in Fig.~\ref{fig:11}. In particular, it shows the maps spanned by the \ce{Pt} and \ce{Co} thickness, the corresponding roughness values, the FWHM of the magnetic depth profile along the $z$-direction (distance from the substrate) $\text{FWHM}_\text{Pt,mag}$ and the roughness of its lower boundary $\sigma_\text{Pt,mag}$ as well as the magnetooptic parameters $\Delta\delta$ and $\Delta\beta$. Similar to the sectioned map shown in Fig.~\ref{fig:3} for the Pt/Fe//MgO bilayer system, we see many local minima obligating us to an asymmetry ratio analysis starting with heuristic algorithms. 

\begin{figure*}[t!]
\centering
\includegraphics[width=\linewidth]{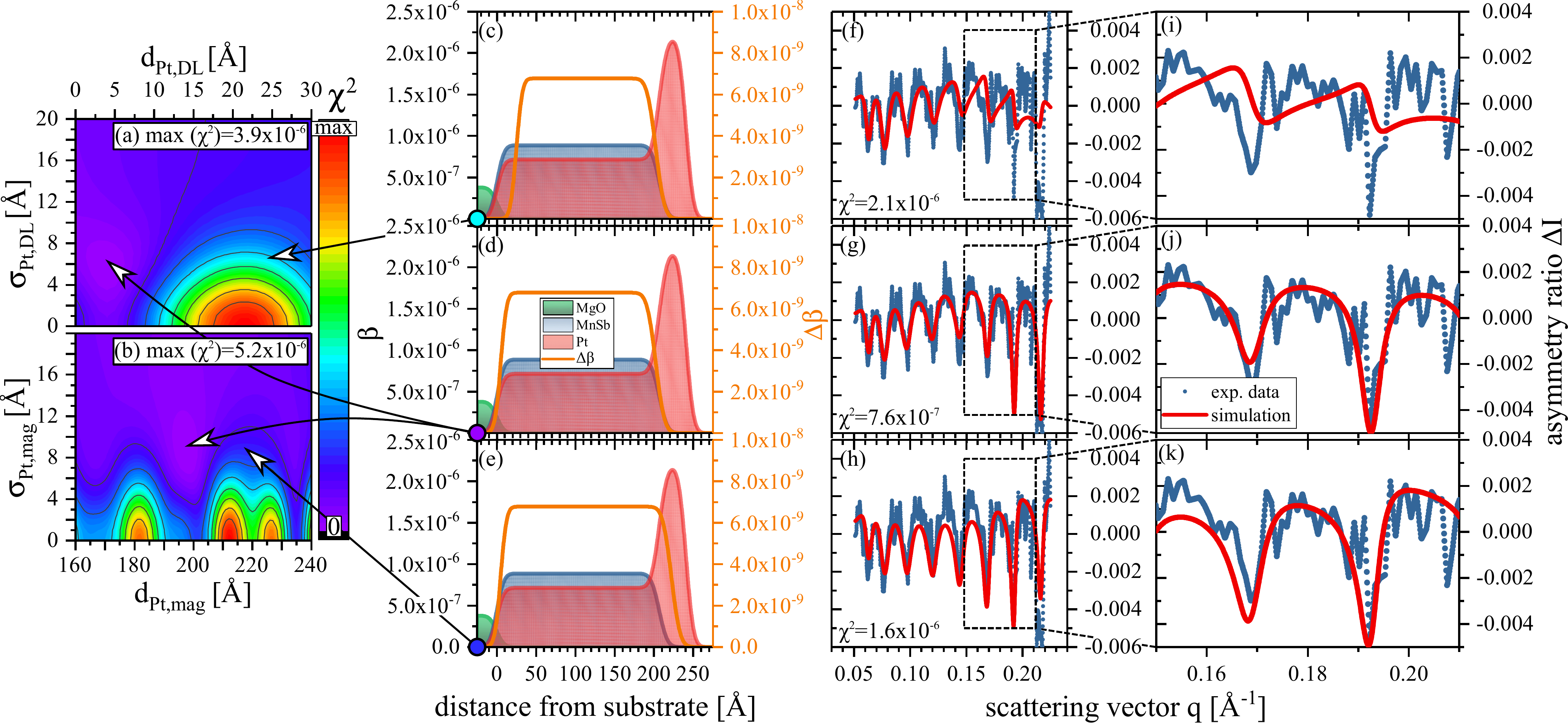}
\caption{Figures (a) and (b) show sections of the general 2D $\chi^2$ landscape of a \ce{Pt}(\SI{3.4}{nm})/\ce{PtMnSb}(\SI{20.5}{nm})//\ce{MgO} bilayer. In (a) the thickness of the magnetic dead layer $d_\text{Pt,DL}$ is plotted against the roughness of the lower magnetic interface $\sigma_\text{Pt,DL}$, (b) shows the thickness of the magnetic layer $d_\text{Pt,mag}$ vs. the roughness of the upper magnetic interface $\sigma_\text{Pt,mag}$. We chose two distinct spots of the 2D maps to compare to the corresponding asymmetry ratio simulations to the optimal fit (g). 
Figure (d) shows the optic $\beta$ and magnetooptic $\Delta \beta$ parameter profiles of the optimal simulation of the asymmetry ratio $\Delta I$ (g). The profile of a simulation with an additional magnetic dead layer of \SI{20}{\angstrom} and an extended magnetic depth profile into the \ce{Pt} layer resembling an MPE is plotted in Figs.~(c) and (e), respectively. The asymmetry ratio plots (f) to (h) show the corresponding ratios $\Delta I$ and simulations, (i) - (k) are close-ups of crucial parts of the asymmetry ratio simulation.}
\label{fig:13}
\end{figure*} 

The 2D landscape presented in Fig.~\ref{fig:11}(a) shows the same parameter space probed and discussed in terms of Fig.~\ref{fig:5}(b). However, in this selection the axes are inverted to fit the map compilation. Map Fig.~\ref{fig:11}(b) is discussed in Fig.~\ref{fig:10} with regard to the analysis of the XRR simulations. The $\chi^2$ map of Fig.~\ref{fig:11}(c) is again highly anisotropic. In this case, the thickness $d_\text{Co}$ is defined precisely by the global minimum while $\chi^2$ is only slightly increasing when the global minimum is left probing the parameter $\sigma_\text{Co}$. The $\chi^2$ map of Fig.~\ref{fig:11}(d) based on both $\sigma_\text{Co}$ and $\sigma_\text{Pt}$ roughness parameters reveals a global minimum steeply sloping towards lower roughness values. It is however particularly spread-out towards rougher interfaces, in general implying higher levels of uncertainty regarding these structural roughness parameters.
As stated in the main text, additional information on the interfaces is required for a precise analysis of the structural parameters. By combining various $\chi^2$ maps, as shown in Fig.~\ref{fig:11}, we are able to make an educated assessment of the real roughness parameter of the \ce{Co}/\ce{Pt} interface which is crucial when investigating the MPE in the multilayer system as outlined above.

The $\chi^2$ landscape of map Fig.~\ref{fig:11}(e) is created by varying the magnetic depth profile roughness $\sigma_\text{Pt,mag}$ in relation to the FWHM of the magnetic depth profile $\text{FWHM}_\text{Pt,mag}$. This effective thickness of the magnetic Pt depends on the thickness $d_\text{Pt,mag}$ of the initially used magnetic layer that becomes the magnetic depth profile $\Delta\beta$ of Fig.~\ref{fig:4}(e) when the magnetic roughness $\sigma_\text{Pt,mag}$ is considered. Here, the dependence is $\text{FWHM}_\text{Pt,mag} = \SI{0.0285}{nm^{-1}} \cdot d^2_\text{Pt,mag} + 0.191 \cdot d_\text{Pt,mag} + \SI{5.87}{nm}$ for the \ce{TaO_x}/\allowbreak\ce{MgO}(\SI{2.6}{nm})/\allowbreak\ce{Ta}(\SI{3.3}{nm})/\allowbreak\ce{Co}(\SI{1.8}{nm})/\allowbreak\ce{Pt}(\SI{2.9}{nm})//\allowbreak\ce{SiO_x} multilayer.
This $\chi^2$ map features a global minimum, which increases marginally when leaving the global minimum along both axes. However, the maximum value of $\chi^2$ on this map is \SI{3.7E-4}{}, which is one order of magnitude lower than the maximum of the comparable parameter space presented in Fig.~\ref{fig:3}(c) for the Pt/Fe bilayer. Independent of the overall $\chi^2$ gradient, the map shows a distinct global minimum and no local minima enabling us to refine the magnetooptic depth profile utilizing a simple downhill fitting algorithm. In Fig.~\ref{fig:11}(f), the $\chi^2$ map is spanned by the magnetooptic parameters $\Delta\delta$ and $\Delta\beta$
demonstrating a $\chi^2$ well with a distinct global minimum.  It is similar to the map created with the same parameters for the asymmetry ratio simulation of the Pt/Fe bilayer.

\section*{Appendix D. Experimental data and Simulation Analysis of the $\textsc{\ce{Pt}/\ce{PtMnSb}}$ bilayer}

Twin samples of the \ce{PtMnSb} thin film have been prepared replacing the \ce{AlO_x}/\ce{MgO} capping layer (see Fig.~\ref{fig:6}) by Pt as a top layer to investigate possible static MPEs in \ce{Pt} via XRMR. There is an excellent agreement between the simulation and the reflectivity scan of this bilayer depicted in Fig.~\ref{fig:12}(a).
The corresponding magnetooptic $\delta$ and $\beta$ depth profiles are shown in Fig.~\ref{fig:12}(b) clearly exhibiting the depth profile of the interface transitions within the bilayer.
The best-fit of the asymmetry ratio $\Delta I$ generated with the ESF mode accurately reproduces the main features and the oscillations, as illustrated in Fig.~\ref{fig:12}(c).
This fit is based on the density depth profile of Fig.~\ref{fig:12}(d) and results in the magnetooptic depth profile illustrated in the same plot as a function of the distance from the substrate. The magnetooptic depth profile involves no additional spin polarization at the interface in addition to the magnetization inherent to the half-Heusler material. However, the presence of an MPE is highly dependent on the ferromagnetic material and, more importantly, the quality of the interface \cite{geissler2002interplay, kuschel2015static}. A detailed study of the MPE in \ce{Pt} on \ce{PtMnSb} will be discussed elsewhere.

Figure~\ref{fig:13} illustrates again a detailed feature analysis regarding the asymmetry ratio simulation of the \ce{Pt}(\SI{3.4}{nm})/\ce{PtMnSb}(\SI{20.5}{nm})//\ce{MgO} bilayer. In the middle row we show the global best-fit of the $\chi^2$ landscapes shown in Figs.~\ref{fig:13}(a) and (b). In the top row, the simulation of an identical magnetic Pt depth profile with a modification of the lower boundary is shown. Here, we model a magnetic dead layer of \SI{20}{\angstrom} starting at the interface of the substrate into the \ce{PtMnSb} layer (see Fig.~\ref{fig:13}(c)). 
The Pt spin depth profile without this dead layer shown in Fig.~\ref{fig:13}(d) can similarly be extended at the top boundary of the magnetic depth profile by \SI{20}{\angstrom} reaching into the Pt cover layer shown in the bottom row in Fig.~\ref{fig:13}(e).

We now focus on the feature details of the asymmetry ratio simulations in Figs.~\ref{fig:13}(f) - (h).
Here, we can rule out a pronounced magnetic dead layer at the substrate interface due to a clear misfit of the main periodic features of the asymmetry ratio $\Delta I$, highlighted in the magnified part of the scan (see Fig.~\ref{fig:13}(i)). In spite of a high noise level, due to the relatively tiny asymmetry ratio of below \SI{0.4}{\percent}, we are able to define a realistic magnetooptic depth profile when carefully analyzing the $\chi^2$ maps. 
The extension of the magnetic depth profile into the \ce{Pt} layer is accompanied by an increase of the simulated asymmetry ratio (see Figs.~\ref{fig:13}(j) and (k)) and, more importantly, by an offset between data and fit most profound in the center of the asymmetry ratio scan. Comparing the simulations in Figs.~\ref{fig:13}(g) and (h), we see a shift in the overall shape of the fit from an upward to a downward directed course of the average curve.

This close look at the simulated asymmetry ratio is essential to investigate the magnetic depth profile with high accuracy. Again, the asymmetry ratio simulation of the best-fit depth profile and the extended magnetic depth profile (see Figs.~\ref{fig:13}(g) and (h)) are very similar and  show only small, however significant, differences as described above. We have to exercise caution when we evaluate these small deviations in the oscillatory behavior since an adequate fit here is crucial for the identification of an MPE. A close analysis of multiple simulations provides us with the real best-fit and most precise representation of a realistic magnetic depth profile in this system.

\begin{figure}[tb!]
\centering
\includegraphics[width=\linewidth]{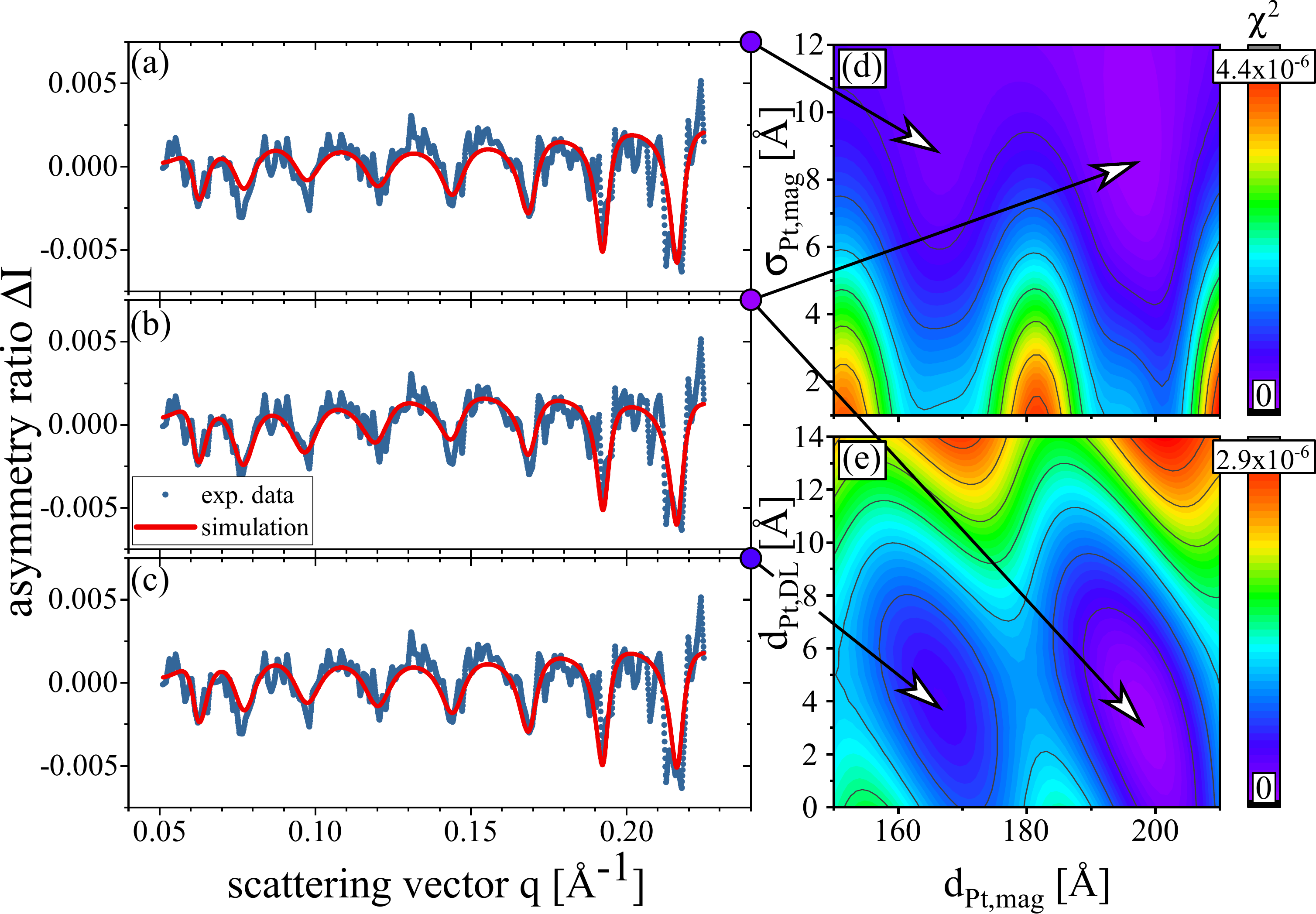}
\caption{The asymmetry ratio plots (a) to (c) show the ratios $\Delta I$ and simulations
of the global best-fit (b) and two good fits representing close neighboring local minima for the \ce{Pt}(\SI{3.4}{nm})/\ce{PtMnSb}(\SI{20.5}{nm})//\ce{MgO} bilayer. 
(d)~2D plot of the $\chi^2$ landscape for the magnetic depth profile of the XRMR asymmetry ratios. The thickness of the magnetic depth profile $d_\text{Pt,mag}$ is varied as well as the roughness $\sigma_\text{Pt,mag}$. (e) Corresponding 2D $\chi^2$ plot of the thickness of the magnetic depth profile $d_\text{Pt,mag}$ and the magnetic dead layer $d_\text{Pt,DL}$ and.}
\label{fig:14}
\end{figure} 

Figure~\ref{fig:14} shows a graphical compilation of highly similar simulations of the asymmetry ratio which yield significantly different magnetic depth profiles. These fits are a result of the XRMR analysis for the \ce{Pt}(\SI{3.4}{nm})/\ce{PtMnSb}(\SI{20.5}{nm})//\ce{MgO} bilayer. This specific example shows where a simple downhill algorithms fails to obtain the correct best-fit solution in case of unfavorable starting parameters. When a simple Simplex algorithm is utilized, the simulation can converge to a local minimum representing a decent fit. As illustrated in Figs.~\ref{fig:14}(a) and (c), these fits can be almost indistinguishable from the global best-fit solution shown in Fig.~\ref{fig:14}(b). Here, only an advanced optimization algorithms such as genetic
algorithms or simulated annealing, sampling the whole configuration space, is reliably converging to the global minimum. Once in the overall vicinity of this minimum, a simple downhill algorithms works fine as the final step of the optimization process. In the displayed situation, the use of the genetic, evolution based, fit routine can separate the two plausible values which yield a highly similar fit profile as illustrated in the direct comparison of all three fits presented here.
The corresponding $\chi^2$ maps presented in Figs.~\ref{fig:14}(d) and (e) illustrate the separation of the obtained solutions on the error landscape and visualize the potential barriers encountered by the optimization process.
This example illustrates where the use of a genetic
algorithms can separate two plausible solutions, which yield highly similar fits yet significantly different magnetic depth profiles, as shown by the distance on these $\chi^2$ maps as well as the range of profiles visualized in Figs.~\ref{fig:13}(c) to (e).

\bibliography{bibliography}

\begin{thebibliography}{64}%
\makeatletter
\providecommand \@ifxundefined [1]{%
 \@ifx{#1\undefined}
}%
\providecommand \@ifnum [1]{%
 \ifnum #1\expandafter \@firstoftwo
 \else \expandafter \@secondoftwo
 \fi
}%
\providecommand \@ifx [1]{%
 \ifx #1\expandafter \@firstoftwo
 \else \expandafter \@secondoftwo
 \fi
}%
\providecommand \natexlab [1]{#1}%
\providecommand \enquote  [1]{``#1''}%
\providecommand \bibnamefont  [1]{#1}%
\providecommand \bibfnamefont [1]{#1}%
\providecommand \citenamefont [1]{#1}%
\providecommand \href@noop [0]{\@secondoftwo}%
\providecommand \href [0]{\begingroup \@sanitize@url \@href}%
\providecommand \@href[1]{\@@startlink{#1}\@@href}%
\providecommand \@@href[1]{\endgroup#1\@@endlink}%
\providecommand \@sanitize@url [0]{\catcode `\\12\catcode `\$12\catcode
  `\&12\catcode `\#12\catcode `\^12\catcode `\_12\catcode `\%12\relax}%
\providecommand \@@startlink[1]{}%
\providecommand \@@endlink[0]{}%
\providecommand \url  [0]{\begingroup\@sanitize@url \@url }%
\providecommand \@url [1]{\endgroup\@href {#1}{\urlprefix }}%
\providecommand \urlprefix  [0]{URL }%
\providecommand \Eprint [0]{\href }%
\providecommand \doibase [0]{http://dx.doi.org/}%
\providecommand \selectlanguage [0]{\@gobble}%
\providecommand \bibinfo  [0]{\@secondoftwo}%
\providecommand \bibfield  [0]{\@secondoftwo}%
\providecommand \translation [1]{[#1]}%
\providecommand \BibitemOpen [0]{}%
\providecommand \bibitemStop [0]{}%
\providecommand \bibitemNoStop [0]{.\EOS\space}%
\providecommand \EOS [0]{\spacefactor3000\relax}%
\providecommand \BibitemShut  [1]{\csname bibitem#1\endcsname}%
\let\auto@bib@innerbib\@empty
\bibitem [{\citenamefont {Bragg}\ and\ \citenamefont
  {Bragg}(1913)}]{bragg1913reflection}%
  \BibitemOpen
  \bibfield  {author} {\bibinfo {author} {\bibfnamefont {W.~H.}\ \bibnamefont
  {Bragg}}\ and\ \bibinfo {author} {\bibfnamefont {W.~L.}\ \bibnamefont
  {Bragg}},\ }\href@noop {} {\bibfield  {journal} {\bibinfo  {journal} {Proc.
  R. Soc. Lond. A}\ }\textbf {\bibinfo {volume} {88}},\ \bibinfo {pages} {428}
  (\bibinfo {year} {1913})}\BibitemShut {NoStop}%
\bibitem [{\citenamefont {Holy}\ \emph {et~al.}(1993)\citenamefont {Holy},
  \citenamefont {Kubena}, \citenamefont {Ohlidal}, \citenamefont {Lischka},\
  and\ \citenamefont {Plotz}}]{holy1993x}%
  \BibitemOpen
  \bibfield  {author} {\bibinfo {author} {\bibfnamefont {V.}~\bibnamefont
  {Holy}}, \bibinfo {author} {\bibfnamefont {J.}~\bibnamefont {Kubena}},
  \bibinfo {author} {\bibfnamefont {I.}~\bibnamefont {Ohlidal}}, \bibinfo
  {author} {\bibfnamefont {K.}~\bibnamefont {Lischka}}, \ and\ \bibinfo
  {author} {\bibfnamefont {W.}~\bibnamefont {Plotz}},\ }\href@noop {}
  {\bibfield  {journal} {\bibinfo  {journal} {Phys. Rev. B}\ }\textbf {\bibinfo
  {volume} {47}},\ \bibinfo {pages} {15896} (\bibinfo {year}
  {1993})}\BibitemShut {NoStop}%
\bibitem [{\citenamefont {Tolan}(1999)}]{tolan1999x}%
  \BibitemOpen
  \bibfield  {author} {\bibinfo {author} {\bibfnamefont {M.}~\bibnamefont
  {Tolan}},\ }\href@noop {} {\emph {\bibinfo {title} {X-ray Scattering from
  Soft-Matter Thin Films: Materials Science and Basic Research}}}\ (\bibinfo
  {publisher} {Springer, Berlin},\ \bibinfo {year} {1999})\BibitemShut
  {NoStop}%
\bibitem [{\citenamefont {Daillant}\ and\ \citenamefont
  {Gibaud}(2008)}]{daillant2008x}%
  \BibitemOpen
  \bibfield  {author} {\bibinfo {author} {\bibfnamefont {J.}~\bibnamefont
  {Daillant}}\ and\ \bibinfo {author} {\bibfnamefont {A.}~\bibnamefont
  {Gibaud}},\ }\href@noop {} {\emph {\bibinfo {title} {X-ray and Neutron
  Reflectivity: Principles and Applications}}},\ Vol.\ \bibinfo {volume} {770}\
  (\bibinfo  {publisher} {Springer, Berlin},\ \bibinfo {year}
  {2008})\BibitemShut {NoStop}%
\bibitem [{\citenamefont {Parratt}(1954)}]{parratt1954surface}%
  \BibitemOpen
  \bibfield  {author} {\bibinfo {author} {\bibfnamefont {L.~G.}\ \bibnamefont
  {Parratt}},\ }\href@noop {} {\bibfield  {journal} {\bibinfo  {journal} {Phys.
  Rev.}\ }\textbf {\bibinfo {volume} {95}},\ \bibinfo {pages} {359} (\bibinfo
  {year} {1954})}\BibitemShut {NoStop}%
\bibitem [{\citenamefont {Dane}\ \emph {et~al.}(1998)\citenamefont {Dane},
  \citenamefont {Veldhuis}, \citenamefont {De~Boer}, \citenamefont {Leenaers},\
  and\ \citenamefont {Buydens}}]{dane1998application}%
  \BibitemOpen
  \bibfield  {author} {\bibinfo {author} {\bibfnamefont {A.}~\bibnamefont
  {Dane}}, \bibinfo {author} {\bibfnamefont {A.}~\bibnamefont {Veldhuis}},
  \bibinfo {author} {\bibfnamefont {D.}~\bibnamefont {De~Boer}}, \bibinfo
  {author} {\bibfnamefont {A.}~\bibnamefont {Leenaers}}, \ and\ \bibinfo
  {author} {\bibfnamefont {L.}~\bibnamefont {Buydens}},\ }\href@noop {}
  {\bibfield  {journal} {\bibinfo  {journal} {Physica B}\ }\textbf {\bibinfo
  {volume} {253}},\ \bibinfo {pages} {254} (\bibinfo {year}
  {1998})}\BibitemShut {NoStop}%
\bibitem [{\citenamefont {van~der Laan}\ \emph {et~al.}(1986)\citenamefont
  {van~der Laan}, \citenamefont {Thole}, \citenamefont {Sawatzky},
  \citenamefont {Goedkoop}, \citenamefont {Fuggle}, \citenamefont {Esteva},
  \citenamefont {Karnatak}, \citenamefont {Remeika},\ and\ \citenamefont
  {Dabkowska}}]{van1986experimental}%
  \BibitemOpen
  \bibfield  {author} {\bibinfo {author} {\bibfnamefont {G.}~\bibnamefont
  {van~der Laan}}, \bibinfo {author} {\bibfnamefont {B.~T.}\ \bibnamefont
  {Thole}}, \bibinfo {author} {\bibfnamefont {G.~A.}\ \bibnamefont {Sawatzky}},
  \bibinfo {author} {\bibfnamefont {J.~B.}\ \bibnamefont {Goedkoop}}, \bibinfo
  {author} {\bibfnamefont {J.~C.}\ \bibnamefont {Fuggle}}, \bibinfo {author}
  {\bibfnamefont {J.-M.}\ \bibnamefont {Esteva}}, \bibinfo {author}
  {\bibfnamefont {R.}~\bibnamefont {Karnatak}}, \bibinfo {author}
  {\bibfnamefont {J.}~\bibnamefont {Remeika}}, \ and\ \bibinfo {author}
  {\bibfnamefont {H.~A.}\ \bibnamefont {Dabkowska}},\ }\href@noop {} {\bibfield
   {journal} {\bibinfo  {journal} {Phys. Rev. B}\ }\textbf {\bibinfo {volume}
  {34}},\ \bibinfo {pages} {6529} (\bibinfo {year} {1986})}\BibitemShut
  {NoStop}%
\bibitem [{\citenamefont {Sch{\"u}tz}\ \emph {et~al.}(1987)\citenamefont
  {Sch{\"u}tz}, \citenamefont {Wagner}, \citenamefont {Wilhelm}, \citenamefont
  {Kienle}, \citenamefont {Zeller}, \citenamefont {Frahm},\ and\ \citenamefont
  {Materlik}}]{schutz1987absorption}%
  \BibitemOpen
  \bibfield  {author} {\bibinfo {author} {\bibfnamefont {G.}~\bibnamefont
  {Sch{\"u}tz}}, \bibinfo {author} {\bibfnamefont {W.}~\bibnamefont {Wagner}},
  \bibinfo {author} {\bibfnamefont {W.}~\bibnamefont {Wilhelm}}, \bibinfo
  {author} {\bibfnamefont {P.}~\bibnamefont {Kienle}}, \bibinfo {author}
  {\bibfnamefont {R.}~\bibnamefont {Zeller}}, \bibinfo {author} {\bibfnamefont
  {R.}~\bibnamefont {Frahm}}, \ and\ \bibinfo {author} {\bibfnamefont
  {G.}~\bibnamefont {Materlik}},\ }\href@noop {} {\bibfield  {journal}
  {\bibinfo  {journal} {Phys. Rev. Lett.}\ }\textbf {\bibinfo {volume} {58}},\
  \bibinfo {pages} {737} (\bibinfo {year} {1987})}\BibitemShut {NoStop}%
\bibitem [{\citenamefont {Chen}\ \emph {et~al.}(1990)\citenamefont {Chen},
  \citenamefont {Sette}, \citenamefont {Ma},\ and\ \citenamefont
  {Modesti}}]{chen1990soft}%
  \BibitemOpen
  \bibfield  {author} {\bibinfo {author} {\bibfnamefont {C.}~\bibnamefont
  {Chen}}, \bibinfo {author} {\bibfnamefont {F.}~\bibnamefont {Sette}},
  \bibinfo {author} {\bibfnamefont {Y.}~\bibnamefont {Ma}}, \ and\ \bibinfo
  {author} {\bibfnamefont {S.}~\bibnamefont {Modesti}},\ }\href@noop {}
  {\bibfield  {journal} {\bibinfo  {journal} {Phys. Rev. B}\ }\textbf {\bibinfo
  {volume} {42}},\ \bibinfo {pages} {7262} (\bibinfo {year}
  {1990})}\BibitemShut {NoStop}%
\bibitem [{\citenamefont {van~der Laan}\ and\ \citenamefont
  {Thole}(1991)}]{van1991strong}%
  \BibitemOpen
  \bibfield  {author} {\bibinfo {author} {\bibfnamefont {G.}~\bibnamefont
  {van~der Laan}}\ and\ \bibinfo {author} {\bibfnamefont {B.}~\bibnamefont
  {Thole}},\ }\href@noop {} {\bibfield  {journal} {\bibinfo  {journal} {Phys.
  Rev. B}\ }\textbf {\bibinfo {volume} {43}},\ \bibinfo {pages} {13401}
  (\bibinfo {year} {1991})}\BibitemShut {NoStop}%
\bibitem [{\citenamefont {Thole}\ \emph {et~al.}(1992)\citenamefont {Thole},
  \citenamefont {Carra}, \citenamefont {Sette},\ and\ \citenamefont {van~der
  Laan}}]{thole1992x}%
  \BibitemOpen
  \bibfield  {author} {\bibinfo {author} {\bibfnamefont {B.}~\bibnamefont
  {Thole}}, \bibinfo {author} {\bibfnamefont {P.}~\bibnamefont {Carra}},
  \bibinfo {author} {\bibfnamefont {F.}~\bibnamefont {Sette}}, \ and\ \bibinfo
  {author} {\bibfnamefont {G.}~\bibnamefont {van~der Laan}},\ }\href@noop {}
  {\bibfield  {journal} {\bibinfo  {journal} {Phys. Rev. Lett.}\ }\textbf
  {\bibinfo {volume} {68}},\ \bibinfo {pages} {1943} (\bibinfo {year}
  {1992})}\BibitemShut {NoStop}%
\bibitem [{\citenamefont {Carra}\ \emph {et~al.}(1993)\citenamefont {Carra},
  \citenamefont {Thole}, \citenamefont {Altarelli},\ and\ \citenamefont
  {Wang}}]{carra1993x}%
  \BibitemOpen
  \bibfield  {author} {\bibinfo {author} {\bibfnamefont {P.}~\bibnamefont
  {Carra}}, \bibinfo {author} {\bibfnamefont {B.}~\bibnamefont {Thole}},
  \bibinfo {author} {\bibfnamefont {M.}~\bibnamefont {Altarelli}}, \ and\
  \bibinfo {author} {\bibfnamefont {X.}~\bibnamefont {Wang}},\ }\href@noop {}
  {\bibfield  {journal} {\bibinfo  {journal} {Phys. Rev. Lett.}\ }\textbf
  {\bibinfo {volume} {70}},\ \bibinfo {pages} {694} (\bibinfo {year}
  {1993})}\BibitemShut {NoStop}%
\bibitem [{\citenamefont {St{\"o}hr}(1995)}]{stohr1995x}%
  \BibitemOpen
  \bibfield  {author} {\bibinfo {author} {\bibfnamefont {J.}~\bibnamefont
  {St{\"o}hr}},\ }\href@noop {} {\bibfield  {journal} {\bibinfo  {journal} {J.
  Electron Spectrosc. Relat. Phenom.}\ }\textbf {\bibinfo {volume} {75}},\
  \bibinfo {pages} {253} (\bibinfo {year} {1995})}\BibitemShut {NoStop}%
\bibitem [{\citenamefont {St{\"o}hr}(1999)}]{stohr1999exploring}%
  \BibitemOpen
  \bibfield  {author} {\bibinfo {author} {\bibfnamefont {J.}~\bibnamefont
  {St{\"o}hr}},\ }\href@noop {} {\bibfield  {journal} {\bibinfo  {journal} {J.
  Magn. Magn. Mater.}\ }\textbf {\bibinfo {volume} {200}},\ \bibinfo {pages}
  {470} (\bibinfo {year} {1999})}\BibitemShut {NoStop}%
\bibitem [{\citenamefont {Tonnerre}\ \emph {et~al.}(1998)\citenamefont
  {Tonnerre}, \citenamefont {S{\`e}ve}, \citenamefont {Barbara-Dechelette},
  \citenamefont {Bartolom{\'e}}, \citenamefont {Raoux}, \citenamefont
  {Chakarian}, \citenamefont {Kao}, \citenamefont {Fischer}, \citenamefont
  {Andrieu},\ and\ \citenamefont {Fruchart}}]{tonnerre1998soft}%
  \BibitemOpen
  \bibfield  {author} {\bibinfo {author} {\bibfnamefont {J.}~\bibnamefont
  {Tonnerre}}, \bibinfo {author} {\bibfnamefont {L.}~\bibnamefont {S{\`e}ve}},
  \bibinfo {author} {\bibfnamefont {A.}~\bibnamefont {Barbara-Dechelette}},
  \bibinfo {author} {\bibfnamefont {F.}~\bibnamefont {Bartolom{\'e}}}, \bibinfo
  {author} {\bibfnamefont {D.}~\bibnamefont {Raoux}}, \bibinfo {author}
  {\bibfnamefont {V.}~\bibnamefont {Chakarian}}, \bibinfo {author}
  {\bibfnamefont {C.}~\bibnamefont {Kao}}, \bibinfo {author} {\bibfnamefont
  {H.}~\bibnamefont {Fischer}}, \bibinfo {author} {\bibfnamefont
  {S.}~\bibnamefont {Andrieu}}, \ and\ \bibinfo {author} {\bibfnamefont
  {O.}~\bibnamefont {Fruchart}},\ }\href@noop {} {\bibfield  {journal}
  {\bibinfo  {journal} {J. Appl. Phys.}\ }\textbf {\bibinfo {volume} {83}},\
  \bibinfo {pages} {6293} (\bibinfo {year} {1998})}\BibitemShut {NoStop}%
\bibitem [{\citenamefont {Geissler}\ \emph {et~al.}(2001)\citenamefont
  {Geissler}, \citenamefont {Goering}, \citenamefont {Justen}, \citenamefont
  {Weigand}, \citenamefont {Sch{\"u}tz}, \citenamefont {Langer}, \citenamefont
  {Schmitz}, \citenamefont {Maletta},\ and\ \citenamefont
  {Mattheis}}]{geissler2001pt}%
  \BibitemOpen
  \bibfield  {author} {\bibinfo {author} {\bibfnamefont {J.}~\bibnamefont
  {Geissler}}, \bibinfo {author} {\bibfnamefont {E.}~\bibnamefont {Goering}},
  \bibinfo {author} {\bibfnamefont {M.}~\bibnamefont {Justen}}, \bibinfo
  {author} {\bibfnamefont {F.}~\bibnamefont {Weigand}}, \bibinfo {author}
  {\bibfnamefont {G.}~\bibnamefont {Sch{\"u}tz}}, \bibinfo {author}
  {\bibfnamefont {J.}~\bibnamefont {Langer}}, \bibinfo {author} {\bibfnamefont
  {D.}~\bibnamefont {Schmitz}}, \bibinfo {author} {\bibfnamefont
  {H.}~\bibnamefont {Maletta}}, \ and\ \bibinfo {author} {\bibfnamefont
  {R.}~\bibnamefont {Mattheis}},\ }\href@noop {} {\bibfield  {journal}
  {\bibinfo  {journal} {Phys. Rev. B}\ }\textbf {\bibinfo {volume} {65}},\
  \bibinfo {pages} {020405} (\bibinfo {year} {2001})}\BibitemShut {NoStop}%
\bibitem [{\citenamefont {Lee}\ \emph {et~al.}(2003)\citenamefont {Lee},
  \citenamefont {Sinha}, \citenamefont {Haskel}, \citenamefont {Choi},
  \citenamefont {Lang}, \citenamefont {Stepanov},\ and\ \citenamefont
  {Srajer}}]{lee2003x}%
  \BibitemOpen
  \bibfield  {author} {\bibinfo {author} {\bibfnamefont {D.}~\bibnamefont
  {Lee}}, \bibinfo {author} {\bibfnamefont {S.}~\bibnamefont {Sinha}}, \bibinfo
  {author} {\bibfnamefont {D.}~\bibnamefont {Haskel}}, \bibinfo {author}
  {\bibfnamefont {Y.}~\bibnamefont {Choi}}, \bibinfo {author} {\bibfnamefont
  {J.}~\bibnamefont {Lang}}, \bibinfo {author} {\bibfnamefont {S.}~\bibnamefont
  {Stepanov}}, \ and\ \bibinfo {author} {\bibfnamefont {G.}~\bibnamefont
  {Srajer}},\ }\href@noop {} {\bibfield  {journal} {\bibinfo  {journal} {Phys.
  Rev. B}\ }\textbf {\bibinfo {volume} {68}},\ \bibinfo {pages} {224409}
  (\bibinfo {year} {2003})}\BibitemShut {NoStop}%
\bibitem [{\citenamefont {Roy}\ \emph {et~al.}(2007)\citenamefont {Roy},
  \citenamefont {Sanchez-Hanke}, \citenamefont {Park}, \citenamefont
  {Fitzsimmons}, \citenamefont {Tang}, \citenamefont {Hong}, \citenamefont
  {Smith}, \citenamefont {Taylor}, \citenamefont {Liu}, \citenamefont {Maple},
  \citenamefont {Berkowitz}, \citenamefont {Kao},\ and\ \citenamefont
  {Sinha}}]{roy2007evidence}%
  \BibitemOpen
  \bibfield  {author} {\bibinfo {author} {\bibfnamefont {S.}~\bibnamefont
  {Roy}}, \bibinfo {author} {\bibfnamefont {C.}~\bibnamefont {Sanchez-Hanke}},
  \bibinfo {author} {\bibfnamefont {S.}~\bibnamefont {Park}}, \bibinfo {author}
  {\bibfnamefont {M.}~\bibnamefont {Fitzsimmons}}, \bibinfo {author}
  {\bibfnamefont {Y.}~\bibnamefont {Tang}}, \bibinfo {author} {\bibfnamefont
  {J.}~\bibnamefont {Hong}}, \bibinfo {author} {\bibfnamefont {D.~J.}\
  \bibnamefont {Smith}}, \bibinfo {author} {\bibfnamefont {B.}~\bibnamefont
  {Taylor}}, \bibinfo {author} {\bibfnamefont {X.}~\bibnamefont {Liu}},
  \bibinfo {author} {\bibfnamefont {M.}~\bibnamefont {Maple}}, \bibinfo
  {author} {\bibfnamefont {A.~E.}\ \bibnamefont {Berkowitz}}, \bibinfo {author}
  {\bibfnamefont {C.-C.}\ \bibnamefont {Kao}}, \ and\ \bibinfo {author}
  {\bibfnamefont {S.~K.}\ \bibnamefont {Sinha}},\ }\href@noop {} {\bibfield
  {journal} {\bibinfo  {journal} {Phys. Rev. B}\ }\textbf {\bibinfo {volume}
  {75}},\ \bibinfo {pages} {014442} (\bibinfo {year} {2007})}\BibitemShut
  {NoStop}%
\bibitem [{\citenamefont {Macke}\ and\ \citenamefont
  {Goering}(2014)}]{macke2014magnetic}%
  \BibitemOpen
  \bibfield  {author} {\bibinfo {author} {\bibfnamefont {S.}~\bibnamefont
  {Macke}}\ and\ \bibinfo {author} {\bibfnamefont {E.}~\bibnamefont
  {Goering}},\ }\href@noop {} {\bibfield  {journal} {\bibinfo  {journal} {J.
  Phys. Condens. Matter}\ }\textbf {\bibinfo {volume} {26}},\ \bibinfo {pages}
  {363201} (\bibinfo {year} {2014})}\BibitemShut {NoStop}%
\bibitem [{\citenamefont {Kuschel}\ \emph {et~al.}(2015)\citenamefont
  {Kuschel}, \citenamefont {Klewe}, \citenamefont {Schmalhorst}, \citenamefont
  {Bertram}, \citenamefont {Kuschel}, \citenamefont {Schemme}, \citenamefont
  {Wollschl{\"a}ger}, \citenamefont {Francoual}, \citenamefont {Strempfer},
  \citenamefont {Gupta}, \citenamefont {Meinert}, \citenamefont {G{\"o}tz},
  \citenamefont {Meier},\ and\ \citenamefont {Reiss}}]{kuschel2015static}%
  \BibitemOpen
  \bibfield  {author} {\bibinfo {author} {\bibfnamefont {T.}~\bibnamefont
  {Kuschel}}, \bibinfo {author} {\bibfnamefont {C.}~\bibnamefont {Klewe}},
  \bibinfo {author} {\bibfnamefont {J.-M.}\ \bibnamefont {Schmalhorst}},
  \bibinfo {author} {\bibfnamefont {F.}~\bibnamefont {Bertram}}, \bibinfo
  {author} {\bibfnamefont {O.}~\bibnamefont {Kuschel}}, \bibinfo {author}
  {\bibfnamefont {T.}~\bibnamefont {Schemme}}, \bibinfo {author} {\bibfnamefont
  {J.}~\bibnamefont {Wollschl{\"a}ger}}, \bibinfo {author} {\bibfnamefont
  {S.}~\bibnamefont {Francoual}}, \bibinfo {author} {\bibfnamefont
  {J.}~\bibnamefont {Strempfer}}, \bibinfo {author} {\bibfnamefont
  {A.}~\bibnamefont {Gupta}}, \bibinfo {author} {\bibfnamefont
  {M.}~\bibnamefont {Meinert}}, \bibinfo {author} {\bibfnamefont
  {G.}~\bibnamefont {G{\"o}tz}}, \bibinfo {author} {\bibfnamefont
  {D.}~\bibnamefont {Meier}}, \ and\ \bibinfo {author} {\bibfnamefont
  {G.}~\bibnamefont {Reiss}},\ }\href@noop {} {\bibfield  {journal} {\bibinfo
  {journal} {Phys. Rev. Lett.}\ }\textbf {\bibinfo {volume} {115}},\ \bibinfo
  {pages} {097401} (\bibinfo {year} {2015})}\BibitemShut {NoStop}%
\bibitem [{\citenamefont {Klewe}\ \emph {et~al.}(2016)\citenamefont {Klewe},
  \citenamefont {Kuschel}, \citenamefont {Schmalhorst}, \citenamefont
  {Bertram}, \citenamefont {Kuschel}, \citenamefont {Wollschl\"ager},
  \citenamefont {Strempfer}, \citenamefont {Meinert},\ and\ \citenamefont
  {Reiss}}]{PhysRevB.93.214440}%
  \BibitemOpen
  \bibfield  {author} {\bibinfo {author} {\bibfnamefont {C.}~\bibnamefont
  {Klewe}}, \bibinfo {author} {\bibfnamefont {T.}~\bibnamefont {Kuschel}},
  \bibinfo {author} {\bibfnamefont {J.-M.}\ \bibnamefont {Schmalhorst}},
  \bibinfo {author} {\bibfnamefont {F.}~\bibnamefont {Bertram}}, \bibinfo
  {author} {\bibfnamefont {O.}~\bibnamefont {Kuschel}}, \bibinfo {author}
  {\bibfnamefont {J.}~\bibnamefont {Wollschl\"ager}}, \bibinfo {author}
  {\bibfnamefont {J.}~\bibnamefont {Strempfer}}, \bibinfo {author}
  {\bibfnamefont {M.}~\bibnamefont {Meinert}}, \ and\ \bibinfo {author}
  {\bibfnamefont {G.}~\bibnamefont {Reiss}},\ }\href {\doibase
  10.1103/PhysRevB.93.214440} {\bibfield  {journal} {\bibinfo  {journal} {Phys.
  Rev. B}\ }\textbf {\bibinfo {volume} {93}},\ \bibinfo {pages} {214440}
  (\bibinfo {year} {2016})}\BibitemShut {NoStop}%
\bibitem [{\citenamefont {Seve}\ \emph {et~al.}(1999)\citenamefont {Seve},
  \citenamefont {Jaouen}, \citenamefont {Tonnerre}, \citenamefont {Raoux},
  \citenamefont {Bartolom{\'e}}, \citenamefont {Arend}, \citenamefont {Felsch},
  \citenamefont {Rogalev}, \citenamefont {Goulon}, \citenamefont {Gautier},\
  and\ \citenamefont {B{\'e}rar}}]{seve1999profile}%
  \BibitemOpen
  \bibfield  {author} {\bibinfo {author} {\bibfnamefont {L.}~\bibnamefont
  {Seve}}, \bibinfo {author} {\bibfnamefont {N.}~\bibnamefont {Jaouen}},
  \bibinfo {author} {\bibfnamefont {J.}~\bibnamefont {Tonnerre}}, \bibinfo
  {author} {\bibfnamefont {D.}~\bibnamefont {Raoux}}, \bibinfo {author}
  {\bibfnamefont {F.}~\bibnamefont {Bartolom{\'e}}}, \bibinfo {author}
  {\bibfnamefont {M.}~\bibnamefont {Arend}}, \bibinfo {author} {\bibfnamefont
  {W.}~\bibnamefont {Felsch}}, \bibinfo {author} {\bibfnamefont
  {A.}~\bibnamefont {Rogalev}}, \bibinfo {author} {\bibfnamefont
  {J.}~\bibnamefont {Goulon}}, \bibinfo {author} {\bibfnamefont
  {C.}~\bibnamefont {Gautier}}, \ and\ \bibinfo {author} {\bibfnamefont
  {J.~F.}\ \bibnamefont {B{\'e}rar}},\ }\href@noop {} {\bibfield  {journal}
  {\bibinfo  {journal} {Phys. Rev. B}\ }\textbf {\bibinfo {volume} {60}},\
  \bibinfo {pages} {9662} (\bibinfo {year} {1999})}\BibitemShut {NoStop}%
\bibitem [{\citenamefont {Kuschel}\ \emph {et~al.}(2016)\citenamefont
  {Kuschel}, \citenamefont {Klewe}, \citenamefont {Bougiatioti}, \citenamefont
  {Kuschel}, \citenamefont {Wollschl\"ager}, \citenamefont {Bouchenoire},
  \citenamefont {Brown}, \citenamefont {Schmalhorst}, \citenamefont {Meier},\
  and\ \citenamefont {Reiss}}]{kuschelstatic}%
  \BibitemOpen
  \bibfield  {author} {\bibinfo {author} {\bibfnamefont {T.}~\bibnamefont
  {Kuschel}}, \bibinfo {author} {\bibfnamefont {C.}~\bibnamefont {Klewe}},
  \bibinfo {author} {\bibfnamefont {P.}~\bibnamefont {Bougiatioti}}, \bibinfo
  {author} {\bibfnamefont {O.}~\bibnamefont {Kuschel}}, \bibinfo {author}
  {\bibfnamefont {J.}~\bibnamefont {Wollschl\"ager}}, \bibinfo {author}
  {\bibfnamefont {L.}~\bibnamefont {Bouchenoire}}, \bibinfo {author}
  {\bibfnamefont {S.}~\bibnamefont {Brown}}, \bibinfo {author} {\bibfnamefont
  {J.}~\bibnamefont {Schmalhorst}}, \bibinfo {author} {\bibfnamefont
  {D.}~\bibnamefont {Meier}}, \ and\ \bibinfo {author} {\bibfnamefont
  {G.}~\bibnamefont {Reiss}},\ }\href@noop {} {\bibfield  {journal} {\bibinfo
  {journal} {IEEE Trans. Magn.}\ }\textbf {\bibinfo {volume} {52}},\ \bibinfo
  {pages} {4500104} (\bibinfo {year} {2016})}\BibitemShut {NoStop}%
\bibitem [{\citenamefont {Althammer}\ \emph {et~al.}(2013)\citenamefont
  {Althammer}, \citenamefont {Meyer}, \citenamefont {Nakayama}, \citenamefont
  {Schreier}, \citenamefont {Altmannshofer}, \citenamefont {Weiler},
  \citenamefont {Huebl}, \citenamefont {Gepr\"ags}, \citenamefont {Opel},
  \citenamefont {Gross}, \citenamefont {Meier}, \citenamefont {Klewe},
  \citenamefont {Kuschel}, \citenamefont {Schmalhorst}, \citenamefont {Reiss},
  \citenamefont {Shen}, \citenamefont {Gupta}, \citenamefont {Chen},
  \citenamefont {Bauer}, \citenamefont {Saitoh},\ and\ \citenamefont
  {Goennenwein}}]{althammer2013quantitative}%
  \BibitemOpen
  \bibfield  {author} {\bibinfo {author} {\bibfnamefont {M.}~\bibnamefont
  {Althammer}}, \bibinfo {author} {\bibfnamefont {S.}~\bibnamefont {Meyer}},
  \bibinfo {author} {\bibfnamefont {H.}~\bibnamefont {Nakayama}}, \bibinfo
  {author} {\bibfnamefont {M.}~\bibnamefont {Schreier}}, \bibinfo {author}
  {\bibfnamefont {S.}~\bibnamefont {Altmannshofer}}, \bibinfo {author}
  {\bibfnamefont {M.}~\bibnamefont {Weiler}}, \bibinfo {author} {\bibfnamefont
  {H.}~\bibnamefont {Huebl}}, \bibinfo {author} {\bibfnamefont
  {S.}~\bibnamefont {Gepr\"ags}}, \bibinfo {author} {\bibfnamefont
  {M.}~\bibnamefont {Opel}}, \bibinfo {author} {\bibfnamefont {R.}~\bibnamefont
  {Gross}}, \bibinfo {author} {\bibfnamefont {D.}~\bibnamefont {Meier}},
  \bibinfo {author} {\bibfnamefont {C.}~\bibnamefont {Klewe}}, \bibinfo
  {author} {\bibfnamefont {T.}~\bibnamefont {Kuschel}}, \bibinfo {author}
  {\bibfnamefont {J.-M.}\ \bibnamefont {Schmalhorst}}, \bibinfo {author}
  {\bibfnamefont {G.}~\bibnamefont {Reiss}}, \bibinfo {author} {\bibfnamefont
  {L.}~\bibnamefont {Shen}}, \bibinfo {author} {\bibfnamefont {A.}~\bibnamefont
  {Gupta}}, \bibinfo {author} {\bibfnamefont {Y.-T.}\ \bibnamefont {Chen}},
  \bibinfo {author} {\bibfnamefont {G.~E.~W.}\ \bibnamefont {Bauer}}, \bibinfo
  {author} {\bibfnamefont {E.}~\bibnamefont {Saitoh}}, \ and\ \bibinfo {author}
  {\bibfnamefont {S.~T.~B.}\ \bibnamefont {Goennenwein}},\ }\href {\doibase
  10.1103/PhysRevB.87.224401} {\bibfield  {journal} {\bibinfo  {journal} {Phys.
  Rev. B}\ }\textbf {\bibinfo {volume} {87}},\ \bibinfo {pages} {224401}
  (\bibinfo {year} {2013})}\BibitemShut {NoStop}%
\bibitem [{\citenamefont {Huang}\ \emph {et~al.}(2012)\citenamefont {Huang},
  \citenamefont {Fan}, \citenamefont {Qu}, \citenamefont {Chen}, \citenamefont
  {Wang}, \citenamefont {Wu}, \citenamefont {Chen}, \citenamefont {Xiao},\ and\
  \citenamefont {Chien}}]{huang2012transport}%
  \BibitemOpen
  \bibfield  {author} {\bibinfo {author} {\bibfnamefont {S.-Y.}\ \bibnamefont
  {Huang}}, \bibinfo {author} {\bibfnamefont {X.}~\bibnamefont {Fan}}, \bibinfo
  {author} {\bibfnamefont {D.}~\bibnamefont {Qu}}, \bibinfo {author}
  {\bibfnamefont {Y.}~\bibnamefont {Chen}}, \bibinfo {author} {\bibfnamefont
  {W.}~\bibnamefont {Wang}}, \bibinfo {author} {\bibfnamefont {J.}~\bibnamefont
  {Wu}}, \bibinfo {author} {\bibfnamefont {T.}~\bibnamefont {Chen}}, \bibinfo
  {author} {\bibfnamefont {J.}~\bibnamefont {Xiao}}, \ and\ \bibinfo {author}
  {\bibfnamefont {C.}~\bibnamefont {Chien}},\ }\href@noop {} {\bibfield
  {journal} {\bibinfo  {journal} {Phys. Rev. Lett.}\ }\textbf {\bibinfo
  {volume} {109}},\ \bibinfo {pages} {107204} (\bibinfo {year}
  {2012})}\BibitemShut {NoStop}%
\bibitem [{\citenamefont {Bougiatioti}\ \emph {et~al.}(2017)\citenamefont
  {Bougiatioti}, \citenamefont {Klewe}, \citenamefont {Meier}, \citenamefont
  {Manos}, \citenamefont {Kuschel}, \citenamefont {Wollschl\"ager},
  \citenamefont {Bouchenoire}, \citenamefont {Brown}, \citenamefont
  {Schmalhorst}, \citenamefont {Reiss},\ and\ \citenamefont
  {Kuschel}}]{bougiatioti2017quantitative}%
  \BibitemOpen
  \bibfield  {author} {\bibinfo {author} {\bibfnamefont {P.}~\bibnamefont
  {Bougiatioti}}, \bibinfo {author} {\bibfnamefont {C.}~\bibnamefont {Klewe}},
  \bibinfo {author} {\bibfnamefont {D.}~\bibnamefont {Meier}}, \bibinfo
  {author} {\bibfnamefont {O.}~\bibnamefont {Manos}}, \bibinfo {author}
  {\bibfnamefont {O.}~\bibnamefont {Kuschel}}, \bibinfo {author} {\bibfnamefont
  {J.}~\bibnamefont {Wollschl\"ager}}, \bibinfo {author} {\bibfnamefont
  {L.}~\bibnamefont {Bouchenoire}}, \bibinfo {author} {\bibfnamefont {S.~D.}\
  \bibnamefont {Brown}}, \bibinfo {author} {\bibfnamefont {J.-M.}\ \bibnamefont
  {Schmalhorst}}, \bibinfo {author} {\bibfnamefont {G.}~\bibnamefont {Reiss}},
  \ and\ \bibinfo {author} {\bibfnamefont {T.}~\bibnamefont {Kuschel}},\ }\href
  {\doibase 10.1103/PhysRevLett.119.227205} {\bibfield  {journal} {\bibinfo
  {journal} {Phys. Rev. Lett.}\ }\textbf {\bibinfo {volume} {119}},\ \bibinfo
  {pages} {227205} (\bibinfo {year} {2017})}\BibitemShut {NoStop}%
\bibitem [{\citenamefont {Macke}()}]{ReMagX}%
  \BibitemOpen
  \bibfield  {author} {\bibinfo {author} {\bibfnamefont {S.}~\bibnamefont
  {Macke}},\ }\href@noop {} {\bibinfo  {journal} {ReMagX - x-ray magnetic
  reflectivity tool, www.remagx.org}\ }\BibitemShut {NoStop}%
\bibitem [{\citenamefont {Rowan-Robinson}\ \emph {et~al.}(2017)\citenamefont
  {Rowan-Robinson}, \citenamefont {Stashkevich}, \citenamefont {Roussign{\'e}},
  \citenamefont {Belmeguenai}, \citenamefont {Ch{\'e}rif}, \citenamefont
  {Thiaville}, \citenamefont {Hase}, \citenamefont {Hindmarch},\ and\
  \citenamefont {Atkinson}}]{rowan2017interfacial}%
  \BibitemOpen
\bibfield  {journal} {  }\bibfield  {author} {\bibinfo {author} {\bibfnamefont
  {R.~M.}\ \bibnamefont {Rowan-Robinson}}, \bibinfo {author} {\bibfnamefont
  {A.}~\bibnamefont {Stashkevich}}, \bibinfo {author} {\bibfnamefont
  {Y.}~\bibnamefont {Roussign{\'e}}}, \bibinfo {author} {\bibfnamefont
  {M.}~\bibnamefont {Belmeguenai}}, \bibinfo {author} {\bibfnamefont {S.-M.}\
  \bibnamefont {Ch{\'e}rif}}, \bibinfo {author} {\bibfnamefont
  {A.}~\bibnamefont {Thiaville}}, \bibinfo {author} {\bibfnamefont
  {T.}~\bibnamefont {Hase}}, \bibinfo {author} {\bibfnamefont {A.}~\bibnamefont
  {Hindmarch}}, \ and\ \bibinfo {author} {\bibfnamefont {D.}~\bibnamefont
  {Atkinson}},\ }\href@noop {} {\bibfield  {journal} {\bibinfo  {journal} {Sci.
  Rep.}\ }\textbf {\bibinfo {volume} {7}},\ \bibinfo {pages} {16835} (\bibinfo
  {year} {2017})}\BibitemShut {NoStop}%
\bibitem [{\citenamefont {Mukhopadhyay}\ \emph {et~al.}(2019)\citenamefont
  {Mukhopadhyay}, \citenamefont {Vayalil}, \citenamefont {Graulich},
  \citenamefont {Ahmed}, \citenamefont {Francoual}, \citenamefont {Kashyap},
  \citenamefont {Kuschel},\ and\ \citenamefont {Kumar}}]{mukhopadhyay2019}%
  \BibitemOpen
  \bibfield  {author} {\bibinfo {author} {\bibfnamefont {A.}~\bibnamefont
  {Mukhopadhyay}}, \bibinfo {author} {\bibfnamefont {S.~K.}\ \bibnamefont
  {Vayalil}}, \bibinfo {author} {\bibfnamefont {D.}~\bibnamefont {Graulich}},
  \bibinfo {author} {\bibfnamefont {I.}~\bibnamefont {Ahmed}}, \bibinfo
  {author} {\bibfnamefont {S.}~\bibnamefont {Francoual}}, \bibinfo {author}
  {\bibfnamefont {A.}~\bibnamefont {Kashyap}}, \bibinfo {author} {\bibfnamefont
  {T.}~\bibnamefont {Kuschel}}, \ and\ \bibinfo {author} {\bibfnamefont
  {P.~S.~A.}\ \bibnamefont {Kumar}},\ }\href@noop {} {\bibfield  {journal}
  {\bibinfo  {journal} {arXiv:1911.12187}\ } (\bibinfo {year}
  {2019})}\BibitemShut {NoStop}%
\bibitem [{\citenamefont {Moskaltsova}\ \emph {et~al.}(2020)\citenamefont
  {Moskaltsova}, \citenamefont {Krieft}, \citenamefont {Graulich},
  \citenamefont {Matalla-Wagner}, \citenamefont {Francoual},\ and\
  \citenamefont {Kuschel}}]{moskaltsova2019}%
  \BibitemOpen
  \bibfield  {author} {\bibinfo {author} {\bibfnamefont {A.}~\bibnamefont
  {Moskaltsova}}, \bibinfo {author} {\bibfnamefont {J.}~\bibnamefont {Krieft}},
  \bibinfo {author} {\bibfnamefont {D.}~\bibnamefont {Graulich}}, \bibinfo
  {author} {\bibfnamefont {T.}~\bibnamefont {Matalla-Wagner}}, \bibinfo
  {author} {\bibfnamefont {S.}~\bibnamefont {Francoual}}, \ and\ \bibinfo
  {author} {\bibfnamefont {T.}~\bibnamefont {Kuschel}},\ }\href@noop {}
  {\bibfield  {journal} {\bibinfo  {journal} {AIP Advances}\ }\textbf {\bibinfo
  {volume} {10}},\ \bibinfo {pages} {015154} (\bibinfo {year}
  {2020})}\BibitemShut {NoStop}%
\bibitem [{\citenamefont {Krieft}\ \emph {et~al.}(2017)\citenamefont {Krieft},
  \citenamefont {Mendil}, \citenamefont {Aguirre}, \citenamefont {Avci},
  \citenamefont {Klewe}, \citenamefont {Rott}, \citenamefont {Schmalhorst},
  \citenamefont {Reiss}, \citenamefont {Gambardella},\ and\ \citenamefont
  {Kuschel}}]{krieft2017co}%
  \BibitemOpen
  \bibfield  {author} {\bibinfo {author} {\bibfnamefont {J.}~\bibnamefont
  {Krieft}}, \bibinfo {author} {\bibfnamefont {J.}~\bibnamefont {Mendil}},
  \bibinfo {author} {\bibfnamefont {M.~H.}\ \bibnamefont {Aguirre}}, \bibinfo
  {author} {\bibfnamefont {C.~O.}\ \bibnamefont {Avci}}, \bibinfo {author}
  {\bibfnamefont {C.}~\bibnamefont {Klewe}}, \bibinfo {author} {\bibfnamefont
  {K.}~\bibnamefont {Rott}}, \bibinfo {author} {\bibfnamefont {J.-M.}\
  \bibnamefont {Schmalhorst}}, \bibinfo {author} {\bibfnamefont
  {G.}~\bibnamefont {Reiss}}, \bibinfo {author} {\bibfnamefont
  {P.}~\bibnamefont {Gambardella}}, \ and\ \bibinfo {author} {\bibfnamefont
  {T.}~\bibnamefont {Kuschel}},\ }\href@noop {} {\bibfield  {journal} {\bibinfo
   {journal} {Phys. Status Solidi Rapid Res. Lett.}\ }\textbf {\bibinfo
  {volume} {11}},\ \bibinfo {pages} {1600439} (\bibinfo {year}
  {2017})}\BibitemShut {NoStop}%
\bibitem [{\citenamefont {Geissler}\ \emph {et~al.}(2002)\citenamefont
  {Geissler}, \citenamefont {Goering}, \citenamefont {Weigand},\ and\
  \citenamefont {Sch{\"u}tz}}]{geissler2002interplay}%
  \BibitemOpen
  \bibfield  {author} {\bibinfo {author} {\bibfnamefont {J.}~\bibnamefont
  {Geissler}}, \bibinfo {author} {\bibfnamefont {E.}~\bibnamefont {Goering}},
  \bibinfo {author} {\bibfnamefont {F.}~\bibnamefont {Weigand}}, \ and\
  \bibinfo {author} {\bibfnamefont {G.}~\bibnamefont {Sch{\"u}tz}},\
  }\href@noop {} {\bibfield  {journal} {\bibinfo  {journal} {Z. Metallkd.}\
  }\textbf {\bibinfo {volume} {93}},\ \bibinfo {pages} {946} (\bibinfo {year}
  {2002})}\BibitemShut {NoStop}%
\bibitem [{\citenamefont {Awaji}\ \emph {et~al.}(2007)\citenamefont {Awaji},
  \citenamefont {Noma}, \citenamefont {Nomura}, \citenamefont {Doi},
  \citenamefont {Hirono}, \citenamefont {Kimura},\ and\ \citenamefont
  {Nakamura}}]{awaji2007soft}%
  \BibitemOpen
  \bibfield  {author} {\bibinfo {author} {\bibfnamefont {N.}~\bibnamefont
  {Awaji}}, \bibinfo {author} {\bibfnamefont {K.}~\bibnamefont {Noma}},
  \bibinfo {author} {\bibfnamefont {K.}~\bibnamefont {Nomura}}, \bibinfo
  {author} {\bibfnamefont {S.}~\bibnamefont {Doi}}, \bibinfo {author}
  {\bibfnamefont {T.}~\bibnamefont {Hirono}}, \bibinfo {author} {\bibfnamefont
  {H.}~\bibnamefont {Kimura}}, \ and\ \bibinfo {author} {\bibfnamefont
  {T.}~\bibnamefont {Nakamura}},\ }\href@noop {} {\bibfield  {journal}
  {\bibinfo  {journal} {J. Phys. Conf. Ser.}\ }\textbf {\bibinfo {volume}
  {83}},\ \bibinfo {pages} {012034} (\bibinfo {year} {2007})}\BibitemShut
  {NoStop}%
\bibitem [{\citenamefont {Ederer}\ \emph {et~al.}(2002)\citenamefont {Ederer},
  \citenamefont {Komelj}, \citenamefont {F{\"a}hnle},\ and\ \citenamefont
  {Sch{\"u}tz}}]{ederer2002theory}%
  \BibitemOpen
  \bibfield  {author} {\bibinfo {author} {\bibfnamefont {C.}~\bibnamefont
  {Ederer}}, \bibinfo {author} {\bibfnamefont {M.}~\bibnamefont {Komelj}},
  \bibinfo {author} {\bibfnamefont {M.}~\bibnamefont {F{\"a}hnle}}, \ and\
  \bibinfo {author} {\bibfnamefont {G.}~\bibnamefont {Sch{\"u}tz}},\
  }\href@noop {} {\bibfield  {journal} {\bibinfo  {journal} {Phys. Rev. B}\
  }\textbf {\bibinfo {volume} {66}},\ \bibinfo {pages} {094413} (\bibinfo
  {year} {2002})}\BibitemShut {NoStop}%
\bibitem [{\citenamefont {Bertinshaw}\ \emph {et~al.}(2014)\citenamefont
  {Bertinshaw}, \citenamefont {Br{\"u}ck}, \citenamefont {Lott}, \citenamefont
  {Fritzsche}, \citenamefont {Khaydukov}, \citenamefont {Soltwedel},
  \citenamefont {Keller}, \citenamefont {Goering}, \citenamefont {Audehm},
  \citenamefont {Cortie}, \citenamefont {Hutchison}, \citenamefont {Ramasse},
  \citenamefont {Arredondo}, \citenamefont {Maran}, \citenamefont {Nagarajan},
  \citenamefont {Klose},\ and\ \citenamefont {Ulrich}}]{bertinshaw2014element}%
  \BibitemOpen
  \bibfield  {author} {\bibinfo {author} {\bibfnamefont {J.}~\bibnamefont
  {Bertinshaw}}, \bibinfo {author} {\bibfnamefont {S.}~\bibnamefont
  {Br{\"u}ck}}, \bibinfo {author} {\bibfnamefont {D.}~\bibnamefont {Lott}},
  \bibinfo {author} {\bibfnamefont {H.}~\bibnamefont {Fritzsche}}, \bibinfo
  {author} {\bibfnamefont {Y.}~\bibnamefont {Khaydukov}}, \bibinfo {author}
  {\bibfnamefont {O.}~\bibnamefont {Soltwedel}}, \bibinfo {author}
  {\bibfnamefont {T.}~\bibnamefont {Keller}}, \bibinfo {author} {\bibfnamefont
  {E.}~\bibnamefont {Goering}}, \bibinfo {author} {\bibfnamefont
  {P.}~\bibnamefont {Audehm}}, \bibinfo {author} {\bibfnamefont {D.~L.}\
  \bibnamefont {Cortie}}, \bibinfo {author} {\bibfnamefont {W.~D.}\
  \bibnamefont {Hutchison}}, \bibinfo {author} {\bibfnamefont {Q.~M.}\
  \bibnamefont {Ramasse}}, \bibinfo {author} {\bibfnamefont {M.}~\bibnamefont
  {Arredondo}}, \bibinfo {author} {\bibfnamefont {R.}~\bibnamefont {Maran}},
  \bibinfo {author} {\bibfnamefont {V.}~\bibnamefont {Nagarajan}}, \bibinfo
  {author} {\bibfnamefont {F.}~\bibnamefont {Klose}}, \ and\ \bibinfo {author}
  {\bibfnamefont {C.}~\bibnamefont {Ulrich}},\ }\href@noop {} {\bibfield
  {journal} {\bibinfo  {journal} {Phys. Rev. B}\ }\textbf {\bibinfo {volume}
  {90}},\ \bibinfo {pages} {041113} (\bibinfo {year} {2014})}\BibitemShut
  {NoStop}%
\bibitem [{\citenamefont {Blackburn}\ \emph {et~al.}(2008)\citenamefont
  {Blackburn}, \citenamefont {Sanchez-Hanke}, \citenamefont {Roy},
  \citenamefont {Smith}, \citenamefont {Hong}, \citenamefont {Chan},
  \citenamefont {Berkowitz},\ and\ \citenamefont
  {Sinha}}]{blackburn2008pinned}%
  \BibitemOpen
  \bibfield  {author} {\bibinfo {author} {\bibfnamefont {E.}~\bibnamefont
  {Blackburn}}, \bibinfo {author} {\bibfnamefont {C.}~\bibnamefont
  {Sanchez-Hanke}}, \bibinfo {author} {\bibfnamefont {S.}~\bibnamefont {Roy}},
  \bibinfo {author} {\bibfnamefont {D.}~\bibnamefont {Smith}}, \bibinfo
  {author} {\bibfnamefont {J.-I.}\ \bibnamefont {Hong}}, \bibinfo {author}
  {\bibfnamefont {K.}~\bibnamefont {Chan}}, \bibinfo {author} {\bibfnamefont
  {A.}~\bibnamefont {Berkowitz}}, \ and\ \bibinfo {author} {\bibfnamefont
  {S.}~\bibnamefont {Sinha}},\ }\href@noop {} {\bibfield  {journal} {\bibinfo
  {journal} {Phys. Rev. B}\ }\textbf {\bibinfo {volume} {78}},\ \bibinfo
  {pages} {180408} (\bibinfo {year} {2008})}\BibitemShut {NoStop}%
\bibitem [{\citenamefont {Przybylski}\ \emph {et~al.}(2012)\citenamefont
  {Przybylski}, \citenamefont {Tonnerre}, \citenamefont {Yildiz}, \citenamefont
  {Tolentino},\ and\ \citenamefont {Kirschner}}]{przybylski2012non}%
  \BibitemOpen
  \bibfield  {author} {\bibinfo {author} {\bibfnamefont {M.}~\bibnamefont
  {Przybylski}}, \bibinfo {author} {\bibfnamefont {J.-M.}\ \bibnamefont
  {Tonnerre}}, \bibinfo {author} {\bibfnamefont {F.}~\bibnamefont {Yildiz}},
  \bibinfo {author} {\bibfnamefont {H.}~\bibnamefont {Tolentino}}, \ and\
  \bibinfo {author} {\bibfnamefont {J.}~\bibnamefont {Kirschner}},\ }\href@noop
  {} {\bibfield  {journal} {\bibinfo  {journal} {J. of Appl. Phys.}\ }\textbf
  {\bibinfo {volume} {111}},\ \bibinfo {pages} {07C103} (\bibinfo {year}
  {2012})}\BibitemShut {NoStop}%
\bibitem [{\citenamefont {Haskel}\ \emph {et~al.}(2001)\citenamefont {Haskel},
  \citenamefont {Srajer}, \citenamefont {Lang}, \citenamefont {Pollmann},
  \citenamefont {Nelson}, \citenamefont {Jiang},\ and\ \citenamefont
  {Bader}}]{haskel2001enhanced}%
  \BibitemOpen
  \bibfield  {author} {\bibinfo {author} {\bibfnamefont {D.}~\bibnamefont
  {Haskel}}, \bibinfo {author} {\bibfnamefont {G.}~\bibnamefont {Srajer}},
  \bibinfo {author} {\bibfnamefont {J.}~\bibnamefont {Lang}}, \bibinfo {author}
  {\bibfnamefont {J.}~\bibnamefont {Pollmann}}, \bibinfo {author}
  {\bibfnamefont {C.}~\bibnamefont {Nelson}}, \bibinfo {author} {\bibfnamefont
  {J.}~\bibnamefont {Jiang}}, \ and\ \bibinfo {author} {\bibfnamefont
  {S.}~\bibnamefont {Bader}},\ }\href@noop {} {\bibfield  {journal} {\bibinfo
  {journal} {Phys. Rev. Lett.}\ }\textbf {\bibinfo {volume} {87}},\ \bibinfo
  {pages} {207201} (\bibinfo {year} {2001})}\BibitemShut {NoStop}%
\bibitem [{\citenamefont {Gibert}\ \emph {et~al.}(2016)\citenamefont {Gibert},
  \citenamefont {Viret}, \citenamefont {Zubko}, \citenamefont {Jaouen},
  \citenamefont {Tonnerre}, \citenamefont {Torres-Pardo}, \citenamefont
  {Catalano}, \citenamefont {Gloter}, \citenamefont {St{\'e}phan},\ and\
  \citenamefont {Triscone}}]{gibert2016interlayer}%
  \BibitemOpen
  \bibfield  {author} {\bibinfo {author} {\bibfnamefont {M.}~\bibnamefont
  {Gibert}}, \bibinfo {author} {\bibfnamefont {M.}~\bibnamefont {Viret}},
  \bibinfo {author} {\bibfnamefont {P.}~\bibnamefont {Zubko}}, \bibinfo
  {author} {\bibfnamefont {N.}~\bibnamefont {Jaouen}}, \bibinfo {author}
  {\bibfnamefont {J.-M.}\ \bibnamefont {Tonnerre}}, \bibinfo {author}
  {\bibfnamefont {A.}~\bibnamefont {Torres-Pardo}}, \bibinfo {author}
  {\bibfnamefont {S.}~\bibnamefont {Catalano}}, \bibinfo {author}
  {\bibfnamefont {A.}~\bibnamefont {Gloter}}, \bibinfo {author} {\bibfnamefont
  {O.}~\bibnamefont {St{\'e}phan}}, \ and\ \bibinfo {author} {\bibfnamefont
  {J.-M.}\ \bibnamefont {Triscone}},\ }\href@noop {} {\bibfield  {journal}
  {\bibinfo  {journal} {Nat. Commun.}\ }\textbf {\bibinfo {volume} {7}},\
  \bibinfo {pages} {11227} (\bibinfo {year} {2016})}\BibitemShut {NoStop}%
\bibitem [{\citenamefont {Hosoito}\ \emph {et~al.}(2014)\citenamefont
  {Hosoito}, \citenamefont {Ohkochi}, \citenamefont {Kodama},\ and\
  \citenamefont {Suzuki}}]{hosoito2014charge}%
  \BibitemOpen
  \bibfield  {author} {\bibinfo {author} {\bibfnamefont {N.}~\bibnamefont
  {Hosoito}}, \bibinfo {author} {\bibfnamefont {T.}~\bibnamefont {Ohkochi}},
  \bibinfo {author} {\bibfnamefont {K.}~\bibnamefont {Kodama}}, \ and\ \bibinfo
  {author} {\bibfnamefont {M.}~\bibnamefont {Suzuki}},\ }\href@noop {}
  {\bibfield  {journal} {\bibinfo  {journal} {J. Phys. Soc. Jpn.}\ }\textbf
  {\bibinfo {volume} {83}},\ \bibinfo {pages} {024704} (\bibinfo {year}
  {2014})}\BibitemShut {NoStop}%
\bibitem [{\citenamefont {Felcher}\ \emph {et~al.}(1987)\citenamefont
  {Felcher}, \citenamefont {Hilleke}, \citenamefont {Crawford}, \citenamefont
  {Haumann}, \citenamefont {Kleb},\ and\ \citenamefont
  {Ostrowski}}]{felcher1987polarized}%
  \BibitemOpen
  \bibfield  {author} {\bibinfo {author} {\bibfnamefont {G.}~\bibnamefont
  {Felcher}}, \bibinfo {author} {\bibfnamefont {R.}~\bibnamefont {Hilleke}},
  \bibinfo {author} {\bibfnamefont {R.}~\bibnamefont {Crawford}}, \bibinfo
  {author} {\bibfnamefont {J.}~\bibnamefont {Haumann}}, \bibinfo {author}
  {\bibfnamefont {R.}~\bibnamefont {Kleb}}, \ and\ \bibinfo {author}
  {\bibfnamefont {G.}~\bibnamefont {Ostrowski}},\ }\href@noop {} {\bibfield
  {journal} {\bibinfo  {journal} {Rev. Sci. Instrum.}\ }\textbf {\bibinfo
  {volume} {58}},\ \bibinfo {pages} {609} (\bibinfo {year} {1987})}\BibitemShut
  {NoStop}%
\bibitem [{\citenamefont {Majkrzak}(1996)}]{majkrzak1996neutron}%
  \BibitemOpen
  \bibfield  {author} {\bibinfo {author} {\bibfnamefont {C.}~\bibnamefont
  {Majkrzak}},\ }\href@noop {} {\bibfield  {journal} {\bibinfo  {journal}
  {Physica B}\ }\textbf {\bibinfo {volume} {221}},\ \bibinfo {pages} {342}
  (\bibinfo {year} {1996})}\BibitemShut {NoStop}%
\bibitem [{\citenamefont {Roy}\ \emph {et~al.}(2005)\citenamefont {Roy},
  \citenamefont {Fitzsimmons}, \citenamefont {Park}, \citenamefont {Dorn},
  \citenamefont {Petracic}, \citenamefont {Roshchin}, \citenamefont {Li},
  \citenamefont {Batlle}, \citenamefont {Morales}, \citenamefont {Misra},
  \citenamefont {Zhang}, \citenamefont {Chesnel}, \citenamefont {Kortright},
  \citenamefont {Sinha},\ and\ \citenamefont {Schuller}}]{roy2005depth}%
  \BibitemOpen
  \bibfield  {author} {\bibinfo {author} {\bibfnamefont {S.}~\bibnamefont
  {Roy}}, \bibinfo {author} {\bibfnamefont {M.}~\bibnamefont {Fitzsimmons}},
  \bibinfo {author} {\bibfnamefont {S.}~\bibnamefont {Park}}, \bibinfo {author}
  {\bibfnamefont {M.}~\bibnamefont {Dorn}}, \bibinfo {author} {\bibfnamefont
  {O.}~\bibnamefont {Petracic}}, \bibinfo {author} {\bibfnamefont {I.~V.}\
  \bibnamefont {Roshchin}}, \bibinfo {author} {\bibfnamefont {Z.-P.}\
  \bibnamefont {Li}}, \bibinfo {author} {\bibfnamefont {X.}~\bibnamefont
  {Batlle}}, \bibinfo {author} {\bibfnamefont {R.}~\bibnamefont {Morales}},
  \bibinfo {author} {\bibfnamefont {A.}~\bibnamefont {Misra}}, \bibinfo
  {author} {\bibfnamefont {X.}~\bibnamefont {Zhang}}, \bibinfo {author}
  {\bibfnamefont {K.}~\bibnamefont {Chesnel}}, \bibinfo {author} {\bibfnamefont
  {J.~B.}\ \bibnamefont {Kortright}}, \bibinfo {author} {\bibfnamefont {S.~K.}\
  \bibnamefont {Sinha}}, \ and\ \bibinfo {author} {\bibfnamefont {I.~K.}\
  \bibnamefont {Schuller}},\ }\href@noop {} {\bibfield  {journal} {\bibinfo
  {journal} {Phys. Rev. Lett.}\ }\textbf {\bibinfo {volume} {95}},\ \bibinfo
  {pages} {047201} (\bibinfo {year} {2005})}\BibitemShut {NoStop}%
\bibitem [{\citenamefont {Bj{\"o}rck}\ \emph {et~al.}(2009)\citenamefont
  {Bj{\"o}rck}, \citenamefont {Andersson}, \citenamefont {Sanyal},
  \citenamefont {Hedlund},\ and\ \citenamefont
  {Wildes}}]{bjorck2009segregation}%
  \BibitemOpen
  \bibfield  {author} {\bibinfo {author} {\bibfnamefont {M.}~\bibnamefont
  {Bj{\"o}rck}}, \bibinfo {author} {\bibfnamefont {G.}~\bibnamefont
  {Andersson}}, \bibinfo {author} {\bibfnamefont {B.}~\bibnamefont {Sanyal}},
  \bibinfo {author} {\bibfnamefont {M.}~\bibnamefont {Hedlund}}, \ and\
  \bibinfo {author} {\bibfnamefont {A.}~\bibnamefont {Wildes}},\ }\href@noop {}
  {\bibfield  {journal} {\bibinfo  {journal} {Phys. Rev. B}\ }\textbf {\bibinfo
  {volume} {79}},\ \bibinfo {pages} {085428} (\bibinfo {year}
  {2009})}\BibitemShut {NoStop}%
\bibitem [{\citenamefont {Kravtsov}\ \emph {et~al.}(2009)\citenamefont
  {Kravtsov}, \citenamefont {Haskel}, \citenamefont {Te~Velthuis},
  \citenamefont {Jiang},\ and\ \citenamefont
  {Kirby}}]{kravtsov2009complementary}%
  \BibitemOpen
  \bibfield  {author} {\bibinfo {author} {\bibfnamefont {E.}~\bibnamefont
  {Kravtsov}}, \bibinfo {author} {\bibfnamefont {D.}~\bibnamefont {Haskel}},
  \bibinfo {author} {\bibfnamefont {S.}~\bibnamefont {Te~Velthuis}}, \bibinfo
  {author} {\bibfnamefont {J.}~\bibnamefont {Jiang}}, \ and\ \bibinfo {author}
  {\bibfnamefont {B.}~\bibnamefont {Kirby}},\ }\href@noop {} {\bibfield
  {journal} {\bibinfo  {journal} {Phys. Rev. B}\ }\textbf {\bibinfo {volume}
  {79}},\ \bibinfo {pages} {134438} (\bibinfo {year} {2009})}\BibitemShut
  {NoStop}%
\bibitem [{\citenamefont {Zafar}\ \emph {et~al.}(2011)\citenamefont {Zafar},
  \citenamefont {Audehm}, \citenamefont {Sch{\"u}tz}, \citenamefont {Goering},
  \citenamefont {Pathak}, \citenamefont {Chetry}, \citenamefont {LeClair},\
  and\ \citenamefont {Gupta}}]{zafar2011cr}%
  \BibitemOpen
  \bibfield  {author} {\bibinfo {author} {\bibfnamefont {K.}~\bibnamefont
  {Zafar}}, \bibinfo {author} {\bibfnamefont {P.}~\bibnamefont {Audehm}},
  \bibinfo {author} {\bibfnamefont {G.}~\bibnamefont {Sch{\"u}tz}}, \bibinfo
  {author} {\bibfnamefont {E.}~\bibnamefont {Goering}}, \bibinfo {author}
  {\bibfnamefont {M.}~\bibnamefont {Pathak}}, \bibinfo {author} {\bibfnamefont
  {K.}~\bibnamefont {Chetry}}, \bibinfo {author} {\bibfnamefont
  {P.}~\bibnamefont {LeClair}}, \ and\ \bibinfo {author} {\bibfnamefont
  {A.}~\bibnamefont {Gupta}},\ }\href@noop {} {\bibfield  {journal} {\bibinfo
  {journal} {Phys. Rev. B}\ }\textbf {\bibinfo {volume} {84}},\ \bibinfo
  {pages} {134412} (\bibinfo {year} {2011})}\BibitemShut {NoStop}%
\bibitem [{\citenamefont {Sch{\"u}tz}\ \emph {et~al.}(1990)\citenamefont
  {Sch{\"u}tz}, \citenamefont {Wienke}, \citenamefont {Wilhelm}, \citenamefont
  {Zeper}, \citenamefont {Ebert},\ and\ \citenamefont
  {Sp{\"o}rl}}]{schutz1990spin}%
  \BibitemOpen
  \bibfield  {author} {\bibinfo {author} {\bibfnamefont {G.}~\bibnamefont
  {Sch{\"u}tz}}, \bibinfo {author} {\bibfnamefont {R.}~\bibnamefont {Wienke}},
  \bibinfo {author} {\bibfnamefont {W.}~\bibnamefont {Wilhelm}}, \bibinfo
  {author} {\bibfnamefont {W.}~\bibnamefont {Zeper}}, \bibinfo {author}
  {\bibfnamefont {H.}~\bibnamefont {Ebert}}, \ and\ \bibinfo {author}
  {\bibfnamefont {K.}~\bibnamefont {Sp{\"o}rl}},\ }\href@noop {} {\bibfield
  {journal} {\bibinfo  {journal} {‎J. Appl. Phys.}\ }\textbf {\bibinfo
  {volume} {67}},\ \bibinfo {pages} {4456} (\bibinfo {year}
  {1990})}\BibitemShut {NoStop}%
\bibitem [{\citenamefont {Nevot}\ and\ \citenamefont
  {Croce}(1980)}]{nevot1980caracterisation}%
  \BibitemOpen
  \bibfield  {author} {\bibinfo {author} {\bibfnamefont {L.}~\bibnamefont
  {Nevot}}\ and\ \bibinfo {author} {\bibfnamefont {P.}~\bibnamefont {Croce}},\
  }\href@noop {} {\bibfield  {journal} {\bibinfo  {journal} {Rev. Phys. Appl.}\
  }\textbf {\bibinfo {volume} {15}},\ \bibinfo {pages} {761} (\bibinfo {year}
  {1980})}\BibitemShut {NoStop}%
\bibitem [{\citenamefont {Zak}\ \emph {et~al.}(1990)\citenamefont {Zak},
  \citenamefont {Moog}, \citenamefont {Liu},\ and\ \citenamefont
  {Bader}}]{zak1990universal}%
  \BibitemOpen
  \bibfield  {author} {\bibinfo {author} {\bibfnamefont {J.}~\bibnamefont
  {Zak}}, \bibinfo {author} {\bibfnamefont {E.}~\bibnamefont {Moog}}, \bibinfo
  {author} {\bibfnamefont {C.}~\bibnamefont {Liu}}, \ and\ \bibinfo {author}
  {\bibfnamefont {S.}~\bibnamefont {Bader}},\ }\href@noop {} {\bibfield
  {journal} {\bibinfo  {journal} {J. Magn. Magn. Mater.}\ }\textbf {\bibinfo
  {volume} {89}},\ \bibinfo {pages} {107} (\bibinfo {year} {1990})}\BibitemShut
  {NoStop}%
\bibitem [{\citenamefont {Lagarias}\ \emph {et~al.}(1998)\citenamefont
  {Lagarias}, \citenamefont {Reeds}, \citenamefont {Wright},\ and\
  \citenamefont {Wright}}]{lagarias1998convergence}%
  \BibitemOpen
  \bibfield  {author} {\bibinfo {author} {\bibfnamefont {J.~C.}\ \bibnamefont
  {Lagarias}}, \bibinfo {author} {\bibfnamefont {J.~A.}\ \bibnamefont {Reeds}},
  \bibinfo {author} {\bibfnamefont {M.~H.}\ \bibnamefont {Wright}}, \ and\
  \bibinfo {author} {\bibfnamefont {P.~E.}\ \bibnamefont {Wright}},\
  }\href@noop {} {\bibfield  {journal} {\bibinfo  {journal} {SIAM J. Optim.}\
  }\textbf {\bibinfo {volume} {9}},\ \bibinfo {pages} {112} (\bibinfo {year}
  {1998})}\BibitemShut {NoStop}%
\bibitem [{\citenamefont {Tiilikainen}\ \emph {et~al.}(2007)\citenamefont
  {Tiilikainen}, \citenamefont {Bosund}, \citenamefont {Tilli}, \citenamefont
  {Sormunen}, \citenamefont {Mattila}, \citenamefont {Hakkarainen},\ and\
  \citenamefont {Lipsanen}}]{tiilikainen2007genetic}%
  \BibitemOpen
  \bibfield  {author} {\bibinfo {author} {\bibfnamefont {J.}~\bibnamefont
  {Tiilikainen}}, \bibinfo {author} {\bibfnamefont {V.}~\bibnamefont {Bosund}},
  \bibinfo {author} {\bibfnamefont {J.-M.}\ \bibnamefont {Tilli}}, \bibinfo
  {author} {\bibfnamefont {J.}~\bibnamefont {Sormunen}}, \bibinfo {author}
  {\bibfnamefont {M.}~\bibnamefont {Mattila}}, \bibinfo {author} {\bibfnamefont
  {T.}~\bibnamefont {Hakkarainen}}, \ and\ \bibinfo {author} {\bibfnamefont
  {H.}~\bibnamefont {Lipsanen}},\ }\href@noop {} {\bibfield  {journal}
  {\bibinfo  {journal} {J. Phys. D: Appl. Phys.}\ }\textbf {\bibinfo {volume}
  {40}},\ \bibinfo {pages} {6000} (\bibinfo {year} {2007})}\BibitemShut
  {NoStop}%
\bibitem [{\citenamefont {Tiilikainen}\ \emph {et~al.}(2006)\citenamefont
  {Tiilikainen}, \citenamefont {Tilli}, \citenamefont {Bosund}, \citenamefont
  {Mattila}, \citenamefont {Hakkarainen}, \citenamefont {Airaksinen},\ and\
  \citenamefont {Lipsanen}}]{tiilikainen2006nonlinear}%
  \BibitemOpen
  \bibfield  {author} {\bibinfo {author} {\bibfnamefont {J.}~\bibnamefont
  {Tiilikainen}}, \bibinfo {author} {\bibfnamefont {J.}~\bibnamefont {Tilli}},
  \bibinfo {author} {\bibfnamefont {V.}~\bibnamefont {Bosund}}, \bibinfo
  {author} {\bibfnamefont {M.}~\bibnamefont {Mattila}}, \bibinfo {author}
  {\bibfnamefont {T.}~\bibnamefont {Hakkarainen}}, \bibinfo {author}
  {\bibfnamefont {V.-M.}\ \bibnamefont {Airaksinen}}, \ and\ \bibinfo {author}
  {\bibfnamefont {H.}~\bibnamefont {Lipsanen}},\ }\href@noop {} {\bibfield
  {journal} {\bibinfo  {journal} {J. Phys. D: Appl. Phys.}\ }\textbf {\bibinfo
  {volume} {40}},\ \bibinfo {pages} {215} (\bibinfo {year} {2006})}\BibitemShut
  {NoStop}%
\bibitem [{\citenamefont {Ulyanenkov}\ and\ \citenamefont
  {Sobolewski}(2005)}]{ulyanenkov2005extended}%
  \BibitemOpen
  \bibfield  {author} {\bibinfo {author} {\bibfnamefont {A.}~\bibnamefont
  {Ulyanenkov}}\ and\ \bibinfo {author} {\bibfnamefont {S.}~\bibnamefont
  {Sobolewski}},\ }\href@noop {} {\bibfield  {journal} {\bibinfo  {journal} {J.
  Phys. D: Appl. Phys.}\ }\textbf {\bibinfo {volume} {38}},\ \bibinfo {pages}
  {A235} (\bibinfo {year} {2005})}\BibitemShut {NoStop}%
\bibitem [{\citenamefont {Hannon}\ \emph {et~al.}(1988)\citenamefont {Hannon},
  \citenamefont {Trammell}, \citenamefont {Blume},\ and\ \citenamefont
  {Gibbs}}]{hannon1988x}%
  \BibitemOpen
  \bibfield  {author} {\bibinfo {author} {\bibfnamefont {J.}~\bibnamefont
  {Hannon}}, \bibinfo {author} {\bibfnamefont {G.}~\bibnamefont {Trammell}},
  \bibinfo {author} {\bibfnamefont {M.}~\bibnamefont {Blume}}, \ and\ \bibinfo
  {author} {\bibfnamefont {D.}~\bibnamefont {Gibbs}},\ }\href@noop {}
  {\bibfield  {journal} {\bibinfo  {journal} {Phys. Rev. Lett.}\ }\textbf
  {\bibinfo {volume} {61}},\ \bibinfo {pages} {1245} (\bibinfo {year}
  {1988})}\BibitemShut {NoStop}%
\bibitem [{\citenamefont
  {Kiessig}(1931{\natexlab{a}})}]{kiessig1931untersuchungen}%
  \BibitemOpen
  \bibfield  {author} {\bibinfo {author} {\bibfnamefont {H.}~\bibnamefont
  {Kiessig}},\ }\href@noop {} {\bibfield  {journal} {\bibinfo  {journal} {Ann.
  Phys.}\ }\textbf {\bibinfo {volume} {402}},\ \bibinfo {pages} {715} (\bibinfo
  {year} {1931}{\natexlab{a}})}\BibitemShut {NoStop}%
\bibitem [{\citenamefont
  {Kiessig}(1931{\natexlab{b}})}]{kiessig1931interferenz}%
  \BibitemOpen
  \bibfield  {author} {\bibinfo {author} {\bibfnamefont {H.}~\bibnamefont
  {Kiessig}},\ }\href@noop {} {\bibfield  {journal} {\bibinfo  {journal} {Ann.
  Phys.}\ }\textbf {\bibinfo {volume} {402}},\ \bibinfo {pages} {769} (\bibinfo
  {year} {1931}{\natexlab{b}})}\BibitemShut {NoStop}%
\bibitem [{\citenamefont {Kim}\ \emph {et~al.}(2016)\citenamefont {Kim},
  \citenamefont {Song}, \citenamefont {Choi}, \citenamefont {Min},
  \citenamefont {Kim}, \citenamefont {Choi},\ and\ \citenamefont
  {Lee}}]{kim2016asymmetric}%
  \BibitemOpen
  \bibfield  {author} {\bibinfo {author} {\bibfnamefont {D.-O.}\ \bibnamefont
  {Kim}}, \bibinfo {author} {\bibfnamefont {K.~M.}\ \bibnamefont {Song}},
  \bibinfo {author} {\bibfnamefont {Y.}~\bibnamefont {Choi}}, \bibinfo {author}
  {\bibfnamefont {B.-C.}\ \bibnamefont {Min}}, \bibinfo {author} {\bibfnamefont
  {J.-S.}\ \bibnamefont {Kim}}, \bibinfo {author} {\bibfnamefont {J.~W.}\
  \bibnamefont {Choi}}, \ and\ \bibinfo {author} {\bibfnamefont {D.~R.}\
  \bibnamefont {Lee}},\ }\href@noop {} {\bibfield  {journal} {\bibinfo
  {journal} {Sci. Rep.}\ }\textbf {\bibinfo {volume} {6}},\ \bibinfo {pages}
  {25391} (\bibinfo {year} {2016})}\BibitemShut {NoStop}%
\bibitem [{\citenamefont {Henke}\ \emph {et~al.}(1993)\citenamefont {Henke},
  \citenamefont {Gullikson},\ and\ \citenamefont {Davis}}]{henke1993x}%
  \BibitemOpen
  \bibfield  {author} {\bibinfo {author} {\bibfnamefont {B.~L.}\ \bibnamefont
  {Henke}}, \bibinfo {author} {\bibfnamefont {E.~M.}\ \bibnamefont
  {Gullikson}}, \ and\ \bibinfo {author} {\bibfnamefont {J.~C.}\ \bibnamefont
  {Davis}},\ }\href@noop {} {\bibfield  {journal} {\bibinfo  {journal} {At.
  Data Nucl. Tables}\ }\textbf {\bibinfo {volume} {54}},\ \bibinfo {pages}
  {181} (\bibinfo {year} {1993})}\BibitemShut {NoStop}%
\bibitem [{\citenamefont {Chantler}(2000)}]{chantler2000detailed}%
  \BibitemOpen
  \bibfield  {author} {\bibinfo {author} {\bibfnamefont {C.~T.}\ \bibnamefont
  {Chantler}},\ }\href@noop {} {\bibfield  {journal} {\bibinfo  {journal} {J.
  Phys. Chem. Ref. Data}\ }\textbf {\bibinfo {volume} {29}},\ \bibinfo {pages}
  {597} (\bibinfo {year} {2000})}\BibitemShut {NoStop}%
\bibitem [{\citenamefont {Brown}\ \emph {et~al.}(2001)\citenamefont {Brown},
  \citenamefont {Bouchenoire}, \citenamefont {Bowyer}, \citenamefont {Kervin},
  \citenamefont {Laundy}, \citenamefont {Longfield}, \citenamefont {Mannix},
  \citenamefont {Paul}, \citenamefont {Stunault}, \citenamefont {Thompson},
  \citenamefont {Cooper}, \citenamefont {Lucas},\ and\ \citenamefont
  {Stirling}}]{brown2001xmas}%
  \BibitemOpen
  \bibfield  {author} {\bibinfo {author} {\bibfnamefont {S.}~\bibnamefont
  {Brown}}, \bibinfo {author} {\bibfnamefont {L.}~\bibnamefont {Bouchenoire}},
  \bibinfo {author} {\bibfnamefont {D.}~\bibnamefont {Bowyer}}, \bibinfo
  {author} {\bibfnamefont {J.}~\bibnamefont {Kervin}}, \bibinfo {author}
  {\bibfnamefont {D.}~\bibnamefont {Laundy}}, \bibinfo {author} {\bibfnamefont
  {M.}~\bibnamefont {Longfield}}, \bibinfo {author} {\bibfnamefont
  {D.}~\bibnamefont {Mannix}}, \bibinfo {author} {\bibfnamefont
  {D.}~\bibnamefont {Paul}}, \bibinfo {author} {\bibfnamefont {A.}~\bibnamefont
  {Stunault}}, \bibinfo {author} {\bibfnamefont {P.}~\bibnamefont {Thompson}},
  \bibinfo {author} {\bibfnamefont {M.~J.}\ \bibnamefont {Cooper}}, \bibinfo
  {author} {\bibfnamefont {C.~A.}\ \bibnamefont {Lucas}}, \ and\ \bibinfo
  {author} {\bibfnamefont {W.~G.}\ \bibnamefont {Stirling}},\ }\href@noop {}
  {\bibfield  {journal} {\bibinfo  {journal} {J. Synchrotron Radiat.}\ }\textbf
  {\bibinfo {volume} {8}},\ \bibinfo {pages} {1172} (\bibinfo {year}
  {2001})}\BibitemShut {NoStop}%
\bibitem [{\citenamefont {Bouchenoire}(2003)}]{bouchenoire2003}%
  \BibitemOpen
  \bibfield  {author} {\bibinfo {author} {\bibfnamefont {L.}~\bibnamefont
  {Bouchenoire}},\ }\href@noop {} {\bibfield  {journal} {\bibinfo  {journal}
  {J. Synchrotron Radiat.}\ }\textbf {\bibinfo {volume} {10}},\ \bibinfo
  {pages} {172} (\bibinfo {year} {2003})}\BibitemShut {NoStop}%
\bibitem [{\citenamefont {Bouchenoire}\ \emph {et~al.}(2007)\citenamefont
  {Bouchenoire}, \citenamefont {Brown}, \citenamefont {Thompson}, \citenamefont
  {Cain}, \citenamefont {Stewart},\ and\ \citenamefont
  {Cooper}}]{bouchenoire2007development}%
  \BibitemOpen
  \bibfield  {author} {\bibinfo {author} {\bibfnamefont {L.}~\bibnamefont
  {Bouchenoire}}, \bibinfo {author} {\bibfnamefont {S.}~\bibnamefont {Brown}},
  \bibinfo {author} {\bibfnamefont {P.}~\bibnamefont {Thompson}}, \bibinfo
  {author} {\bibfnamefont {M.}~\bibnamefont {Cain}}, \bibinfo {author}
  {\bibfnamefont {M.}~\bibnamefont {Stewart}}, \ and\ \bibinfo {author}
  {\bibfnamefont {M.}~\bibnamefont {Cooper}},\ }\href@noop {} {\bibfield
  {journal} {\bibinfo  {journal} {AIP Conf. Proc.}\ }\textbf {\bibinfo {volume}
  {879}},\ \bibinfo {pages} {1679} (\bibinfo {year} {2007})}\BibitemShut
  {NoStop}%
\bibitem [{\citenamefont {Bougiatioti}\ \emph {et~al.}(2018)\citenamefont
  {Bougiatioti}, \citenamefont {Manos}, \citenamefont {Kuschel}, \citenamefont
  {Wollschl{\"a}ger}, \citenamefont {Tolkiehn}, \citenamefont {Francoual},\
  and\ \citenamefont {Kuschel}}]{bougiatioti2018impact}%
  \BibitemOpen
  \bibfield  {author} {\bibinfo {author} {\bibfnamefont {P.}~\bibnamefont
  {Bougiatioti}}, \bibinfo {author} {\bibfnamefont {O.}~\bibnamefont {Manos}},
  \bibinfo {author} {\bibfnamefont {O.}~\bibnamefont {Kuschel}}, \bibinfo
  {author} {\bibfnamefont {J.}~\bibnamefont {Wollschl{\"a}ger}}, \bibinfo
  {author} {\bibfnamefont {M.}~\bibnamefont {Tolkiehn}}, \bibinfo {author}
  {\bibfnamefont {S.}~\bibnamefont {Francoual}}, \ and\ \bibinfo {author}
  {\bibfnamefont {T.}~\bibnamefont {Kuschel}},\ }\href@noop {} {\bibfield
  {journal} {\bibinfo  {journal} {arXiv:1807.09032}\ } (\bibinfo {year}
  {2018})}\BibitemShut {NoStop}%
\bibitem [{\citenamefont {Olsson}\ and\ \citenamefont
  {Nelson}(1975)}]{olsson1975nelder}%
  \BibitemOpen
  \bibfield  {author} {\bibinfo {author} {\bibfnamefont {D.~M.}\ \bibnamefont
  {Olsson}}\ and\ \bibinfo {author} {\bibfnamefont {L.~S.}\ \bibnamefont
  {Nelson}},\ }\href@noop {} {\bibfield  {journal} {\bibinfo  {journal}
  {Technometrics}\ }\textbf {\bibinfo {volume} {17}},\ \bibinfo {pages} {45}
  (\bibinfo {year} {1975})}\BibitemShut {NoStop}%
\end{thebibliography}%

\end{document}